\ifdefined\DoubleSided
  \documentclass[twoside,openright,a4paper]{nusthesis}
\else
  \documentclass[a4paper]{nusthesis}
\fi

\dsp 

\usepackage{indentfirst} 
\usepackage{ebgaramond}
\usepackage{silence} 
\usepackage{todonotes}
\usepackage{bookmark}
\usepackage{cleveref}
\usepackage[
  backend=biber,
  maxnames=999, 
  minnames=1,
  style=ieee,
  citestyle=numeric-comp,
  giveninits=true,
  sorting=none,
  sortcites=true,
  maxbibnames=99,
  dashed=false,
  doi=false,
  url=false]{biblatex}

\DeclareFieldFormat*{title}{``#1''\newunitpunct} 
\usepackage{microtype} 
\usepackage{tabulary} 
\usepackage{hhline}
\usepackage{braket}
\usepackage{enumitem}
\usepackage{booktabs} 
\usepackage{makecell} 

\usepackage{hyperref}

\usepackage{listings}

\usepackage{amsthm}
\theoremstyle{definition}
 
\theoremstyle{plain}

\usepackage{mathtools}

\usepackage{interval}
\intervalconfig{soft open fences}

\usepackage[linesnumbered,ruled,vlined]{algorithm2e}
\SetAlgorithmName{Algorithm}{Algorithm}{Algorithm}
\IncMargin{0.5em}
\SetCommentSty{textnormal}
\SetNlSty{}{}{:}
\SetAlgoNlRelativeSize{0}
\SetKwInput{KwGlobal}{Global}
\SetKwInput{KwPrecondition}{Precondition}
\SetKwProg{Proc}{Procedure}{:}{}
\SetKwProg{Func}{Function}{:}{}
\SetKw{And}{and}
\SetKw{Or}{or}
\SetKw{To}{to}
\SetKw{DownTo}{downto}
\SetKw{Break}{break}
\SetKw{Continue}{continue}
\SetKw{SuchThat}{\textit{s.t.}}
\SetKw{WithRespectTo}{\textit{wrt}}
\SetKw{Iff}{\textit{iff.}}
\SetKw{MaxOf}{\textit{max of}}
\SetKw{MinOf}{\textit{min of}}
\SetKwBlock{Match}{match}{}{}

\usepackage{array}
\newcolumntype{L}[1]{>{\raggedright\let\newline\\\arraybackslash\hspace{0pt}}m{#1}}
\newcolumntype{C}[1]{>{\centering\let\newline\\\arraybackslash\hspace{0pt}}m{#1}}
\newcolumntype{R}[1]{>{\raggedleft\let\newline\\\arraybackslash\hspace{0pt}}m{#1}}

\usepackage{color}
\usepackage[usenames,dvipsnames]{xcolor}
\usepackage{soul}
\soulregister\cite7
\soulregister\ref7
\soulregister\pageref7
\soulregister\autoref7
\soulregister\eqref7

\renewcommand{\eqref}{Equation~\ref}

\newcommand*{\RootPicDir}{pic}
\newcommand*{\PicDir}{\RootPicDir}

\newcommand*{\SetPicSubDir}[1]{\renewcommand*{\PicDir}{\RootPicDir /#1}}

\newcommand*{\RootExpDir}{exp}
\newcommand*{\ExpDir}{\RootExpDir}

\newcommand*{\SetExpSubDir}[1]{\renewcommand*{\ExpDir}{\RootExpDir /#1}}

\usepackage{lipsum} 

\usepackage{fancyhdr}
\usepackage{etoolbox}

\makeatletter

\newcommand{\nusth@currentchaptertitle}{}
\newcommand{\nusth@lottablechapter}{} 

\apptocmd{\@chapter}{%
  \gdef\nusth@currentchaptertitle{#2}%
  \gdef\nusth@lottablechapter{}%
  \addtocontents{lof}{\protect\addvspace{0.7\baselineskip}}%
  \addtocontents{lof}{%
    \protect\contentsline{chapter}{%
      {\headingfont\bfseries\protect\numberline{\thechapter}#2}%
    }{}{}%
  }%
}{}{}

\newcommand{\nusth@maybeaddchaptertolot}{%
  \begingroup
    \edef\nusth@thischapter{\thechapter}%
    \ifx\nusth@thischapter\nusth@lottablechapter
    \else
      \addtocontents{lot}{\protect\addvspace{0.7\baselineskip}}%
      \addtocontents{lot}{%
        \protect\contentsline{chapter}{%
          {\headingfont\bfseries\protect\numberline{\thechapter}\nusth@currentchaptertitle}%
        }{}{}%
      }%
      \xdef\nusth@lottablechapter{\nusth@thischapter}%
    \fi
  \endgroup
}

\let\nusth@origtable\table
\let\nusth@endorigtable\endtable
\renewenvironment{table}[1][]{%
  \nusth@maybeaddchaptertolot
  \nusth@origtable[#1]%
}{%
  \nusth@endorigtable
}

\makeatother

\declaredate{19 February 2026}
\begin{document}

\title{Dynamical light-matter coupling with protected superconducting oscillators}

\author{Fernando Valadares Calheiros de Siqueira}

\degree{Doctor of Philosophy}
\field{Quantum Physics}
\degreeyear{February 2026}
\supervisor{Assistant Professor Yvonne Y. Gao}

\examiners{%
  Professor Alexander Ling\\
  Associate Professor Murray Barrett \\
  Dr Yao Lu, Fermilab}


\pagestyle{plain}

\maketitle

\begin{frontmatter}
  \tableofcontents  
  \begin{abstract}

Superconducting circuits hosting long-lived bosonic modes are an excellent platform for emulating the dynamics of light and matter and exploiting them for quantum information processing. The space of accessible dynamics is greatly extended by introducing the frequency tunability of artificial atoms, whereby the interactions with the light field of a microwave cavity can be modified on demand. However, using external flux bias to control the frequency of transmons introduces significant losses to coupled long-lived cavities, reducing their single-photon lifetimes to a few tens of $\mu\textnormal{s}$. These imperfections compromise the usefulness of the bosonic mode, which loses its ability to store and manipulate the quantum information of many-photon states. This thesis investigates the issue, and demonstrates how to enhance the flux control of microwave cavities without compromising protection. Several hardware solutions were explored, including off-plane remote flux sources routed by a magnetic hose, and lithographically printed on-chip lines equipped with multiple stages of filtering. These developments equipped the systems with broadband flux control and photon lifetimes of several hundreds of $\mu\textnormal{s}$. They enabled the real time manipulation of the two central parameters of light-matter interactions, as described by the Rabi model: the detuning between the two subsystems and their coupling strength. This added flexibility translated into operational primitives of interest for quantum information processing, such as the calibration of nonlinearities to different interaction regimes, the activation of parametric resonances with adjustable coupling strengths, and the use of number-selective transmon-cavity transitions for the state preparation and control of a qudit encoded in the cavity. Together, these results provide a stepping stone for the exploration of more complex synthetic light-matter systems. Among many other topics, the strategies produced in this work are a promising route for harnessing more complex nonlinear circuits for the cavity control, as well as for creating driven-dissipative systems with applications in quantum computation and sensing.

\end{abstract}

  \begin{acknowledgments}

Working towards a PhD is meant to be challenging. But the greatest difficulties are sometimes those we least expect. For me, that was moving across the globe, far away from my family and home, with no return date. Fortunately, I found a community in Singapore that made this journey much more pleasant, and within which I could grow both as a researcher and as a person.

First and foremost, I would like to thank Professor \textbf{Yvonne Y. Gao}. She believed in my potential and gave me the opportunity to join her group. In turn, I believed in her vision of building a new research laboratory, which she has accomplished masterfully. She assembled an exceptional group of people, provided us with the resources and support we needed, and continuously encouraged us to be the best version of ourselves. To me, she exemplifies what a PhD supervisor should be.

The early days in the new laboratory were some of the most difficult. There was a lot to learn. But soon I found in \textbf{Jonathan Schwinger} and \textbf{Atharv Joshi} a shared excitement for getting the most out of those first qubits. More importantly, I learned from them how to think like a physicist - rationally but intuitively, top-down as well as bottom-up - rather than relying excessively on maths. Many of the results reported in this thesis can be traced back to that way of thinking that I got from those two.

Likewise, I learned much from \textbf{Adrian Copetudo}. His eagerness to learn all about circuit QED and his knack for research made us great partners in discussing science and in figuring out what it means to be a researcher.

Speaking of research partners, I must thank \textbf{Aleksandr Dorogov}, \textbf{Candace Ni-Ni Huang} and \textbf{Kyle Chu} for their contributions to the results in this thesis. Before starting my PhD, I had little experience working in a group. Working with them taught me how valuable it is to have trustworthy people to catch my mistakes, challenge my assumptions, and contribute with diverse skills to the project. As I leave the group, I am happy to see that our line of research is in good hands and will be smoothly continued by great new students such as \textbf{Héctor Calero Mas}. These people are an example of hard work and utmost commitment, and I’m excited to see what they will do next.

I owe much of what I learned about teamwork to \textbf{Pengtao Song} and \textbf{Tanjung Krisnanda}. Each in their own way, they are exactly what a PhD student needs from a postdoc: they are dependable, experienced, and generous people. I am thankful for the guidance and for anchoring the group when lab work seemed most difficult.

My time with my colleagues was much more than just work. \textbf{Weipin Chua} was the person who taught me the most about Singapore and Singlish, and how to live in this new place. At the same time, he was always curious about my home in Recife and our strange customs and stories. Thanks to him, Brazil and Singapore don’t seem that far away anymore.

Out of everyone, \textbf{Clara Fontaine} is the one most similar to me. Even before we met, we could predict what was in each other’s minds in virtual meetings, or later with just an exchange of glances. Still, she was always the person I could turn to for fresh perspectives and for finding rational solutions when emotions ran high.

Whether things were going well or not, \textbf{Celine Trieu} and \textbf{Amon Kasper} were always there to make PhD life more fun. From parties and trips we took together, down to our daily chit-chatting and coffee breaks, I will miss being around them.

There were many other people who contributed to make this a unique experience. The group wouldn't be the same without \textbf{May Chee Loke} and \textbf{Yingshan Zhang}. And it wouldn't have become what it did without many of our colleagues that have left, but still come back now and then. I hope to be back for a visit soon as well.

I would also like to acknowledge the institutional support I received from the Centre for Quantum Technologies. The open-door policy of the director \textbf{José Ignacio Latorre} and other professors allowed us to reach out at any time for support and collaboration, even for projects outside our primary research. The same can be said about \textbf{Jenny Hogan} and \textbf{Puah Xin Yi} from the Outreach team, always doing an incredible job supporting the organization of QCamp and other events for students to participate in. I also would like to thank \textbf{Teo Kok Seng} for his excellent work at the CQT machine shop, consistently providing us with high-quality machined aluminum packages for our experiments.

Finally, I want to thank my family, who have always given me unconditional support in following my passions. They always did their best to be with me, no matter where I was. I was very fortunate to have so many of them in Singapore for my thesis defense. And, of course, I am grateful to my wife \textbf{Chin Chai Yee} for her loving and unwavering support, and for her understanding during the many evenings I worked late. I am excited for the next steps in our life together, starting today.

\end{acknowledgments}

  \begin{refsection}
\pagestyle{plain}
\chapter*{List of preprints and publications}
\addcontentsline{toc}{chapter}{List of preprints and publications}

\nocite{%
  valadares_flux_2025,
  valadares_ondemand_2024,
  copetudo_shaping_2024,
  krisnanda_direct_2025,
}

\subsection*{Publications constituting the main results of this thesis}


\begin{enumerate}
\item \fullcite{valadares_ondemand_2024}
\item \fullcite{valadares_flux_2026}
\end{enumerate}

\subsection*{Related publications with major contribution}

\begin{enumerate}
\setcounter{enumi}{2}  
\item \fullcite{copetudo_shaping_2024}
\item \fullcite{krisnanda_direct_2025}
\end{enumerate}

\subsection*{Other contributions}

\begin{enumerate}
\setcounter{enumi}{4}  
\item \fullcite{pan_protecting_2023}
\item \fullcite{krisnanda_demonstrating_2025}
\end{enumerate}

\thispagestyle{plain}
\end{refsection}
  \listoffigures
  \listoftables
\end{frontmatter}

\pagestyle{fancy}
\SetPicSubDir{ch-Intro}
\chapter{Introduction}

Every act of exploration starts from what is known. Drawing inspiration from nature or from the works of our predecessors, we build theories, models and devices that reproduce the phenomena we wish to understand. It follows that, somewhat paradoxically, the first step of any scientific endeavor is to copy. However, each new iteration brings new twists and perspectives, enabling us to take one step further at a time and explore ideas that were previously out of reach.

When the field of cavity quantum electrodynamics (QED) emerged during the 1980s and the 1990s~\cite{haroche_cavity_1989, haroche_short_2007}, its central idea was to encapsulate (quite literally) interactions between light and matter. Making use of a cavity built from reflective surfaces, cavity QED confined the continuum of propagating light modes of free space into a finite set of stationary oscillations. Atoms could be inserted into this structure one at a time, coupling with the light field without interference from interatomic effects. The atom and cavity could then interact cleanly, free of the complexities of many-body phenomena, and following the simple yet rich dynamics of the Rabi model~\cite{rabi_space_1937}, with Hamiltonian
\begin{equation}
\label{eq:rabi}
    \hat{H}_{int} = \hbar g\hat{\sigma}_x \left(\hat{a}^\dagger + \hat{a}\right).
\end{equation}
This model is the blueprint of light-matter interactions. Atomic transitions are represented by the Pauli matrix $\hat{\sigma}_x$, which couple to the cavity field given by the ladder operators $\hat{a}$, $\hat{a}^\dagger$. This model may describe the interaction of a dipole moment with an electric field, as is the case in cavity QED; and may just as well describe the interaction of a magnetic field impinging upon an atomic nuclear spin. Equipped with such rich dynamics, the clever setup of cavity QED allowed the observation of emblematic predictions of the quantum theory of radiation, such as the Purcell effect~\cite{goy_observation_1983, hulet_inhibited_1985} and the Lamb shift~\cite{heinzen_vacuum_1987}. 

As the field progressed, improvements in cavity quality and design allowed the atom to interact more strongly with the confined light. Interactions could be enhanced to be faster than the energy relaxation rates of the system, so as to imprint the effects of coherent quantum dynamics onto the state evolution before it could degrade. From here on, cavity QED moved beyond the domain of macroscopic setups and semi-classical formulations, and entered the single-photon regime. This opened the way to demonstrating coherent exchange of energy between the atom and the cavity with vacuum Rabi oscillations~\cite{meschede_one_1985}, strong photon-photon interactions through photon blockade~\cite{birnbaum_photon_2005}, and the creation of light field states with well-defined excitation numbers, known as Fock states~\cite{brattke_generation_2001}. So, by capturing the natural phenomena of light-matter interactions while abstracting its complexities, cavity QED provided the means to observe quantum dynamics with unprecedented resolution: one photon at a time.

The unprecedented fine-grained control of cavity QED enabled the exploration of quantum phenomena from the perspective of information theory. The fundamental unit of quantum information, the ``quantum bit'' (or \textit{qubit}, for short), could be implemented using a pair of energy levels of the atom. And qubit operations, described in terms of Pauli operators $\hat{\sigma}_{x,y,z}$, could be intermediated by the light-matter interactions of the Rabi model. The existing technology allowed the early demonstration of interesting quantum computational primitives, such as the creation of entanglement between two~\cite{hagley_generation_1997} or more atoms~\cite{rauschenbeutel_step_2000}, the use of light field states as memory~\cite{maitre_quantum_1997}, the quantum phase gate between the atomic state and the photon~\cite{rauschenbeutel_coherent_1999}, and the non-demolition detection of photons in a cavity~\cite{nogues_seeing_1999}. Despite this track record, the growth of cavity QED systems toward implementing more complex quantum information processing (QIP) tasks was limited. Coherent quantum dynamics fight an uphill battle against the decoherence and relaxation induced by interactions with the environment~\cite{unruh_maintaining_1995, landauer_physical_1996}. And the natural dipole coupling of cavity QED provided a low ceiling for the speed at which these dynamics could be enacted~\cite{haroche_cavity_2020}. Thus, further progress required a qualitative shift in the field of quantum technologies.

In the meantime, superconducting systems were being explored as a platform that could demonstrate quantum effects at a macroscopic scale~\cite{leggett_prospects_1978}. In 1985, Devoret et al. reported the first observation of macroscopic quantum tunneling between the states of a superconducting circuit~\cite{devoret_measurements_1985}. The potential barrier was implemented by a Josephson junction~\cite{josephson_possible_1962}, a circuit element formed by an oxide layer bridging two superconducting terminals. Later, in 1999, Nakamura et al. used two levels of a Cooper-pair box, a Josephson junction circuit, to demonstrate coherent qubit control for the first time in a superconducting platform~\cite{nakamura_coherent_1999}. In light of the growing interest in QIP, the question was posed: could such circuits be employed as artificial atoms, reshaped so as to emulate the light-matter dynamics, and then be used to build large-sized quantum computers?~\cite{blais_cavity_2004}

\begin{figure}
    \centering
    \includegraphics[width=\linewidth]{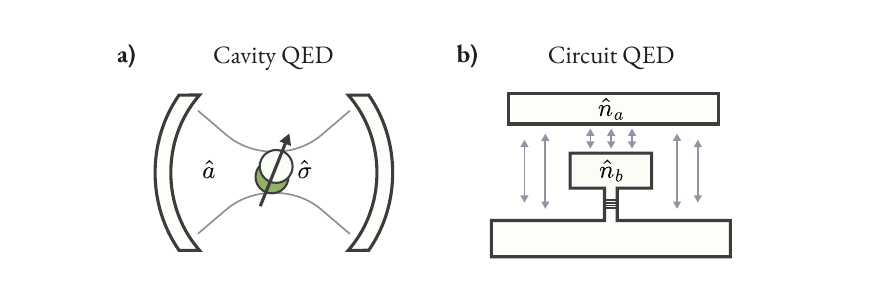}
    \caption[Cavity and circuit quantum electrodynamics]{\textbf{Cavity and circuit quantum electrodynamics}. a) Cavity QED directly probes light-matter interactions by using single atoms inside the confined light fields of cavities. b) Circuit QED emulates this physics by using superconducting artificial atoms strongly coupled to microwave resonators.}
    \label{fig:cavity_and_circuit}
\end{figure}

The pursuit of this question marked the early 2000s with the emergence of the field of circuit quantum electrodynamics. Circuit QED employed superconducting circuits to recreate the ideas of cavity QED with improved engineerability~\cite{blais_circuit_2021}. In this platform, the qubits and light fields of microwave resonators coupled through their charge variables $\hat{n}_a$, $\hat{n}_b$ as
\begin{equation}
    \hat{H}_{int} = \hbar g_C\hat{n}_a\hat{n}_b,
\end{equation}
which can be recast into Equation~\ref{eq:rabi} to evoke the dynamics of light and matter. The coupling factor $g_C$ could be enhanced by resizing the capacitor pads of the qubit to increase the field overlap with the cavity, taking their coupling to levels not trivially achievable with cavity QED. Together with the continuous improvement of the lifetimes of these systems, leading to the development of the transmon qubit~\cite{koch_charge_2007}, circuit QED provided the ability to perform increasingly complex quantum information processing tasks. 

Owing to this change in paradigm, the last decade has seen an impressive evolution of superconducting quantum technologies. The arbitrary control of microwave light fields is now part of the standard experimental toolkit~\cite{ma_quantum_2021}, with techniques that are often expanded to two~\cite{wang_schrodinger_2016, gao_programmable_2018, chapman_high_2023, lu_high_2023} or more modes~\cite{naik_random_2017, huang_fast_2025, schwinger_thesis_2025}. Logical qubits are encoded in the cavities with coherences surpassing those of the host device itself~\cite{sivak_real_2023, ni_beating_2023, ofek_extending_2016, ni_autonomous_2025, sun_extending_2025}, providing clear demonstrations of quantum error correction and offering pathways for fault-tolerant computing. On a separate track, transmon qubits are assembled in lattices containing tens to hundreds of interconnected units, taking great industrial and academic efforts to control progressively larger Hilbert spaces. These systems have been the stage of important demonstrations such as the surface error-correction code~\cite{krinner_realizing_2022, acharya_quantum_2024} and advantageous computational tasks~\cite{abanin_observation_2025}.

With the aim of representing observed physics, cavity and circuit QED have thus become tangible examples of abstraction and idealization processes in science~\cite{chakravartty_truth_2010}. They show how the attempts to copy nature have in fact empowered experiments to not only better understand the dynamics of light and matter, but to take apart its constituent elements and recombine them into synthetic systems beyond those initially known. The objective of the present work is to try and add one more line to this story.

In this thesis, I will take one more step to increase the quality of control over light-matter dynamics in superconducting systems. I will use transmons imbued with Superconducting Quantum Interference Devices (SQUIDs) as tunable artificial atoms. The SQUIDs will be biased by fast-response magnetic flux sources, introducing the real-time control over the circuit eigenstates and transition frequencies. By coupling such circuits to the bosonic modes of microwave cavities, the flux control introduces the possibility of tuning light-matter interactions on demand, significantly expanding the practical uses of the Rabi model for quantum information processing. I will demonstrate how this extra axis of control unfolds into several primitive operations that are useful building blocks for complex quantum dynamics. On one hand, it allows switching between qualitatively distinct interaction regimes, from resonance to decoupling, each offering their own operational advantages. On the other hand, the periodic flux driving of the system allows the in situ modulation of parameters such as the coupling factor $g$, expanding even further the space of accessible dynamics. With these capabilities, I will implement control protocols on the light field of the cavity that are not trivially accessible in fixed-frequency devices. 

Fast frequency tunability by itself is not the central point here. Or at least not all of it. After all, tunable systems involving superconducting resonators are as old as circuit QED itself. But these systems are infamously plagued by large cavity photon loss of one or two orders of magnitude larger than devices coupled to fixed-frequency transmons~\cite{hofheinz_generation_2008, hofheinz_synthesizing_2009, reed_entanglement_2013, gargiulo_fast_2021, majumder_fast_2022}. The effects of loss scale with the number of photons, imposing a significant barrier to harnessing the large Hilbert space of the cavity for quantum information processing. This contradiction must be solved to achieve the full potential of tunable devices. My work will thus be guided by two principles of equal importance: adding fast tunability of interactions by means of controlling the transmon frequency on short timescales, and simultaneously protecting the cavity from the photon loss due to the extra control channels required by tunable bosonic architectures.

By the time this research had started, this was a relatively underexplored topic, much owing to the challenges involved. The final results open the way for exciting new opportunities in tunable bosonic circuit QED experiments, even if many of them could not be addressed during the extent of this Ph.D. candidature. For example, here I work with SQUID-transmons due to their simplicity. However, these results can be extended to other superconducting circuits such as the SNAIL~\cite{frattini_wave_2017}, the fluxonium~\cite{manucharyan_fluxonium_2009}, and the LINC~\cite{maiti_linear_2025}, exploring different nonlinear interactions with the cavity. Similarly, the results are not fundamentally limited to single-cavity systems, and can be adapted to circuits featuring multiple bosonic modes to implement more complex quantum dynamics. Along this thesis, I will discuss these and other possible directions for future research, to which this work should provide a good starting point.

But before moving to the details, let us discuss the overall structure of this thesis.

\section{Thesis Overview}

This thesis starts with the background knowledge necessary to understand and reproduce this research, given in Chapters~\ref{ch:bosonic} and~\ref{ch:flux}. Then, Chapters~\ref{ch:somerset} and~\ref{ch:lakeside} describe its main results.  

Chapter~\ref{ch:bosonic} presents the experimental platform. Here I explain the principles of superconducting circuits in the context of  circuit QED, specifically applied to building high-quality and controllable microwave cavities. I review some different cavity architectures from the literature, focusing on the coaxial stub cavities that will be used in the forthcoming experiments, highlighting experimental considerations. Their theoretical description as harmonic oscillators is explained in detail, and their Hilbert space is reformulated in terms of the phase space. This description provides the grounds for discussing concepts of bosonic quantum information such as the Wigner function and non-gaussianity. Then, I will move on to explaining the physics of Josephson junctions and transmons. In bosonic setups, the transmon takes an auxiliary role. Its purpose is not to be the main holder of quantum information, but instead to provide the nonlinearities that are used for state preparation, control and tomography of the cavity state. I will therefore focus on its relationship with the cavity and how they interact. The last circuit element addressed is the readout resonator, which measures the transmon state and provides the link between the quantum system and the classical control electronics. The chapter will conclude by explaining the simulation and experimental methodologies involved in the design and characterization of bosonic circuit QED devices.

Then I go into the more particular techniques involved in this research. Chapter~\ref{ch:flux} focuses on the physics of SQUIDs and flux-tunable transmons. I describe the behavior of the system in a time-dependent field, providing a good starting point for simplifying approximations and for master equation simulations. I analyze the dynamics in the specific case of parametrically driving the transmon with a periodic flux bias $\propto \cos(\omega t)$. Under these conditions, the effective coupling to the cavity can be modulated, providing an extra degree of control of light-matter interactions. Unfortunately, the transmon flux sensitivity also introduces vulnerability to flux noise, which decoheres the transmon state; to mitigate this issue, I will explain the different types of flux sweet spot and how they can in principle be leveraged to increase transmon coherence times. I also explore how noise and other issues such as heating can be mitigated by thoughtfully designing the microwave signal chain of the flux control, and how distortions from this signal chain can be reverted using digital signal processing. By the end of the chapter, I will elaborate on the difficulties of adding flux tunability to bosonic devices, showing why this is a particularly challenging task. The critical factors discussed here will lay down the problems to be solved in experiment.

Chapter~\ref{ch:somerset} presents the first solution. I experimentally show tunable light-matter interactions in a long-lived bosonic circuit QED device, and apply it to quantum information processing. To mitigate cavity photon loss induced by the flux line, I use a magnetic flux hose~\cite{gargiulo_fast_2021}. The hose's purpose is to conduct the field generated by a remote magnetic field source to bias the transmon while using built-in filters to attenuate cavity and transmon leakage. The optimized device, which I refer to as Somerset throughout this thesis, incorporates the hose while balancing loss, flux delivery, and transmon-cavity coupling strength. Somerset is used for QIP experiments mixing different interaction regimes, fully leveraging the on-demand frequency tunability. In the first experiment, fast activation of transmon-cavity resonance is used to swap excitations one at a time to create Fock states in the light field. Then, the prepared states are characterized with two different tomography methods: the Wigner function, which uses the parity-mapping protocol of the strong-dispersive interaction regime, and the characteristic function, which uses echo-conditional displacement gates that are only accessible in the weak dispersive regime. Lastly, I employ dynamical detuning to turn off interactions, suppressing the harmful nonlinear dynamics that the cavity inherits from the transmon. This chapter's results extend the experimental toolkit of bosonic systems and provide solutions to relevant problems in the field such as Kerr distortions.

These results are as much motivating as they are humbling. Despite its achievements, the device is difficult to reproduce, comes with significant active heating that requires dedicated thermalization strategies, and is still limited in terms of flux strength and bandwidth. Chapter~\ref{ch:lakeside} delivers a prototype that improves over Somerset in all aspects. The new device, which I refer to as Lakeside, uses a flux source printed on-chip designed with a multi-step filtering strategy to protect the cavity against photon loss. The flux source can carry high-frequency pulses, which I use to activate parametrically resonant interactions with adjustable strength and duration. With the upgraded features of Lakeside, the transmon nonlinearities are harnessed to their full extent to implement qudit operations on the cavity. I use number-selective transitions in the resonant regime to create arbitrary bosonic mode states within a well-defined photon truncation. I use multi-tone control to optimize these transitions and implement Givens rotations between two Fock levels of the cavity. These results expand the available tools for the universal control of a light field, offering the possibility of exact, analytical operations that are not readily available in fixed-frequency setups.  

The results of this thesis put forward dynamical light-matter interactions as a feasible resource in superconducting circuits, overcoming long-standing challenges in the field. Still, the prototypes and techniques shown here are just a subset of what should be possible. Further research should be encouraged in the pursuit of even higher-quality devices, more refined control techniques, and creative ways to employ light-matter interactions. Let us prepare the foundations for this future work, starting from the physics of bosonic circuit QED.
\chapter{Bosonic circuit QED}
\label{ch:bosonic}
\vspace{2em}

The physics at the single-particle level put forward by cavity QED kindled interest in building quantum systems into larger, well-controlled assemblies. These scaled systems are of great scientific and technological interest, since they offer several advantages over their classical counterparts. For example, they enable direct access to fundamental many-body phenomena such as information scrambling~\cite{kukuljan_weak_2017, lantagnehurtubise_diagnosing_2020, karamlou_probing_2024}, localization~\cite{smith_localization_2016, rosen_flat_2024}, and quantum phase transitions~\cite{greiner_quantum_2002, greif_site_2016, hofstetter_quantum_2018}. Quantum states with high average excitation number $N$ are a resource for quantum parameter estimation, a central element of sensing: the estimation uncertainty can scale up to the Heisenberg limit~$\propto1/N$~\cite{degen_quantum_2017}, overcoming the classical bound of~$\propto1/\sqrt{N}$. The large Hilbert spaces of multi-particle systems are also useful to encode information nonlocally, protecting it from dominant errors. Error correction codes can then be used to build the highly coherent logical qubits~\cite{shor_scheme_1995, acharya_quantum_2024} required for advantageous quantum computation~\cite{dalzell_quantum_2025}. 

Circuit QED provides a promising platform to test these ideas in multi-qubit setups. The highly adaptable transmon design provides qubit-qubit interactions with high on-off ratios and site-resolved gates and readout. The use of standard nanofabrication techniques and integration with off-the-shelf microwave electronics also facilitates scaling, leading to setups that orchestrate the dynamics of tens to hundreds of nodes. Still, building such systems is a demanding engineering challenge. The wiring overhead, heat load, and leakage channels increase quickly with system size, and on-chip crosstalk becomes a concern.

Microwave cavities stand out as a hardware-efficient approach to realize large Hilbert spaces while circumventing these issues~\cite{copetudo_shaping_2024}. Instead of packing together many low-dimensional elements such as qubits, it is possible to use the multi-level structure of the harmonic oscillator to store high-dimensional information. That is the approach of the branch of superconducting circuits named \textit{bosonic circuit QED}: to first scale ``vertically" by increasing the quality and control over a single cavity module before assembling additional modules together. With this approach, single- or few-cavity bosonic cQED systems display features that would otherwise require large qubit assemblies. Quantum error correction beyond break-even was first achieved with bosonic codes~\cite{sivak_real_2023, ni_beating_2023, ofek_extending_2016, ni_autonomous_2025, sun_extending_2025}. Quantum metrology can reach the Heisenberg limit by only using intra-mode correlations~\cite{sahota_quantum_2015, wang_heisenberg_2019, deng_heisenberg_2024, pan_realization_2025},  without requiring entanglement. Single cavities also demonstrate finite-component phase transitions~\cite{dicandia_critical_2023} with uses in quantum sensing~\cite{dicandia_critical_2023, beaulieu_criticality_2025, petrovnin_microwave_2024}.

This thesis shares the goals of bosonic circuit QED. My objective is to improve single-cavity modules by introducing tunable light-matter interactions for quantum information processing. As a first step, this chapter will introduce the minimal, non-tunable device, which is the starting point for these improvements.

This minimal device contains three main elements: the cavity, which stores quantum information but is not fully controllable on its own; the transmon circuit, which provides nonlinear controls to the cavity; and the readout resonator, through which it is possible to do nondestructive measurements of the transmon state. Let us explore them one at a time.

\section{The basic elements}

\subsection{Microwave cavities}

In reference to cavity QED, the word \textit{cavity} usually refers to long-lived resonators. And resonators, in microwave systems, are essentially any structure that can host confined vibrational modes of the electromagnetic field.

\begin{figure}[ht!]
    \centering
    \includegraphics[width=\linewidth]{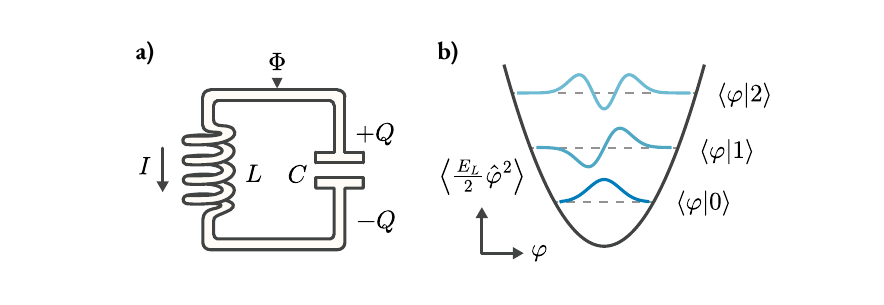}
    \caption[The quantum LC circuit]{\textbf{The quantum LC circuit.} a) Circuit diagram of a resonator composed of an inductance $L$ and a capacitance $C$. The dynamical variables of charge $Q$, flux $\Phi$ of the common node, and electrical current $I$ are shown. b) The quantum LC oscillator has a quadratic potential $\frac{E_L}{2}\hat{\varphi}^2$, leading to a ladder-like eigenstate structure with evenly-spaced eigenenergies. The picture shows the three first eigenstates and their eigenfunctions in the phase basis. }
    \label{fig:LC_circuit}
\end{figure}
The simplest example is the lumped-element LC oscillator (Figure~\ref{fig:LC_circuit}a). The dynamical variables of this system follow the classical Hamiltonian
\begin{equation}
    \label{eq:harmonic_H}
    \mathcal{H} = \frac{Q^2}{2C} + \frac{\Phi^2}{2L},
\end{equation}
representing the sum of the energies stored in the capacitance $C$ and the inductance $L$. The charge $Q$ and flux $\Phi$ variables form a canonical pair in classical mechanics. In circuit QED, it is common to use the dimensionless variables instead
\begin{equation}
    n = \frac{Q}{2e}, \, \,     \varphi = \frac{2\pi}{\Phi_0}\Phi.
\end{equation}
$n$ is the number of charge carriers, which in superconducting circuits are Cooper pairs with twice the fundamental charge $e$. $\varphi$ is the phase expressed in terms of the flux quantum, $\Phi_0 = h/2e$. The quantized Hamiltonian in terms of these variables has the well-known form
\begin{equation}
\label{eq:linear_oscillator}
    \hat{H} = 4E_C\hat{n}^2 + \frac{E_L}{2}\hat{\varphi}^2 =  \hbar\omega \left(\hat{a}^\dagger \hat{a} + \frac{1}{2}\right),
\end{equation}
with $\left[\hat{n}, \hat{\varphi}\right] = i$, the charging energy $E_C = e^2/2C$ and the inductive energy $E_L = \Phi_0^2/4\pi^2L$. $\omega = \frac{1}{\hbar}\sqrt{8E_CE_L}$ is the resonant frequency of the circuit. From now on, I will ignore the constant offset $\hbar\omega/2$, since it does not contribute to the dynamics of the system, and use the conventional $\hbar = 1$ for simplicity. $\hat{a}$ and $\hat{a}^\dagger$ are the ladder operators that diagonalize the Hamiltonian and describe the dynamical variables as
\begin{align}
    \hat{\varphi} = \varphi_{zpf} (\hat{a}^{\dagger} + \hat{a}), \label{eq:charge_flux_ops} \\
    \hat{n} = in_{zpf} (\hat{a}^{\dagger} - \hat{a}), \label{eq:number_ops}
\end{align}
where the quantities $\varphi_{zpf} = \left(\frac{2E_C}{E_L}\right)^{\frac{1}{4}}$ and $n_{zpf} = \left(\frac{E_L}{32E_C}\right)^{\frac{1}{4}}$ represent the quantum fluctuations in the dynamical variables when the resonator is in its point of minimal energy, the vacuum state $|0\rangle$. The variables of phase and Cooper pair number in the $|0\rangle$ state follow Gaussian distributions with variances
\begin{equation}
\sqrt{\left\langle\hat{\varphi}^2\right\rangle} = \varphi_{zpf}, \, \, \sqrt{\left\langle\hat{n}^2\right\rangle} = n_{zpf}.
\end{equation}

\noindent The energy spectrum of $\hat{H}$ is evenly spaced (Fig.~\ref{fig:LC_circuit}b), forming a ladder of eigenstates that we call Fock states $|z\rangle$. They correspond to having a number $z$ of excitations--or \textit{microwave photons}--in the system. Photons are created or annihilated by applying ladder operators as
\begin{align}
    \hat{a}^\dagger|z\rangle = \sqrt{z+1}|z+1\rangle \\
    \hat{a}|z\rangle = \sqrt{z}|z-1\rangle.
\end{align}
Each excitation contributes with $\omega$ to the total energy as $\hat{H}|z\rangle = \omega z|z\rangle$, which reveals an interesting feature of linear circuits: the photons are non-interacting. That is, adding one photon to the resonator does not increase nor decrease the energy required to add the next one. 

This simple model applies for the quantum description of any microwave resonator. But in practice, most resonators won't have lumped circuit elements with point-like dynamical variables. Instead, $Q$ and $\Phi$ will have continuous distributions that oscillate in space and time, which determines where the microwave photons are located. 

\begin{figure}
    \centering
    \includegraphics[width=\linewidth]{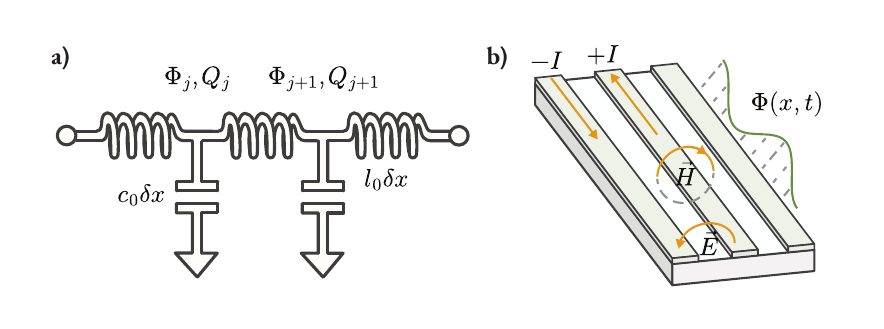}
    \caption[Distributed-element LC circuit]{\textbf{Distributed-element LC circuit.} a) Network of distributed capacitances~$c_0$ and inductances~$l_0$ per unit length, showing the dynamical variables at each node. Such networks can host many independent modes of oscillation, depending on the boundary conditions. b) Coplanar waveguides are an example of distributed LC networks. The picture shows the electric field $\vec{E}$ connecting the center conductor and the surrounding ground plane, the currents $\pm I$ flowing through both and the magnetic field $\hat{H}$ resulting from the currents. All of these variables, as well as $\Phi$ and $Q$, oscillate in space and time.}
    \label{fig:distributed_LC}
\end{figure}

To model this spatial distribution, let us upgrade the simple LC oscillator to the series of capacitors and inductors shown in Figure~\ref{fig:distributed_LC}a. Each of the $J$ nodes is labelled by $j$ and is located at position $j\delta x$, where $\delta x$ is our spatial resolution and $d = J\delta x$ is the total length of the resonator. Taking the limit $\delta x\to 0$ for constant $d$, the Hamiltonian can be described as the integral
\begin{equation}
\label{eq:distributed_H}
    \mathcal{H} = \sum_{j = 0}^{J-1}\left[\frac{Q_j^2}{2C_0} + \frac{1}{2L_0} (\Phi_{j+1}-\Phi_{j})^2\right] \xrightarrow{\delta x \to 0} \int^d_0 \left[\frac{\rho_Q^2(x, t)}{2c_0} + \frac{(\partial_x\Phi(x, t))^2}{2l_0} \right]dx,
\end{equation}
where $\rho_Q(j\delta x) = Q_j/\delta x$, $c_0 = C_0/\delta x$ and $l_0 = L_0/\delta x$ are the charge, capacitance and inductance per unit length. The classical equations of motion lead to the wave equation
\begin{equation}
\label{eq:fluxwave}
    \frac{1}{l_0c_0}\frac{\partial^2 \Phi}{\partial x^2} = \frac{\partial^2 \Phi}{\partial t^2}.
\end{equation}
With given boundary conditions, the wave equation can be solved by separation of variables $\Phi(x,t) = \sum_k\Phi_k(t)u_k(x)$. This gives a set of time-varying functions $\Phi_k(t)$ and linearly independent spatial distributions $u_k(x)$, called the \textit{modes of oscillation}, such that $\int^d_0u_ku_{k'}dx = \delta_{kk'}d$. Inserting the wave solution in Eq.~\ref{eq:distributed_H} with $\rho_Q(x,t) = c_0\dot{\Phi}(x,t)$ leads to a collection of harmonic oscillator modes
\begin{equation}
\label{eq:harmonic_collection_H}
    \mathcal{H} = \sum_{k = 0}^{\infty} \left[\frac{Q_k(t)^2}{2C} + \frac{\Phi_k(t)^2}{2L}\right], 
\end{equation}
with $C =c_0d$, $L = l_0d$ and $Q_k(t) = C\dot{\Phi}_k(t)$. Eq.~\ref{eq:harmonic_collection_H} can be quantized just as Eq.~\ref{eq:harmonic_H}, with each mode having an independent pair of conjugate variables that carry a spatial distribution. Figure~\ref{fig:distributed_LC}b illustrates this model for the case of a coplanar waveguide (CPW), highlighting the dynamical variables and field profiles.

\begin{figure}
    \centering
    \includegraphics[width=\linewidth]{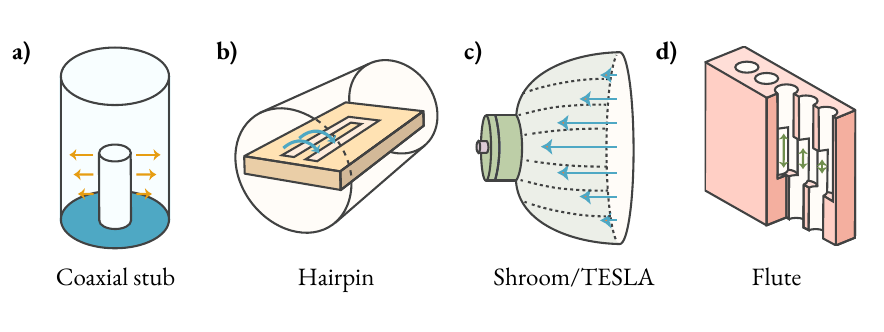}
    \caption[Cavity architectures]{\textbf{Cavity architectures.} There are many different types of cavities, which balance high lifetimes with other advantages and functionalities.}
    \label{fig:cavity_architectures}
\end{figure}

Much of the research around microwave cavities involves reshaping the mode of oscillation. A good cavity design has high capacity to couple with other circuits of interest while preventing field overlap with loss channels. One typical example is the $\lambda/4$ coaxial resonator shown in Figure~\ref{fig:cavity_architectures}a. The electromagnetic boundary conditions are defined by a central pin that is short-circuited to the outer walls of the cavity (the ``ground'') at the bottom surface, while the top end of the pin is an open circuit. The latter point is an antinode of the electric field, providing a region where other circuits such as a transmon can be inserted for strong capacitive coupling. A long stretch of the cavity is left empty between the end of the pin and the top of the structure. This region acts as a cylindrical waveguide with a high cutoff frequency, which attenuates the cavity field and prevents it from overlapping with the lossy seams at the top. 


The capacity of a cavity mode to protecting its quantum state against energy leakage is quantified by the quality factor $Q$, defined as~\footnote{The usual notation clashes with that of the charge variable, but the distinction will always be clear from context.}
\begin{equation}
    Q = \frac{\omega}{\kappa}. 
\end{equation}
$\kappa$ is the photon loss rate, which describes the continuous and irreversible energy relaxation of the resonator. Its reciprocal quantity is the single-photon lifetime $T_1 = 1/\kappa$, a central figure of merit of quantum circuits. If there are more than one photon loss channel, the total rate is given by $\kappa_T = \sum_i\kappa_i$, and the total $Q$ factor is the harmonic mean
\begin{equation}
    \frac{1}{Q} = \sum_i\frac{1}{Q_i}.
\end{equation}
Seam, surface oxide and superconducting AC losses are usually bundled up as the resonator's internal quality factor $Q_{int}$. Throughout this work, I will use coaxial $\lambda/4$ cavities made with $4\textnormal{N}6$ aluminum, etched twice with aluminum etchant type A to remove machining defects present on the top $150\,\mu\textnormal{m}$-deep layer from the cavity walls. Similar cavities can reach $Q_{int}$ between $3\times 10^7$ and $11.5\times 10^7$~\cite{reagor_quantum_2016, kudra_high_2020}, which establish a safe upper bound for the single-photon lifetime at the millisecond range for a mode at frequency $\omega/2\pi \approx 5\,\textnormal{GHz}$.

I chose this cavity design for its simplicity, but alternatives can be considered in future work to leverage different functionalities. For example, the hairpin resonator (Figure~\ref{fig:cavity_architectures}b) can be printed on-chip and is thus conveniently replaceable. Its \textit{U} shape hosts a resonant mode between both arms, leading to little field overlap with the superconducting outer walls. This isolates the structure from the losses at that surface and keeps most of the field's energy in lossless vacuum, reaching up to $Q_{int} = 3\times 10^7$~\cite{ganjam_surpassing_2024}. Other notable designs are the mushroom~\cite{milul_superconducting_2023} and TESLA cavities~\cite{aune_superconducting_2000, romanenko_three_2020}, which have an optimized electrical field distribution to reduce surface losses and use electro-beam welding to remove seams, reaching record-high $Q_{int}\approx10^{10}$. The flute resonator~\cite{chakram_seamless_2021} is ideal for multimode experiments. It can host multiple high-$Q$ resonances evenly spaced in frequency, which can be simultaneously coupled to a single transmon circuit. Lastly, the $Q_{int}$ of coaxial cavities can still be considerably improved up to $10^9$ by replacing aluminum with niobium and using rigorous surface treatment, according to recent developments from the literature~\cite{oriani_niobium_2025, takenaka_three_2025}.

Despite the importance of enhancing single-photon lifetimes, these numbers are only meaningful when compared to the speed and capacity at which photons are controlled to create interesting quantum states. This control can come either through a coupled circuit, which I will talk about in the next section, or a drive line. 

The drive line is the simplest form of control of a cavity field. It consists of a capacitively coupled voltage source $V(t)$, which introduces the Hamiltonian term
\begin{equation}
    \hat{H}_d = \frac{2eC_g}{C+C_g} V(t)\hat{n} = i\Omega(t)\left(\hat{a}^\dagger - \hat{a}\right),
\end{equation}
where $C_g$ is the gate capacitance. The Rabi frequency $\Omega(t)$ establishes the rate at which photons are coherently created and annihilated in the cavity. 

To activate this coherent exchange, $\Omega(t)$ must carry photons with the same frequency $\omega$ as the resonator excitations to respect conservation of energy. Consider a single-tone drive $\Omega(t) = \Omega_0\cos(\omega t)$. The driven resonator Hamiltonian is
\begin{equation}
    \hat{H}  = \omega \hat{a}^\dagger \hat{a} + i\Omega_0\cos\left(\omega t\right)\left( \hat{a}^\dagger - \hat{a}\right).
\end{equation}
Moving into the interaction picture of the bare resonator term $H_0 = \omega \hat{a}^\dagger \hat{a}$ leads to
\begin{equation}
    \hat{H} = i\Omega_0\cos\left(\omega t\right)\left( \hat{a}^\dagger e^{i\omega t} - \hat{a} e^{-i\omega t}\right)
     = i\frac{\Omega_0}{2}\left( \hat{a}^\dagger - \hat{a}+  \hat{a}^\dagger e^{i2\omega t} - \hat{a} e^{-i2\omega t}\right),
\end{equation}
where the cosine is expanded as a sum of complex exponential functions. The Hamiltonian ends up as a sum of two static ladder operator terms, and other two that rotate rapidly as $2\omega$. The former are often called \textit{co-rotating} terms, since the phase evolution of the drive follows the evolution of the resonator and constructively contributes to it. The latter are named \textit{counter-rotating} terms. Their high relative angular speed of $2\omega$ makes their effects negligible to the overall dynamics, which can be seen from the propagator
\begin{equation}
    \hat{U}_{counter}(t) = \exp\left[ \frac{\Omega_0}{2}\hat{a}^\dagger \int_0^te^{i2\omega t' } dt' + \textnormal{h.c.}\right] = \exp\left[\frac{\Omega_0}{4\omega}\hat{a}^\dagger e^{i2\omega t' } + \textnormal{h.c.}\right],
\end{equation}
where $\textnormal{h.c.}$ indicates the Hermitian conjugate of the preceding term. For $2\omega \gg \Omega_0$, the exponent is negligible and the propagator is approximately unitary, only carrying small ripples in the system's evolution. We can therefore use the Rotating Wave Approximation (RWA) to neglect the counter-rotating terms. The co-rotating terms, on the other hand, contribute to the evolution with the propagator
\begin{equation}
\label{eq:cav_disp}
    \hat{U}_{co}(t) = \hat{D}\left(\frac{\Omega_0t}{2}\right) = \exp\left[\frac{\Omega_0t}{2}\left(\hat{a}^\dagger -\hat{a}\right)\right].
\end{equation}
$\hat{D}(\beta)$, where $\beta = \tfrac{1}{2}\Omega_0 t$, is known as the displacement operator, the most common control resource for microwave resonators. It describes the net evolution of the cavity photon population under continuous exchange of photons with a resonant drive.

The displacement operator can be used to create an interesting class of states in the resonator. Consider the action of $\hat{D}$ over the vacuum state:
\begin{equation}
    \hat{D}\left(\beta\right)|0\rangle = e^{\frac{-|\beta|^2}{2}}\sum_{z=0}^\infty\frac{\beta^z}{\sqrt{z!}}|z\rangle = |\beta\rangle.
\end{equation}
$|\beta\rangle$ is called a \textit{coherent state}\footnote{To avoid confusion, I will label Fock and coherent states with Latin and Greek characters, respectively. The nature of the state will also be made clear from context.}. $\beta\in\mathbb{C}$ with the complex phase $\arg{(\beta)}=\theta$ following that of the drive $\Omega(t) = \Omega_0\cos(\omega t+\theta)$. All coherent states have the same $\hat{n}$ and $\hat{\varphi}$ fluctuations as the vacuum state, i.e.,
\begin{equation}
    \Delta n^2 =  \langle \beta|\hat{n}^2 |\beta\rangle - \left( \langle \beta|\hat{n} |\beta\rangle\right)^2 = n_{zpf}^2,
\end{equation}
and analogously for $\Delta\varphi^2 = \varphi_{zpf}^2$. They differ, however, in their mean photon number, quantified by $\hat{Z} = \hat{a}^\dagger \hat{a}$,
\begin{equation}
    \langle\beta|\hat{Z}|\beta\rangle = |\beta|^2,
\end{equation}
and in the photon statistics
\begin{equation}
\label{eq:cohstate_population}
    |\langle z|\beta\rangle|^2 = \frac{|\beta|^{2z}}{z!}e^{-|\beta|^2},
\end{equation}
where $|\beta|^2$ parametrizes a Poissonian distribution. These states are quite interesting for their direct connection with the classical regime. Moving from the interaction to the Schrodinger picture, they evolve in time following the resonator frequency as 
\begin{equation}
    e^{-i\omega\hat{a}^\dagger \hat{a}t}|\beta\rangle = |\beta e^{-i\omega t}\rangle.
\end{equation}
That means the instantaneous values of $\left\langle \hat{n}\right\rangle$ and $\left\langle \hat{\varphi} \right\rangle$ evolve sinusoidally and out-of-phase exactly as in a classical harmonic oscillator, except that these variables now exhibit quantum fluctuations.

Coherent states form a very interesting and useful class of states, but they are also limited compared to the range of possibilities offered by a quantum harmonic oscillator. It is not possible to explore the full Hilbert space having only displacement operations at hand, since these can only produce other coherent states
\begin{equation}
    \hat{D}(\beta)|\alpha\rangle = e^{\frac{1}{2}\left(\beta^*\alpha - \beta\alpha^*\right)}|\alpha + \beta\rangle.
\end{equation}
To visualize this limitation, I will introduce a very useful and graphical way of representing phase space: the Wigner function.

\begin{figure}
    \centering
    \includegraphics[width=\linewidth]{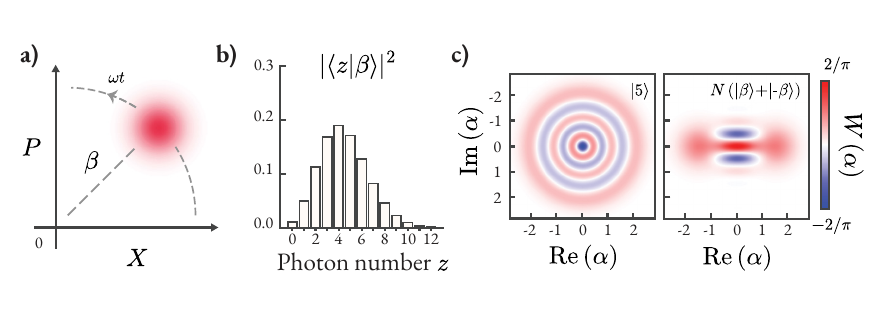}
    \caption[Phase space representation of the cavity state]{\textbf{Phase space representation of the cavity state.} a) Wigner plot of a cavity coherent state. Its Wigner function is a Gaussian distribution centered in $\left(X, P\right) = \left(\textnormal{Re}(\beta),\textnormal{Im}(\beta)\right)$, which evolves in time with a rotation angle $\omega t$. b) Photon statistics of a coherent state for $\beta \approx 2$, following a Poissonian distribution with mean $|\beta|^2$. c) Wigner plots $W(\alpha = X + iP)$ for a Fock state $|5\rangle$ and a cat state given by a superposition of two coherent states $N\left(|\beta\rangle + |-\beta\rangle\right)$ ($N$ a normalization constant). These states cannot be created from Gaussian operations over coherent states due to their Wigner negativity.}
    \label{fig:phase_space}
\end{figure}

We want to represent a general cavity state in the phase space of variables $\hat{\varphi}, \hat{n}$ while abstracting away the zero-point fluctuations that depend on circuit parameters. So I will use the dimensionless variables
\begin{equation}
    \hat{X} = \frac{1}{2}\left( \hat{a}^\dagger + \hat{a}\right), \, \,
    \hat{P} = \frac{i}{2}\left( \hat{a}^\dagger -\hat{a}\right),
\end{equation}
with $\left[\hat{X}, \hat{P}\right] = i/2$. A general cavity state density matrix $\boldsymbol{\rho}$ can be represented in the eigenbasis of $\hat{X}$ as the function $\rho(x,x')$, where
\begin{equation}
    \hat{X}|x\rangle = x|x\rangle \longrightarrow \boldsymbol{\rho} = \int_{-\infty}^\infty\rho(x,x')|x\rangle\langle x'|dxdx'.
\end{equation}
$\rho(x,x')$ makes explicit the $\hat{X}$ information of the quantum state, with $\rho(x,x)$ the probability of finding the value $x$. To also express information about the conjugate variable $\hat{P}$, which translates into periodic oscillations of $x$, we can take the Fourier transform over $\xi = x-x'$:
\begin{equation}
\label{eq:wigner_1}
    W(x, p) = \frac{1}{\pi}\int_{-\infty}^\infty  \rho\left(x-\frac{\xi}{2}, x+\frac{\xi}{2}\right)e^{-2p\xi}d\xi.
\end{equation}
$W(x,p)$ is the Wigner function of state $\boldsymbol{\rho}$, and it treats both quadratures symmetrically. Its marginal distributions are equivalent to the probabilities
\begin{align}
    \rho_x(x) = \langle x|\boldsymbol{\rho}|x\rangle = \int_{-\infty}^{\infty}W(x,p)dp, \\
    \rho_p(p) = \langle p|\boldsymbol{\rho}|p\rangle = \int_{-\infty}^{\infty}W(x,p)dx. 
\end{align}
However, $W$ is not a probability distribution. Instead it is referred as a \textit{quasi-probability distribution} because, although it obeys the normalization rule $\iint Wdxdp = 1$, in general it can still have negative values. Still, this distribution provides a complete, graphical representation of the cavity dynamics. For example, displacement operations show up simply as translations in phase space -- which change the location, but not the form of the Wigner function. Moreover, it is possible to completely reconstruct the density matrix $\boldsymbol{\rho}$ or directly infer some of its features from $W$.

Much physical meaning can be inferred from the phase space representation. The Wigner function of a coherent state $|\alpha\rangle$ is a positive Gaussian distribution centered in $\left\langle \hat{X}\right\rangle = \textnormal{Re}(\alpha)$, $\left\langle \hat{P}\right\rangle = \textnormal{Im}(\alpha)$, with fluctuations $\sqrt{\left\langle\Delta \hat{X}^2\right\rangle} = \sqrt{\left\langle\Delta \hat{P}^2\right\rangle} = 1/2$ (see Figure~\ref{fig:phase_space}a). They are part of a larger class called \textit{Gaussian states}, whose Wigner functions follow a normal distribution $\mathcal{N}(\vec{\mu}, \boldsymbol{\Sigma})$, for $\vec{\mu}$ the vector of average quadratures and $\Sigma$ their covariance matrix. The displacement operation classifies as a Gaussian operation~\cite{giedke_characterization_2002}, since it consists of a mere translation in phase space and transforms Gaussian states into other Gaussian states. This is also the case for squeezing operations $\hat{U}=\exp\left[\frac{1}{2} \left(\zeta^*\hat{a}^{\dagger2} - \zeta\hat{a}^2\right)\right]$ for $\zeta \in \mathbb{C}$ the squeezing parameter, which only reshape the covariance matrix, and phase shifts by an angle $\theta$ described by $\hat{U}=\exp(i\theta \hat{a}^\dagger \hat{a})$, which add overall rotations. The same idea applies to multiple resonators, for which Gaussian operations include beamsplitting $\hat{U} = \exp\left[i\theta\left( \hat{a}_1^\dagger \hat{a}_2 + \hat{a}_1 \hat{a}_2^\dagger \right)\right]$ and two-mode squeezing $\hat{U} = \exp\left(\zeta^*\hat{a}_1 \hat{a}_2 - \zeta\hat{a}_1^\dagger \hat{a}_2^\dagger \right)$. In summary, any operation generated by Hamiltonian terms that are linear or quadratic in $\hat{a}, \, \hat{a}^\dagger$ is Gaussian. And the set of Gaussian operations cannot offer arbitrary manipulation of the Wigner function~\cite{lloyd_quantum_1999}. For example, Fock states are non-Gaussian, as can be seen from the Wigner negativity ($W(\alpha)<0$) in Figure~\ref{fig:phase_space}c. The same applies for coherent state superpositions $N\left(|\beta\rangle + |-\beta\rangle\right)$ (for a normalization constant $N$) known as cat states, which have non-Gaussian ``whiskers'' in the center of the phase space. 

Universal control of the cavity is of central importance for bosonic circuit QED and for this thesis. This is where the \textit{matter} in light-matter interactions comes in: coupling the cavity to a nonlinear system such as a transmon introduces higher-order $\hat{a}, \,\hat{a}^\dagger$ terms in the Hamiltonian, leading to non-Gaussian evolution. This is a sufficient condition for the desired universal control of bosonic quantum information~\cite{krastanov_universal_2015, heeres_cavity_2015}. In the next section, I will talk about transmons in more detail, and how they are used for that objective.

\subsection{Transmons: introducing photon interactions}

The linearity of microwave resonators stems directly from their linear circuit elements, leading to systems of non-interacting photons. To introduce photon-photon interactions that provide non-Gaussian evolution, I will use one of the most important circuit elements of superconducting systems: the Josephson junction~\cite{josephson_possible_1962}.

\begin{figure}
    \centering
    \includegraphics[width=\linewidth]{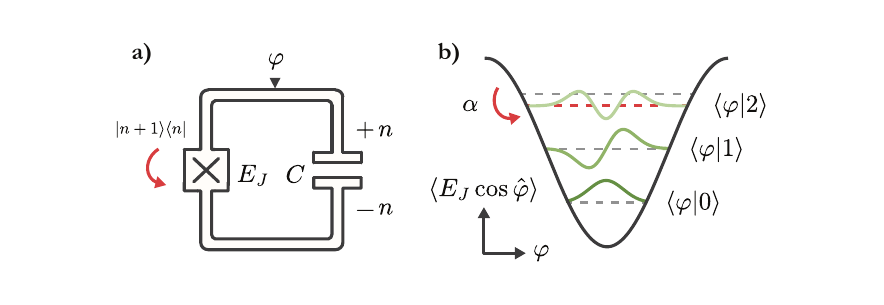}
    \caption[The transmon circuit]{\textbf{The transmon circuit.} a) The transmon circuit diagram is very similar to an LC oscillator, except the inductor is replaced by a Josephson junction (depicted as a cross). The junction allows Cooper pair tunneling with characteristic energy $E_J$. b) The junction transforms the transmon into an anharmonic oscillator with a cosinusoidal potential. This results in an on-site photon-photon interaction that reduces the energy of the second excitation relative to the first one by an anharmonicity factor $\alpha$. }
    \label{fig:transmon_circuit}
\end{figure}

Consider two superconducting terminals that are weakly connected by an oxide barrier or a narrow constriction. While this junction prevents the free propagation of Cooper pairs between the terminals, they can still tunnel across according to the Hamiltonian
\begin{equation}
    \hat{H}_T = -\frac{1}{2}E_J\sum_{n\in\mathbb{Z}}|n\rangle\langle n+1| + |n+1\rangle\langle n |.
\end{equation}
The $|n\rangle$ are eigenvectors of the operator $\hat{n}$ counting the number $n\in\mathbb{Z}$ of Cooper pairs accumulated at each terminal. The Josephson energy $E_J$ defines the tunneling rate. In terms of the conjugate variable $\hat{\varphi}$, this Hamiltonian becomes
\begin{equation}
    \label{eq:JJ_cosine_potential}
    \hat{H}_T = -E_J\cos\hat{\varphi}.
\end{equation}
This is the nonlinear cosine potential of the Josephson junction. In real systems, additional harmonics of the cosine can appear and are relevant for highly-optimized systems~\cite{willsch_observation_2024}, but in our case the simplest cosine form will be a good approximation. Compare this result against the quadratic potential of inductors: the Taylor expansion of Eq.~\ref{eq:JJ_cosine_potential} reveals a quadratic term $\propto\hat{\varphi}^2$, but also comes with higher even-order terms $\hat{\varphi}^4$, $\hat{\varphi}^6$ that introduce nonlinearity. 

The Josephson junction can be embedded in superconducting circuits to make them nonlinear. Consider a junction shunted by a capacitance as shown in Figure~\ref{fig:transmon_circuit}a. The Hamiltonian of this system is
\begin{equation}
\begin{aligned}
    \hat{H} & = 4E_C\hat{n}^2 - E_J\cos\hat{\varphi} \\ &\approx \, 4E_C\hat{n}^2 + \frac{E_J}{2}\hat{\varphi}^2 - \frac{E_J}{4!}\hat{\varphi}^4.
\end{aligned}
\end{equation}
Here, the cosine potential is approximated by a series expansion up to the fourth order in $\hat{\varphi}$. The first two terms of the bottom line correspond to the usual LC circuit (Eq.~\ref{eq:linear_oscillator}) with inductance $L_J = \Phi_0^2/4\pi^2E_J$, and can thus be diagonalized in terms of ladder operators $\hat{b},\hat{b}^\dagger$. In this formulation, the Hamiltonian becomes
\begin{equation}
    \hat{H} = \sqrt{8E_CE_J} \, \hat{b}^\dagger \hat{b} -\frac{E_J}{4!}\varphi_{zpf}^4\left(\hat{b}^\dagger + \hat{b}\right)^4.
\end{equation}
Expanding the product, rewriting the expression in the normal order of ladder operators and applying the RWA, we get
\begin{equation}
\label{eq:H_transmon}
    \hat{H} = \omega_T\hat{b}^\dagger\hat{b} - \frac{\alpha}{2}\hat{b}^\dagger \hat{b}^\dagger \hat{b} \hat{b}.
\end{equation}
Here, the nonlinearity appears as the anharmonicity $\alpha\approx E_C$~\footnote{The anharmonicity $\alpha$ should not to be confused with the often used label for a coherent state $|\alpha\rangle$! The difference should always be clear from context.}~\cite{blais_circuit_2021}. Due to this new term, the energy levels are shifted from the evenly-spaced ladder of the linear LC circuit: whereas the first excitation requires an energy $\omega_T = \sqrt{8E_JE_C}-E_C$, the second excitation requires an energy $\omega_T - \alpha$  (Fig.~\ref{fig:transmon_circuit}b). In other words, $\alpha >0$ describes a circuit where photons have an on-site interaction, whereby adding extra photons requires progressively lower excitation energies.

To use this nonlinear circuit for quantum information processing, we must make it resilient against leading sources of noise. One important cause of noise are the charge offsets caused by quasiparticle tunneling. Consider some of the Cooper pairs in the superconducting state are separated into quasiparticles, which are excitations of the superconducting ground state~\cite{glazman_bogoliubov_2021}. These excitations may either be a result of thermal equilibrium (suppressed at cryogenic temperatures), or due to out-of-equilibrium mechanisms such as incident infrared radiation~\cite{barends_minimizing_2011}. In the event that a quasiparticle crosses the junction, the $\hat{n}$ variable receives a background offset $n_{offset} = 0.5$. This offset shifts the $i$-th level of the circuit by an energy $\delta_i$ given by
\begin{equation}
\label{eq:qp_dephasing}
    \delta_i \propto \frac{1}{i!}\left(8n_{zpf}\right)^{2i}e^{-16n_{zpf}^2}.
\end{equation}
The tunneling events are stochastic, making the transition frequencies of the circuit jitter randomly. Because of this frequency noise, the control electronics lose track of the phase evolution of the circuit's quantum state over a dephasing timescale $T_\phi$, compromising coherent control. In practice, $T_\phi$ will impact the \textit{transverse relaxation time} $T_2$ of superposition states~\cite{krantz_quantum_2019}. These two quantities are related as
\begin{equation}
    \frac{1}{T_2} = \frac{1}{2T_1} + \frac{1}{T_\phi}.
\end{equation}
Together with $T_1$, $T_2$ is an important factor quantifying the quality of superconducting circuits. Unfortunately, it can be significantly compromised by quasiparticle tunneling events. The strategy to solve this problem and achieve high $T_2$ is given by Equation~\ref{eq:qp_dephasing}: to increase the charge zero-point fluctuations $n_{zpf} = (E_J/32E_C)^\frac{1}{4}$, reducing the frequency jittering amplitude $\delta_i$. This is achieved in the \textit{transmon regime} given by $E_J\gg E_C$~\cite{koch_charge_2007}. In a transmon circuit, the charge noise is suppressed beyond the next leading source of noise, providing a highly coherent system for quantum information processing.

The enhanced frequency stability and the anharmonicity of transmons support their use as superconducting qubits. To create a qubit, we can use the subspace spanned by the first two excitation levels, relabeled as $|0\rangle \to |g\rangle$ (the ground state) and $|1\rangle \to |e\rangle = \hat{b}^\dagger |g\rangle$ (the excited state). Universal control over this subspace is possible because, unlike harmonic oscillators, each transition can be individually addressed by external drives. This can be shown from the action of a microwave drive over the transmon Hamiltonian (Eq.~\ref{eq:H_transmon}) truncated up to the third level $|f\rangle = \frac{1}{\sqrt{2}}\hat{b}^\dagger|e\rangle$:
\begin{equation}
    \hat{H} = -\frac{\omega_T}{2}\hat{\sigma}_{z} - \frac{\left(\omega_T-\alpha\right)}{2}\hat{s}_z + \Omega_0\cos(\omega_d t)\left(\hat{\sigma}_y +\sqrt{2}\hat{s}_y\right).
\end{equation}
I split the ladder operators into Pauli operators $\hat{\sigma}_z = |g\rangle\langle g| - |e\rangle\langle e|$ and $\hat{s}_z = |e\rangle\langle e| - |f\rangle\langle f|$, respectively describing the first and second transitions (analogously for $\hat{\sigma}_y, \, \hat{s}_y$). Using the same procedure as before to move into the interaction picture and eliminate counter-rotating terms:
\begin{equation}
\label{eq:transmon_drive}
    \hat{H} = \frac{i\Omega_0}{2}\left(e^{i\left(\omega_T -\omega_d\right)t}|e\rangle\langle g| + \sqrt{2}e^{i\left(\omega_T - \alpha -\omega_d\right)t}|f\rangle\langle e| + \textnormal{h.c.}\right).
\end{equation}
This equation shows that, by choosing the drive frequency $\omega_d = \omega_T$, the $|g\rangle \to|e\rangle$ transition can be activated while leakage to $|f\rangle$ is suppressed due to the detuning $\omega_T - \alpha -\omega_d = \alpha$. In practice, $\Omega(t)$ is finite in time and is thus not spectrally clean in $\omega_d$. So qubit pulses carry other frequency components within a finite bandwidth that can incur into leakage. To mitigate this issue, the bandwidth is limited by using pulses with duration $t\gtrsim 2\pi/\alpha$ so as to not excite the $|e\rangle\to|f\rangle$ transition. Several pulse-shaping techniques exist with the objective of optimizing both leakage and drive length, such as the Derivative Removal by Adiabatic Gate (DRAG)~\cite{motzoi_simple_2009} and its subsequent developments~\cite{hyyppa_reducing_2024}.

\begin{figure}
    \centering
    \includegraphics[width=\linewidth]{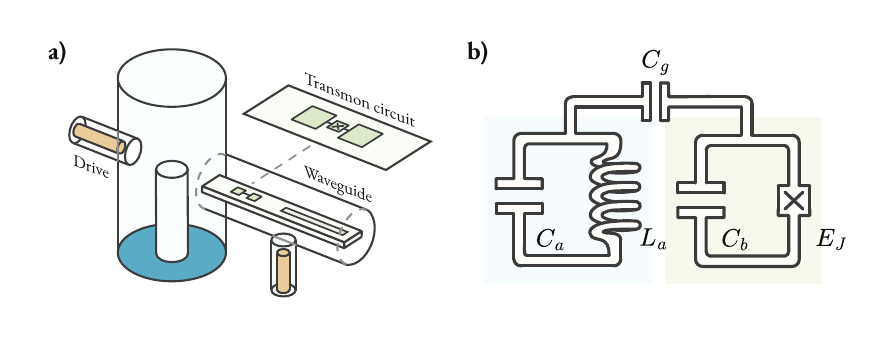}
    \caption[Cavity-transmon coupling]{\textbf{Cavity-transmon coupling.} a) Typical bosonic circuit QED setup consisting of $\lambda/4$ coaxial stub cavity coupled to a transmon chip in a circular waveguide~\cite{axline_architecture_2016}. b) Equivalent lumped-element circuit of the system. The transmon and cavity couple through a gate capacitance $C_g$.}
    \label{fig:cavity_transmon_coupling}
\end{figure}

The transmon nonlinearities grant complete control over its own quantum state, which is a very useful feature for several experiments demonstrated in this thesis. However, our interest lies especially in the microwave cavity. How can the transmon borrow these photon interactions to the cavity field to guarantee universal control?

Commonly, in bosonic circuit QED devices, the transmon is coupled to a coaxial stub cavity as shown in Figure~\ref{fig:cavity_transmon_coupling}a. The transmon is printed on a dielectric chip (in this work a c-plane sapphire) as two open capacitor pads connected by a Josephson junction. The cavity fields overlap with the transmon pads, creating a cross-circuit capacitance $C_g$. The standard quantization procedure~\cite{blais_circuit_2021, rasmussen_superconducting_2021} of the circuit diagram in Figure~\ref{fig:cavity_transmon_coupling}b leads to the Hamiltonian
\begin{equation}
    \hat{H} = 4\tilde{E}_{C,a}\hat{n}_a^2 +\frac{E_{L,a}}{2}\hat{\varphi}_a^2 + 4\tilde{E}_{C,b}\hat{n}_b^2 -E_{J,b}\cos\hat{\varphi}_b + g_C\hat{n}_a\hat{n}_b.
\end{equation}
The subscripts label the cavity ($a$) and transmon ($b$) operators. The capacitive energies are slightly modified by the change in capacitances $\tilde{C}_{a, b} = D/\left(C_{b, a} + C_g\right)$, where $D = C_aC_b + C_gC_b + C_gC_a$. This correction can be ignored when assuming $C_g \ll C_{a,b}$, so I will drop the tilde. The term $g_C = 4e^2C_g/D$ is called the \textit{capacitive coupling factor}, and is purely defined by the geometric capacitances of the circuit. Rewriting in terms of ladder operators:
\begin{equation}
\label{eq:transmon_cavity}
    \hat{H} = \omega_{a,0} \hat{a}_0^\dagger \hat{a}_0 +  g\left(\hat{a}_0^\dagger\hat{b}_0+\hat{b}_0^\dagger\hat{a}_0 \right) + \omega_{T,0}\hat{b}_0^\dagger\hat{b}_0 - E_J\left(\cos\hat{\varphi}_{b} + \frac{\hat{\varphi}_{b}^2}{2}\right).
\end{equation}
This Hamiltonian is divided into two main parts: the first three terms compose the linear Hamiltonian $\hat{H}_l$, while the last term is the nonlinear $\hat{H}_{nl}$. The operator subscript $0$ emphasizes the linear part is written in terms of the bare cavity and transmon modes. They couple with strength~\footnote{Here I assumed $g\ll\omega_a,\,\omega_b$, which is the regime of interest of this work. This allows using the RWA  eliminate the counter-rotating coupling terms $\hat{a}_0\hat{b}_0$, $\hat{a}_0^\dagger\hat{b}_0^\dagger$. This assumption is broken in the deepstrong and ultrastrong coupling regimes~\cite{kockum_ultrastrong_2019}.} $g = g_Cn_{zpf,a}n_{zpf,b}$, which now depends on the zero-point fluctuations, and therefore depends on the junction's $E_J$. Through this coupling, the nonlinear potential of the transmon can influence the cavity mode. However, it is still not obvious how $\hat{H}_{nl}$, which is only a function of the transmon's $\hat{\varphi}_b$, impacts the dynamics of the cavity.

The nonlinearities inherited by the cavity become evident in the picture in which $\hat{H}_l$ is diagonalized. The standard approach is to use a Bogoliubov transformation. First $H_{l}$ is rewritten in the matrix form
\begin{equation}
    \hat{H}_{l} = \begin{pmatrix}
        \hat{a}_0^\dagger & \hat{b}_0^\dagger
    \end{pmatrix}
    \begin{pmatrix}
        \omega_{a,0} & g \\ g & \omega_{T,0}
    \end{pmatrix}
    \begin{pmatrix}
    \hat{a}_0\\ \hat{b}_0
    \end{pmatrix},
\end{equation}
and then we find a rotation $R(\Lambda)$ that diagonalizes the system 
\begin{equation}
R(\Lambda) = 
    \begin{pmatrix}
        \cos\Lambda & \sin\Lambda \\ -\sin\Lambda & \cos\Lambda
    \end{pmatrix} \longrightarrow
    R(\Lambda)
    \begin{pmatrix}
        \omega_{a,0} & g \\ g & \omega_{T,0}
    \end{pmatrix}
    R(\Lambda)^{-1} = 
    \begin{pmatrix}
        \omega_{a,L} & 0 \\ 0 & \omega_{T,L}
    \end{pmatrix}.
\end{equation}
The transformation dresses the operators $\hat{a} = \hat{a}_0\cos\Lambda +  \hat{b}_0\sin\Lambda$, $\hat{b} = \hat{b}_0\cos\Lambda -  \hat{a}_0\sin\Lambda$ and leads to the dressed mode frequencies $\omega_{a,L}$,  $\omega_{T,L}$~\footnote{Note these are still not the final eigenfrequencies of the system. We still need to account for shifts introduced by the nonlinear term! Still, $\omega_{a,L}$ and $\omega_{T,L}$ play an important part in design since they are directly accessible from finite-element simulations.}. Going back to Eq.~\ref{eq:transmon_cavity}, we get the dressed Hamiltonian
\begin{equation}
\label{eq:transmon_cavity_diag}
    \hat{H} = \omega_{a,L} \hat{a}^\dagger \hat{a} + \omega_{T,L}\hat{b}^\dagger\hat{b} - E_J\left(\cos\hat{\varphi}_{b} + \frac{\hat{\varphi}_{b}^2}{2}\right),
\end{equation}
where now the flux over the Josephson junction $\hat{\varphi}_b$ is given by 
\begin{align}
\label{eq:phi_hybridization}
    \hat{\varphi}_{b} & = \varphi_{zpf,b}\cos\Lambda\hat{b} + \varphi_{zpf,b}\sin\Lambda\hat{a} + \textnormal{h.c.} \nonumber \\ &= \varphi^\Lambda_{b}\hat{b} + \varphi^\Lambda_{a}\hat{a} + \textnormal{h.c.}.
\end{align}
The coefficients $\varphi_a^\Lambda$ and $\varphi_b^\Lambda$ are the zero-point fluctuations of each mode over the junction, and quantify the transmon-cavity hybridization. In this formulation, the dynamics of the cavity are directly influenced by the nonlinear potential due to the presence of $\hat{a}$ in the argument of the cosine.

This coupling gives origin to two especially interesting terms. Expanding the cosine up to the fourth order leads to 
\begin{equation}
\label{eq:dispersive_transmon_cav}
    \hat{H} \approx  \omega_a \hat{a}^\dagger \hat{a} + \omega_T\hat{b}^\dagger \hat{b}-\frac{\alpha}{2}\hat{b}^\dagger\hat{b}^\dagger\hat{b}\hat{b} -\frac{K}{2}\hat{a}^\dagger\hat{a}^\dagger\hat{a}\hat{a} - \chi \hat{a}^\dagger \hat{a}\hat{b}^\dagger\hat{b}.
\end{equation}
The approximation is valid in the dispersive regime, where the detuning $\Delta = \omega_a-\omega_T$ between the final frequencies $\omega_a = \omega_{a,L} - K$ and $\omega_T = \omega_{T,L} - \alpha$ is much larger than the coupling $g$. The $-\frac{1}{2}K\hat{a}^\dagger\hat{a}^\dagger\hat{a}\hat{a}$ term is the cavity anharmonicity, or self-Kerr. In the dispersive regime, $\varphi_a^\Lambda \approx\varphi_{zpf, b}\left(\frac{g}{\Delta}\right) \ll \varphi_b^\Lambda$, meaning the cavity nonlinearities are much weaker and $K\ll\alpha$. For this reason, the cavity anharmonicity does not offer meaningful level selectivity by an external drive as in the case of the transmon, but is an important correction to the cavity's dynamics at high photon numbers. The second term is the dispersive shift $- \chi \hat{a}^\dagger \hat{a}\hat{b}^\dagger\hat{b}$, with $\chi\approx \alpha g^2/\Delta^2$. This term describes a photon-number-dependent frequency shift between the two circuits and dominates their interaction in the dispersive regime. 

This discussion on the cavity-transmon coupling gives a solid foundation for the rest of the thesis. First, because taking the transmon in the two-level approximation directly ties the system to the dynamics of light-matter interactions. For example, Eq.~\ref{eq:transmon_cavity} becomes
\begin{equation}
\label{eq:jaynes-cummings}
    \hat{H} \approx \omega_{a} \hat{a}_0^\dagger \hat{a}_0 +  g\left(\hat{a}_0^\dagger\hat{\sigma}_0^-+\hat{a}_0\hat{\sigma}^+_0 \right) - \frac{\omega_{T}}{2}\hat{\sigma}_{z,0}.
\end{equation}
This is the Jaynes-Cummings Hamiltonian~\cite{jaynes_comparison_1963, bina_coherent_2012}, which is equivalent to the Rabi model in the strong coupling regime. Moreover, Equations~\ref{eq:transmon_cavity_diag} and~\ref{eq:phi_hybridization} are the starting point for several other applications: they are directly translatable to other Josephson circuits besides the transmon, such as SNAILs~\cite{frattini_wave_2017}. They can also be used to formulate charge-driven parametric processes such as three- and four-wave mixing beamsplitting~\cite{gao_programmable_2018, chapman_high_2023}. And they directly tie to circuit design approaches such as black-box circuit quantization~\cite{nigg_blackbox_2012} and the energy participation ratio~\cite{minev_energy_2021} methods.

This closes this introduction to the two basic elements of bosonic circuit QED responsible for hosting and manipulating quantum information. But one essential ingredient is still missing: the ability to extract information from this system. In the next section, I will explain how this is possible by dispersively coupling the transmon to a readout resonator.

\subsection{Readout resonators}

Reliably extracting information about the system's dynamics is a basic requirement of any experimental setup. But quantum measurements can go much beyond mere characterization. Mid-circuit measurements can actively correct logical qubits errors~\cite{terhal_quantum_2015}, introduce programmable decoherence in simulations~\cite{burger_digital_2022} and demonstrate interesting effects such as partial measurement back-action~\cite{hatridge_quantum_2013} and reversion of quantum jumps mid-flight~\cite{minev_catch_2019}. 

These functionalities are achieved in circuit QED by dispersively coupling the transmon to a readout resonator. Take the Hamiltonian~\ref{eq:dispersive_transmon_cav} while replacing the cavity $\hat{a}$ with the readout mode $\hat{c}$
\begin{equation}
\label{eq:dispersive_transmon_RR}
    \hat{H} = \omega_T\hat{b}^\dagger \hat{b} +\omega_c \hat{c}^\dagger \hat{c} -\frac{\alpha}{2}\hat{b}^\dagger\hat{b}^\dagger\hat{b}\hat{b} -\frac{K}{2}\hat{c}^\dagger\hat{c}^\dagger\hat{c}\hat{c} - \chi \hat{b}^\dagger\hat{b}\hat{c}^\dagger \hat{c}.
\end{equation}
As discussed before, the self-Kerr $K$ has small impact in the dynamics in the dispersive regime and at low photon population, so it is not a central element in the readout mechanism and I will take it as negligible. The dispersive shift term $-\chi \hat{b}^\dagger\hat{b}\hat{c}^\dagger \hat{c}$, on the other hand, is responsible for impressing the transmon state information onto the readout resonator. This becomes evident by rewriting the readout frequency as 
\begin{equation}
    \omega_c\left(\hat{b}^\dagger\hat{b}\right) = \omega_c - \chi\hat{b}^\dagger\hat{b}.
\end{equation}
That is, each excitation of the transmon shifts the readout frequency by $-\chi$. This shift changes the resonator response to incoming signals, which can be measured at the output of a feed line to infer the transmon state.

\begin{figure}
    \centering
    \includegraphics[width=\linewidth]{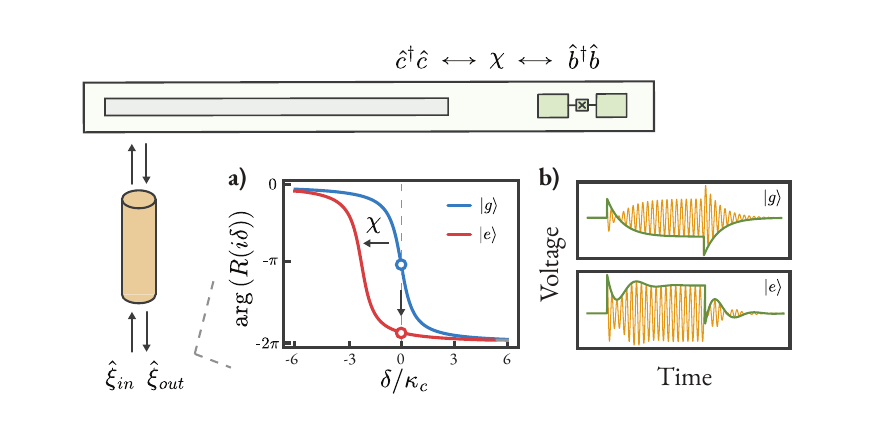}
    \caption[Transmon readout]{\textbf{Transmon readout.} The state of the transmon is inferred from the reflection measurement of a dispersively coupled readout resonator. a) Phase of the reflection coefficient. Transmon excitations shift the readout response in frequency by $-\chi$, leading to a different response to a probe signal at $\delta = 0$. b) Readout output signal for a square input wave with $\delta = 0$ for states $|g\rangle$ and $|e\rangle$. The state assignment is done from the output signal's envelope.}
    \label{fig:readout_response}
\end{figure}

To efficiently transfer the transmon state information to the room-temperature electronics, the readout should be able to quickly exchange photons with a feed line. This photon exhange is characterized by a rate $\kappa_c$, related to the external quality factor as $Q_{ext} = \omega/\kappa_c$. Considering the resonator's internal $Q_{int}$, this leads to a total $Q$ factor
\begin{equation}
    \frac{1}{Q} = \frac{1}{Q_{int}} + \frac{1}{Q_{ext}}.
\end{equation}
Ideally, we want to be in the overcoupled regime $Q_{ext}\ll Q_{int}$, so the information-carrying photons can be picked up by the feed line before being lost to other channels.

The interplay between the transmon-readout dispersive shift and the photon transfer rate to the feed line is described by the open dynamics of a Langevin equation~\cite{blais_circuit_2021}. Considering the propagating feed line modes $\xi_{in}$,  $\xi_{out}$, the readout evolution is given by
\begin{equation}
    \label{eq:langevin_rr}
    \frac{d}{dt}\hat{c} = \left(\frac{\kappa_c}{2} -i\chi b^\dagger b\right)\hat{c}(t) - \sqrt{\kappa_c}\hat{\xi}_{in}(t).
\end{equation}
Here I neglect the constant rotations from $\omega_c\hat{c}^\dagger \hat{c}$. The incident signal $\hat{\xi}_{in}(t)$ is related to the output $\hat{\xi}_{out}(t)$ according to the input-output relation~\cite{yurke_input_2004}
\begin{equation}
    \label{eq:input_output}
    \hat{\xi}_{out}(t) = \hat{\xi}_{in}(t) + \sqrt{\kappa_c}\hat{c}(t).
\end{equation}
Solving Equations~\ref{eq:langevin_rr} and~\ref{eq:input_output} lead to different solutions for $\hat{\xi}_{out}(t)$ for each transmon state. The classical signal envelope of $\hat{\xi}_{out}(t)$ is obtained by using the stiff pump approximation $\hat{\xi}_{in,out}\to\xi_{in,out}(t)$, and $\hat{c}\to c(t)$, which transforms the problem into an ordinary differential equation. Alternatively, we can analyze the transmon-dependent response is from the reflection coefficient, obtained analytically through a Laplace transform
\begin{equation}
\label{eq:reflection}
    R(i\delta) = \frac{\hat{\xi}_{out}\left[i\delta\right]}{\hat{\xi}_{in}\left[i\delta\right]} = \frac{i\left(\delta -\chi\hat{b}^\dagger \hat{b} \right) - \kappa_c/2}{i\left(\delta -\chi\hat{b}^\dagger \hat{b} \right) + \kappa_c/2},
\end{equation}
where $\delta$ is the detuning between the input signal frequency and $\omega_c$. This relation indicates both the output signal amplitude and phase depend on $\hat{b}^\dagger\hat{b}$ (see Figure~\ref{fig:readout_response}a). Interestingly, proper optimization of Eq.~\ref{eq:reflection} can allow the state discrimination of several levels at once, with experimental demonstrations reaching up to 12 transmon levels~\cite{wang_high_2025}.

The resonator has to be carefully designed not to disturb the rest of the circuit while it is not being measured. The readout interaction with the environment, given by $\kappa_c$, is designed in the few-hundreds of $\textnormal{ns}$ range ($Q_{ext}$ from $10^3$ to $10^4$). This is a very strong photon loss channel and might affect the transmon and cavity lifetimes through resonator-mediated Purcell decay~\cite{goy_observation_1983}. To prevent this, it is possible to design Purcell filters before the feed line which reflect the fields of sensitive modes while allowing the readout signal through. There are many ways to implement this filtering, which vary from adding an extra circuit~\cite{jeffrey_fast_2014, yan_broadband_2023} to leveraging symmetries~\cite{yen_interferometric_2025} or field nodes~\cite{sunada_fast_2022}. Later I will make use of similar strategies to isolate the cavity from flux line losses.

\begin{figure}
    \centering
    \includegraphics[width=\linewidth]{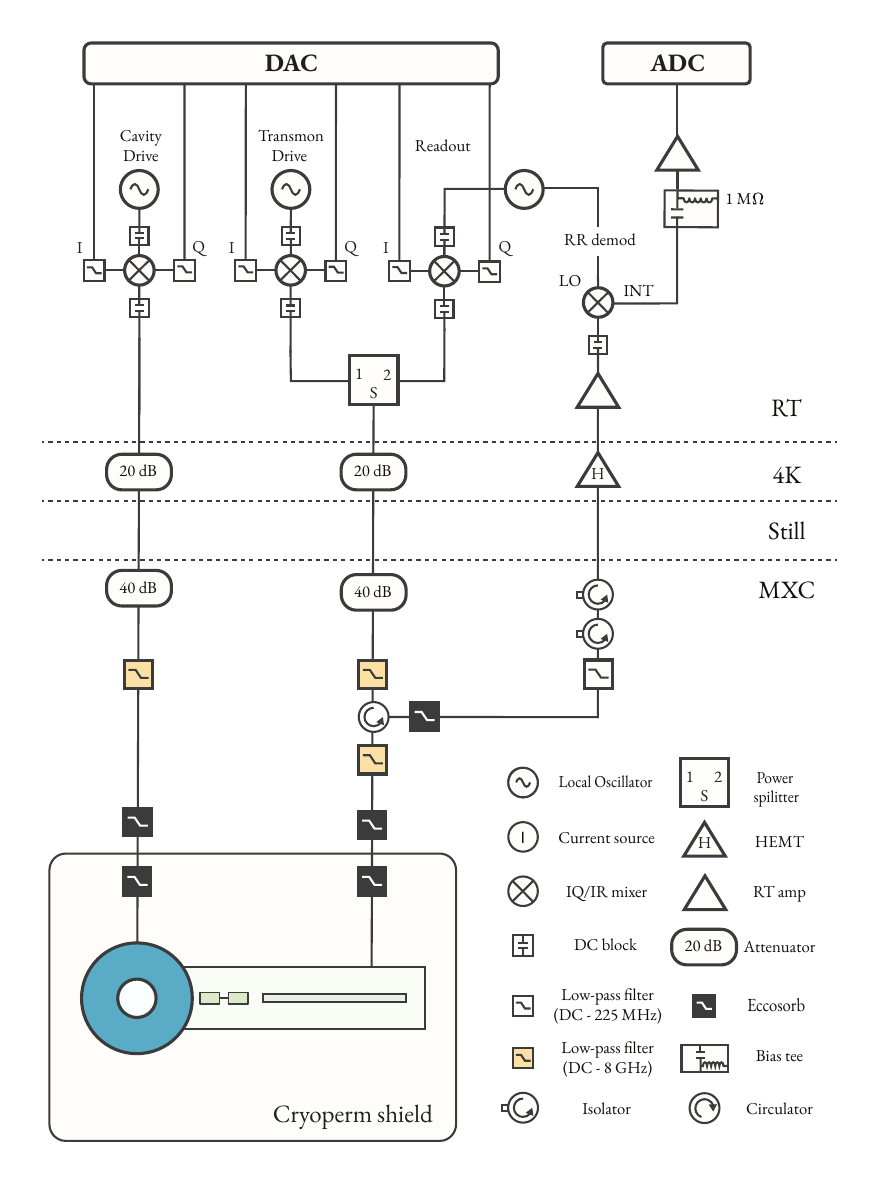}
    \caption[Experimental setup and wiring]{\textbf{Experimental setup and wiring.} The sample is mounted inside a cryoperm shield at the mixing chamber plate of an LD-250 dilution refrigerator. The cavity, transmon and readout control signals are generated at the DAC of a Field-Programmable Gate Array (FPGA) at room temperature. The reflected readout signal goes through multiple steps of amplification before being acquired at the ADC of the FPGA.}
    \label{fig:wiring_diagram}
\end{figure}

Finally, I would like to describe the general signal chain which I will use for control and measurement (see Ref.~\cite{krinner_engineering_2019} for more details). All of the microwave signals for the cavity, transmon and readout drives are generated from a Digital-to-Analog Converter (DAC) with a $\pm\textnormal{250}\,\textnormal{MHz}$ bandwidth (Figure~\ref{fig:wiring_diagram}), and then upconverted to the $\textnormal{GHz}$ range by using a microwave mixer and local oscillators. The lines go into an LD-250 dilution refrigerator, in which the circuit is mounted at the coldest stage (the mixing chamber plate) at about $10\,\textnormal{mK}$. The readout and cavity lines go through a total of $50\,\textnormal{dB}$-$60\,\textnormal{dB}$ attenuation to decrease the temperature of the noise carried by the lines, and the transmon drive signal is sent on the same line as the readout. A circulator at the readout port of the circuit separates $\xi_{in}$ and $\xi_{out}$. The readout output goes through a $40\,\textnormal{dB}$ isolator, which prevents high-temperature noise from reaching the device without requiring the attenuation of the line. The output is then amplified with a High-Electron-Mobility Transistor (HEMT) and two stages of room-temperature amplification before it reaches the Analog-to-Digital Converter (ADC) for acquisition. After the readout output signal $\xi_{out}(t)$ is acquired, the transmon state is inferred using digital signal processing and state discrimination techniques~\cite{krantz_quantum_2019}. 

So these are the basic ingredients that compose a typical bosonic circuit QED system: the cavity, which stores the quantum information, the transmon, which provides the cavity with nonlinearities for universal control, and the readout, which characterizes the system by measuring the transmon state. I also showed how this system of interacting microwave photons maps into the light-matter dynamics of the Jaynes-Cummings model, which will later be applied to quantum information processing. In the next section, I will explain in more detail how to accurately design and simulate circuits that implement the desired dynamics.

\section{Circuit design}
\label{sec:designandsim}

The methodology used to design bosonic circuit QED devices consists of three main steps: (1) draw an electromagnetic structure and obtain the linear resonant modes; (2) calculate the zero-point fluctuations of each mode over the Josephson junction, then (3) define the system Hamiltonian and use it to simulate the dynamics of the system.

The electromagnetic structure defining the circuit is drawn using the \textit{Ansys HFSS} software~\cite{ansysHFSS}. The walls of the cavity and the waveguide are assumed to be perfect conductors, and the sapphire chip is defined as having a homogeneous relative permittivity $\varepsilon_r = 10$. On the top of the chip, the transmon and readout circuits are defined as perfectly conducting planar sheets. The Josephson junction is simply drawn as a linear lumped-element inductance $L_J$ connecting the transmon's capacitor pads. The drive lines are usually defined as copper coupling pins. All coupling pins start at the walls of the model, where $50\,\Omega$ surface resistance is defined to represent impedance-matched boundary conditions.

The finite-element solution of the electromagnetic fields reveals the eigenmodes of the structure. The result is the set of linear mode frequencies $\omega_{\nu,L}$, the field distributions of modes $\hat{\nu} = \hat{a}, \hat{b}, \hat{c}$ (see Eq.~\ref{eq:transmon_cavity_diag}), and the $Q_{ext}$ of each mode with relation to each drive line. I target $Q_{ext}\approx 10^3$ for the readout and $\approx10^7$ for the transmon through the feed line, and $\approx5\times10^8$ for the cavity through its own drive line. This method provides all necessary information about the diagonalized linear Hamiltonian $\hat{H}_l$, so it is necessary to go one step further to include the $\hat{H}_{nl}$ coming from the transmon potential.

To describe $\hat{H}_{nl}$, it suffices to calculate the contribution of each mode to the flux across the Josephson junction as $\hat{\varphi}_b = \sum_{\nu = a,b,c}\varphi^\Lambda_\nu\left(\hat{\nu}^\dagger + \hat{\nu}\right)$. For that we use the Energy Participation Ratio method~\cite{minev_energy_2021}, according to which 
\begin{equation}
    \varphi_\nu^\Lambda = \sqrt{p_\nu\frac{\omega_{\nu,L}}{2E_J}}. 
\end{equation}
The quantities $p_{\nu}$ are the ratio between the inductive energy of mode $\nu$ stored in the junction $U^\nu_{JJ,ind}$ and the total inductive energy $U^\nu_{ind}$. That is:
\begin{equation}
    p_\nu = \frac{U^\nu_{JJ,ind}}{U^\nu_{ind}} = \frac{\langle \frac{1}{2}E_J\hat{\varphi}_b\rangle}{\langle \frac{1}{2}\hat{H}_l\rangle}.
\end{equation}
This expression is fully calculated within classical electromagnetism. For a given field excitation, $U^\nu_{ind}$ comes from integrating the field energy density across the whole volume of the circuit, while  $U^\nu_{JJ, ind} = \frac{1}{4}L_J\left(I_{peak}^{\nu}\right)^2$, where $I^\nu_{peak}$ is the amplitude of the current wave flowing through the junction. These calculations are made with the pyEPR Python package. pyEPR reads the field solutions from HFSS and returns $\varphi_\nu^\Lambda$, which I use to calculate the full system Hamiltonian as
\begin{equation}
    \hat{H} = \sum_{\nu = a,b,c}\omega_{\nu, L}\hat{\nu}^\dagger\hat{\nu} - E_J\left(\cos\hat{\varphi}_{b} + \frac{\hat{\varphi}_{b}^2}{2}\right),  \, \, \hat{\varphi}_b = \sum_{\nu = a,b,c}\varphi^\Lambda_\nu\left(\hat{\nu}^\dagger + \hat{\nu}\right).
\end{equation}
This procedure can in principle be applied to an arbitrary number of modes and Josephson junctions, as well as different nonlinear potentials, with corrections for highly anharmonic circuits~\cite{yilmaz_energy_2024}. 

With the full Hamiltonian at hand, we can calculate the system coefficients in the dispersive regime. Usually bosonic circuit QED experiments are designed with the transmon dispersively coupled to both the cavity and the readout resonator, so we can treat the Hamiltonian similarly to Eq.~\ref{eq:dispersive_transmon_cav} as 
\begin{align}
    \hat{H}_{disp} \approx & \sum_{\nu = a,b,c}\omega_\nu\hat{\nu}^\dagger\hat{\nu}
     -\frac{\alpha}{2}\hat{b}^\dagger\hat{b}^\dagger\hat{b}\hat{b}
     -\chi_{ab}\hat{a}^\dagger\hat{a}\hat{b}^\dagger\hat{b} -\chi_{cb}\hat{c}^\dagger\hat{c}\hat{b}^\dagger\hat{b}
     \notag \\
    & \hspace{1cm}
     -\frac{K_a}{2}\hat{a}^\dagger\hat{a}^\dagger\hat{a}\hat{a}
     -\frac{K_c}{2}\hat{c}^\dagger\hat{c}^\dagger\hat{c}\hat{c}
     -K_{ac}\hat{a}^\dagger\hat{a}\hat{c}^\dagger\hat{c}.
\end{align}
Table~\ref{tab:dispersive_params} summarizes how each parameter is calculated numerically from $\hat{H}$, as well as the analytical formulas from Ref.~\cite{minev_energy_2021}. The numerical values are in general more accurate. However, the analytical values are useful to understand trends and constrains, and are computationally faster to calculate when many modes are considered. The table also presents some of the typical values for each parameter. The readout frequency should be within the HEMT bandwidth between $4$ to $8\,\textnormal{GHz}$ and, although the transmon and cavity can be outside this range, they are usually not too far from it. The transmon anharmonicity is desirably as high as possible, but this comes at the cost of increasing $\omega_T$ or reducing the transmon factor $E_J/E_C$, so $\alpha$ is most often constrained to $\lesssim 300\,\textnormal{MHz}$. The dispersive shift is mostly a design choice, and can be either in the strong dispersive~\cite{schuster_resolving_2007} or the weak dispersive regime~\cite{eickbusch_fast_2022}. The last three terms, namely the cavity and resonator self-Kerr and the cross-Kerr between them, are often undesirable and designed to be as small as possible.

\begin{table}[htp]
\centering
\begin{tabular}{l c c c}

Term & Numerical calculation & Analytical formula & Typ. range ($1/2\pi$) \\
\addlinespace
\toprule
\addlinespace
Eigenfrequency $\omega_\nu$ & 
$\langle \hat{\nu}\hat{H}\hat{\nu}^\dagger\rangle$ & 
$\omega_{\nu, L} - \frac{1}{2} E_J (\varphi^\Lambda_\nu)^4$ & 
$4$ to $8\,\text{GHz}$ \\ 
\addlinespace
\hline
\addlinespace
\makecell[l]{Transmon\\ Anharmonicity $\alpha$}  & 
$2\omega_T -\langle \frac{1}{2}\hat{b}^2\hat{H}\hat{b}^{\dagger2}\rangle$ & 
$\frac{1}{2} E_J (\varphi^\Lambda_b)^4$ & 
$100$ to $300\,\text{MHz}$ \\
\addlinespace
\hline
\addlinespace
Dispersive shift $\chi_{\nu b}$  & 
$  \omega_\nu + \omega_T - \langle\hat{\nu}\hat{b}\hat{H}\hat{b}^{\dagger}\hat{\nu}^{\dagger}\rangle $ & 
$ E_J (\varphi^\Lambda_b\varphi^\Lambda_\nu)^2$ & 
\makecell{$0.5$ to $2.5\,\textnormal{MHz}$ \cite{schuster_resolving_2007} \\ $\lesssim0.5\,\textnormal{MHz}$ \cite{eickbusch_fast_2022} } \\
\addlinespace
\hline
\addlinespace
\makecell[l]{Readout/Cavity\\ self-Kerr $K_\nu$}  & 
$2\omega_\nu -\langle \frac{1}{2}\hat{\nu}^2\hat{H}\hat{\nu}^{\dagger2}\rangle$  & 
$\frac{1}{2} E_J (\varphi^\Lambda_\nu)^4$ & 
$\lesssim \text{kHz}$ \\
\addlinespace
\hline
\addlinespace
Cross-Kerr $K_{ac}$   & 
$ \omega_a + \omega_c - \langle \hat{c}\hat{a}\hat{H}\hat{a}^{\dagger}\hat{c}^{\dagger}\rangle$ & 
$ E_J (\varphi^\Lambda_a\varphi^\Lambda_c)^2$ & 
$\lesssim \textnormal{kHz}$ \\
\addlinespace
\bottomrule

\end{tabular}
\caption[Design guide for a cavity-transmon-readout circuit in the dispersive regime]{\textbf{Design guide for a cavity-transmon-readout circuit in the dispersive regime.} The expectation values are taken for vacuum state. The numerical formulas assume the Hamiltonian offset $\langle\hat{H}\rangle = 0$. }
\label{tab:dispersive_params}
\end{table}

The dispersive approximation suffices for fixed-frequency circuits designed in this regime. However, we are interested in systems with flux-tunable transmons, which operate at many different effective $L_J$s and may not always be in the dispersive regime. A more meaningful figure of merit in this case is the coupling factor $g$ in Eq.~\ref{eq:transmon_cavity}. It is only weakly dependent on $L_J$ as $g\propto n_{zpf,b} \propto\left(1/L_J\right)^{\frac{1}{4}}$ so it can be assumed to be constant over a wide frequency range, and it is directly measurable in avoided-crossing experiments. Fortunately we can de-hybridize the Hamiltonian by realizing that the Bogoliubov transformation angle for the two-mode case is given by
\begin{equation}
    \Lambda = \arctan{\left(\frac{\varphi^\Lambda_a}{\varphi^\Lambda_b}\right)}.
\end{equation}
Reverting the transformation gives the expressions for the de-hybridized parameters
\begin{equation}
    g = \big( \omega_{T, L} - \omega_{a, L} \big) \frac{\varphi^\Lambda_a \varphi^\Lambda_b}{\left(\varphi^\Lambda_a\right)^2 + \left(\varphi^\Lambda_b\right)^2},
\end{equation}
\begin{equation}
    \omega_{a, 0} = \frac{\omega_{a, L} \left(\varphi^\Lambda_a\right)^2+ \omega_{T, L} \left(\varphi^\Lambda_b\right)^2}{\left(\varphi^\Lambda_a\right)^2 + \left(\varphi^\Lambda_b\right)^2}, \, \, \, \, \,  \omega_{T, 0} = \frac{\omega_{a, L} \left(\varphi^\Lambda_b\right)^2+ \omega_{T, L} \left(\varphi^\Lambda_a\right)^2}{\left(\varphi^\Lambda_a\right)^2 + \left(\varphi^\Lambda_b\right)^2}.
\end{equation}
These parameters can be calculated from a single finite-element simulation to produce a Hamiltonian that is valid for a large range of $L_J$s. The $g\propto \left(1/L_J\right)^\frac{1}{4}$ dependence can also be considered analytically for a more precise description. 

These techniques can be used to precisely design the system parameters and enable the experimental manipulation of quantum phenomena. The following section links the design choices to common quantum information processing tasks, highlighting the techniques they enable as well as the limitations of different parameter regimes.

\section{Quantum information processing with light-matter interactions}
\label{sec:LMIinQIP}

When talking about quantum information processing, I refer to the three main steps of any quantum dynamics experiment: the preparation of an initial state, the evolution of the system, be it through always-on dynamics or by active control, and the characterization of the final results. For all the these steps, circuit QED relies mostly on dispersive interactions~\cite{ma_quantum_2021}. 

\begin{figure}
    \centering
    \includegraphics[width=\linewidth]{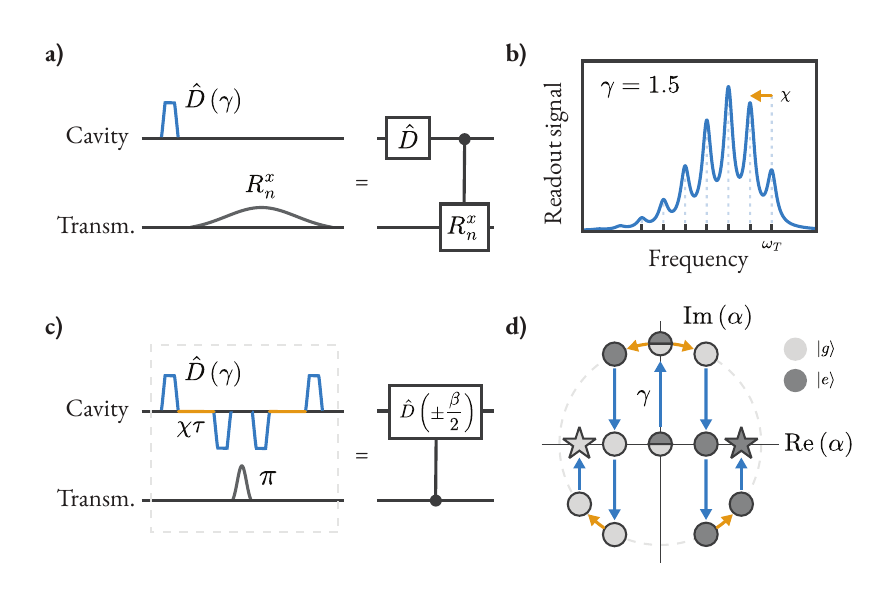}
    \caption[Cavity control primitives in the strong and the weak dispersive regimes]{\textbf{Cavity control primitives in the strong and the weak dispersive regimes.} a) Long qubit drives interact with the dispersive shift Hamiltonian and become selective of the number of photons in the cavity. b) Selective qubit pulses can be used to measure the photon statistics in the cavity in a number-splitting spectroscopy experiment. c) Echo conditional displacement ($\text{ECD}$) gate sequence. The echo $\pi$ qubit pulse in the middle of the sequence makes the result less sensitive to transmon frequency offsets. d) Evolution in phase space. The idle intervals $\tau$ lead to transmon-state rotations $\pm\tfrac{1}{2}\chi\tau$, which generate entanglement between the two circuits.}
    \label{fig:ecd}
\end{figure}

The simplified cavity-transmon dispersive dynamics are given by
\begin{equation}
\label{eq:dispersive_transmon_cav2}
    \hat{H}_{disp} = -\frac{\omega_T}{2}\hat{\sigma}_z + \omega_a \hat{a}^\dagger\hat{a} - \frac{\chi_{ab}}{2}\left(\mathbb{I} -\hat{\sigma}_z\right)\hat{a}^\dagger \hat{a} - \frac{ K_a}{2}\hat{a}^\dagger\hat{a}^\dagger\hat{a}\hat{a}.
\end{equation}
When $\chi_{ab}$ is larger than the fastest decoherence rate of the system (most often the transmon decoherence measured by $T_2$), it is said to be in the \textit{strong dispersive regime}. Under these conditions, the coherent dynamics last long enough that long transmon pulses (with narrow bandwidth) can resolve the transition frequencies $\omega_T - \chi_{ab}\hat{a}^\dagger\hat{a}$ associated with different number of photons in the cavity. The photon-number-splitting spectroscopy (Figure~\ref{fig:ecd}a-b) is the hallmark demonstration of this regime~\cite{schuster_resolving_2007}. It is then possible to apply selective qubit operations as
\begin{equation}
    R^{x}_z(\theta) = |z\rangle\langle z|\otimes \exp\left(-\frac{i\theta}{2}\hat{\sigma}_x\right).
\end{equation}
In other words, the qubit will follow different trajectories depending on the cavity state, which is a powerful primitive for the universal control of the cavity. For example, the selective number-dependent arbitrary phase (SNAP) gate~\cite{krastanov_universal_2015, heeres_cavity_2015} uses selective pulses to add a $z$-dependent Berry phase
\begin{equation}
    S(\theta_0, \theta_1,...) = \sum_{z\in\mathbb{N}} |z\rangle\langle z|e^{i\theta_z}.
\end{equation}
A numerically-optimized sequence of cavity displacements and SNAP gates can create arbitrary cavity states, including non-Gaussian resources such as Fock states $|z\rangle$. 

Universal control can also be achieved by playing both qubit and cavity drives simultaneously. Each drive's waveform is numerically optimized with a procedure known as Gradient-Ascent Pulse Engineering (GRAPE)~\cite{heeres_implementing_2017} to achieve a target evolution for known Hamiltonian parameters. This is a versatile way to build complex unitaries, and it has been continuously improved to consider open systems~\cite{chen_robust_2025} and closed-loop dynamics~\cite{porotti_gradient_2023}, for example.  

One example of the versatility of the strong dispersive regime is the ability to implement arbitrary observable measurements on the cavity state. In a collaboration with Tanjung Krisnanda, we have demonstrated how numerical optimization can map complex cavity observables onto the qubit~\cite{krisnanda_direct_2025}. This procedure compresses information that is distributed across phase space within the single-point measurement of the qubit readout, removing the need to tomographically reconstruct the cavity state. We use this technique to measure several observables of practical interest, including phase-space quadratures, the non-Gaussian rank, and projections onto target states.  

However, the strong dispersive regime has a considerable caveat: the cavity self-Kerr $K_a$. Despite being much slower than the      dispersive shift, it can become a relevant factor of distortion during longer experiments. This is especially harmful for parametric operations such as beamsplitting~\cite{gao_programmable_2018}, which rely on strong coupling.

A recently developed strategy to mitigate self-Kerr is to leverage operations in the weak dispersive regime. Given that $K_a\propto \left(\varphi^\Lambda_a\right)^4$ while $\chi\propto \left(\varphi^\Lambda_a\right)^2$, decreasing the transmon-cavity hybridization by decreasing the $g/\Delta = \varphi^\Lambda_a/\varphi_{zpf,b}$ ratio can help separate the harmful term from the more useful one. The downside is that, at low values of $\chi/2\pi\lesssim 100\,\textnormal{kHz}$, selective qubit rotations become disadvantageous, if not impossible to execute. The workaround is to use techniques specially designed for the weak dispersive regime such as the echo conditional displacement ($\text{ECD}$). The idea is to use a large displacement $\hat{D}\left(\gamma\right)$ as a lever for the dispersive shift, which after a time $\tau$ leads to a rotation arc of $\approx\pm\frac{1}{2}\gamma\chi\tau$ with a direction depending on the qubit state (orange arrows in Fig.~\ref{fig:ecd}d)\footnote{The ($\text{ECD}$) assumes the phase space coordinates rotate with frequency $\omega_a -\chi/2$, so the cavity state rotates with a relative frequency $-\tfrac{1}{2}\chi\hat{\sigma}_z$.}. The ($\text{ECD}$) gate uses this procedure twice, symmetrically in phase space and with a qubit $\pi$ pulse in between to cancel out the effect of spurious detunings (Fig.~\ref{fig:ecd}c). The effective gate is
\begin{equation}
    \textnormal{ECD}\left(\beta\right) = \hat{D}\left(-\frac{\beta}{2}\right)|g\rangle\langle e| +\hat{D}\left(\frac{\beta}{2}\right)|e\rangle\langle g|,
\end{equation}
which displaces the cavity to opposite directions depending on the qubit state. Note there is a residual qubit flip during the protocol, which can be removed with a non-selective $\pi$ pulse at the end of the sequence. Sequences of $\pi$ pulses and $\text{ECD}$ gates can be numerically optimized to grant universal control in the weak dispersive regime~\cite{eickbusch_fast_2022}.

This dichotomy between strong and weak interaction regimes extends to tomography as well. The Wigner function is accessible in the strong dispersive regime with relative ease, but becomes infeasible at small values of $\chi$. To see that, we can rewrite the Wigner function (Eq.~\ref{eq:wigner_1}) in terms of displacements and the parity operator $\hat{\Pi}= e^{i\pi a^\dagger a}$ as
\begin{equation}
    W(\alpha) = \frac{2}{\pi}\textnormal{Tr}\left[\hat{D}(\alpha) \boldsymbol{\rho}\hat{D}(-\alpha)\hat{\Pi}\right]
\end{equation}
This means the Wigner tomography can be measured by first displacing the cavity and then executing the parity measurement protocol 
\begin{equation}
    \hat{\Pi} = R_x\left(\frac{\pi}{2}\right)\exp\left[ i\frac{\pi}{2}\left(\mathbb{I}-\hat{\sigma}_z\right)\hat{a}^\dagger \hat{a}\right]R_x\left(\frac{\pi}{2}\right),
\end{equation}
corresponding to two non-selective qubit $\pi/2$-pulses separated by a free evolution under the dispersive shift interaction for a time $t_W =\pi/\chi_{ab}$. 

This parity measurement procedure becomes highly ineffective in the weak dispersive regime, since $t_W$ becomes large compared with the coherence times. Instead, at low $\chi$ it is more reasonable to measure the characteristic function $C(\beta)$ of $\boldsymbol{\rho}$, which can be defined as the expectation value of the displacement operator
\begin{equation}
    C(\beta) = \textnormal{Tr}\left[ \boldsymbol{\rho} D(\beta)\right].
\end{equation}
The characteristic function is related to the Wigner function by a double Fourier transform
\begin{equation}
    W(\alpha) = \frac{1}{\pi^2} \int C(\beta) e^{\alpha\beta^* - \alpha^*\beta} d^2\beta.
    \label{eq:Char_fourier}
\end{equation}
which means they are complex functions and convey the same information as the Wigner tomography. Turns out it is possible to map the real and imaginary part of $C(\beta)$ onto the qubit coordinate $\langle\hat{\sigma_z}\rangle$ by using an $\textnormal{ECD}\left(\beta\right)$ gate in-between two $\pi/2$ pulses~\cite{fluhmann_direct_2020, eickbusch_fast_2022}. The phase $\theta$ of the second qubit pulse relative to the first determines whether the real or imaginary part of the characteristic function is measured, according to
\begin{equation}
    \langle\hat{\sigma}_z\rangle = \textnormal{Re}\left[C(\beta)\right]\cos\theta 
    + \textnormal{Im}\left[C(\beta)\right]\sin\theta.
\end{equation}

So here we have two distinct ways to do QIP in bosonic circuit QED, each with their own control and tomography techniques. Which one is accessible in experiments depends on the value of $\chi$, which is set and fixed during the design and fabrication of the circuit. This begs the question: can we break from this constraint, and reap the benefits of different interaction regimes at the same time? The answer is yes, as long as $\chi$ can be changed in situ. In the next chapter, I will explain in detail how this is possible by using flux-sensitive devices with real-time control of the Hamiltonian. And how this control opens up much more possibilities for quantum information processing and the emulated dynamics of light and matter.

\SetPicSubDir{ch-Rice}
\SetExpSubDir{ch-Rice}

\chapter{Flux control of superconducting circuits}
\label{ch:flux}
\vspace{2em}

\newcommand{\figpartref}[2]{\cref{#1}#2}

The dispersively coupled cavity-transmon system discussed in the last chapter is an excellent setup for quantum information processing. However, it is still rather limited compared with the broader range of dynamics that bosonic circuit QED can offer.

Let us revisit the Jaynes-Cummings Hamiltonian in Equation~\ref{eq:jaynes-cummings}. There are two main parameters that dictate the dynamics of the system: the coupling factor $g$ and the detuning $\Delta = \omega_{T}-\omega_{a}$. The coupling $g$ is set by the gate capacitance between the transmon pads and the cavity, which is determined by their physical dimensions. The detuning $\Delta$, defined by the resonant frequencies of each circuit, is also ultimately set by the geometrical features such as inductances, capacitances, and the size of the junction's oxide barrier, which defines $E_J$. In other words, all of these features are a priori fixed during the fabrication of the device.

Still, it is highly desirable to manipulate the interaction parameters during experiments. For example, control over $\Delta$ enables changing $\chi\approx \alpha g^2/\Delta^2$ to move between the weak and strong dispersive regimes, or even access resonant interactions at $\Delta = 0$. This is useful in experiments that need to switch quickly between qualitatively different interactions to produce more complex dynamics~\cite{park_efficient_2024} or to dynamically suppress nonlinearities~\cite{terhal_encoding_2016}. Tuning into resonance also activates the direct exchange of energy between the cavity and the transmon to implement fast operations~\cite{vigneau_quantum_2025}, or it can be used to make higher energy levels degenerate and implement two-qubit gates~\cite{strauch_quantum_2003, dicarlo_demonstration_2009}. In larger systems, tuning circuit frequencies helps avoid (or harness) the transition into the chaotic phase of the Bose-Hubbard model, in which the controllability of the system is compromised~\cite{berke_transmon_2022}.

Beyond $\Delta$, the post-fabrication control of $g$ gives the opportunity to explore a much wider range of dynamics. The cavity-transmon coupling can be completely switched off at $g=0$~\cite{wu_efficient_2018}, something that is only approximated in the limit $\Delta \to \infty$. This allows the detection of virtual photons in ultrastrongly-coupled systems~\cite{giannelli_detecting_2024}, which require a fast decoupling from the light field. Fine-tuning $g$ also allows building parity-time symmetric systems~\cite{lu_pt_2023} and exceptional points~\cite{partanen_exceptional_2019}, which require a careful balance of the energy flow within the circuit. Similarly, $g$ may be calibrated to put the system in critical points of driven-dissipative phase transitions, which are highly susceptible to external disturbances and so are useful for quantum sensing~\cite{fernandez_quantum_2017}.

We could make even more radical changes to the system's dynamics by modifying the nonlinear potential $\hat{H}_{nl}$. In the case of the transmon, $\hat{H}_{nl} = - E_J\left(\cos\hat{\varphi}_{b} + \tfrac{1}{2}\hat{\varphi}_{b}^2\right)$ yields only Hamiltonian terms that are even powers of $\hat{\varphi}$, leading to undesirable $\propto \hat{\varphi}^4$ nonlinearities such as self-Kerr. Alternative circuits such as the SNAIL can provide a $\propto\hat{\varphi}^3$ potential, which is a resource for cavity control through three-wave mixing interactions while mitigating nonlinear distortions~\cite{frattini_wave_2017}. With a LINC circuit instead, the nonlinearities may be completely switch off when idle, and at the same time activate clean interactions with one or more cavities~\cite{maiti_controlling_2025, maiti_linear_2025}.

All of these applications have one feature in common: they can be implemented with circuits that are sensitive to magnetic flux. The simplest example is replacing the single Josephson junction of the transmon with a symmetric SQUID loop threaded with an external flux $\Phi_e$. This modifies the potential as
\begin{equation}
    E_J \to E_J\cos\left(\frac{2\pi}{\Phi_0}\Phi_e\right),
\end{equation}
introducing tunability to the system. The SNAIL, the LINC, and other circuits such as the fluxonium~\cite{manucharyan_fluxonium_2009} and the $0-\pi$ qubit~\cite{brooks_protected_2013} all require flux tunability to implement more complex dynamics. Therefore, to unlock such a broad range of applications in bosonic circuit QED, it is necessary to master the flux control of systems featuring high-$Q$ cavities.

This chapter addresses all the details that I found necessary to understand, design, simulate, and set up experiments involving flux-tunable circuits. I will start by explaining the physics underlying the flux sensitivity of SQUID loops. I then discuss the accurate modeling of a SQUID-transmon in a time-dependent field, pointing out useful approximations for numerical simulations. I show how periodic flux drives can be used to activate parametric resonances with other circuits such as a microwave cavity with a modulated coupling factor $g$. I also address other practical concerns including the engineering of flux sweet spots (configurations with minimum sensitivity to flux noise), and the specific wiring of flux lines inside a dilution refrigerator. The final section expands on the problem of adding flux control to bosonic superconducting structures, setting up the challenges to be addressed in the next chapters and giving other researchers the fundamentals for developing their own solutions.

\section{Why are SQUIDs sensitive to flux?}

\begin{figure}
    \centering
    \includegraphics[width=\linewidth]{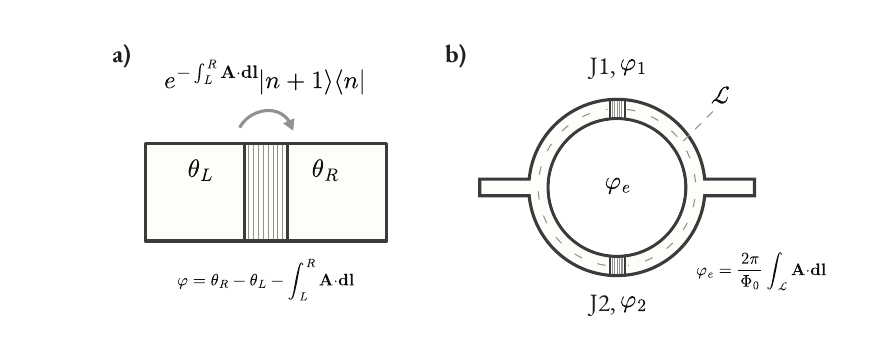}
    \caption[Introducing flux sensitivity with a SQUID]{\textbf{Introducing flux sensitivity with a SQUID.} (a) A Josephson junction has two superconducting terminals, $\textnormal{L}$ and $\textnormal{R}$, separated by a barrier. The number of Cooper pairs that have crossed the junction is given by $\hat{n}$. (b) The SQUID loop is made of two junctions $J_1$ and $J_2$. The external magnetic flux $\Phi_e$ threading the loop generates a curling vector potential $\mathbf{A}$, which interacts with the superconductors phases $\theta_1$, $\theta_2$, This leads to the fluxoid quantization relation between the junctions' gauge-invariant phases $\varphi_1$, $\varphi_2$ and the external flux.}
    \label{fig:JJ_SQUID}
\end{figure}

As we have discussed in the case of transmons, the phase variable $\hat{\varphi}$ across a Josephson junction forms a conjugate pair with the difference in the number of Cooper pairs $\hat{n}$ at each superconducting terminals. Their eigenvectors are related by the Fourier transform
\begin{equation}
    |n\rangle = \frac{1}{2\pi}\int_0^{2\pi}e^{-in\varphi}|\varphi\rangle d\varphi.
\end{equation}
This equation tells us something about the physical meaning of $\hat{\varphi}$. A transformation $\hat{\varphi} \to \hat{\varphi} + \varphi_0$ leads to $|n\rangle \to e^{-in\varphi_0}|n\rangle$ for all $n$. Meaning that tunneling events should add factor of $e^{-i\varphi_0}$ to each Cooper pair crossing the junction so they can accommodate to the phase of the superconducting ground state of the electrode.

This raises a problem. The phases of the superconducting BCS ground states are not gauge-invariant. They depend on the gauge choice of the local vector potential $\mathbf{A}$. But $\hat{\varphi}$ is a physical observable, so it must be gauge-invariant! So $\hat{\varphi}$ cannot be directly equal to the superconducting phase difference, although both quantities are related up to a factor depending on $\mathbf{A}$. The full argument in terms of BCS theory can be found in Ref.~\cite{liao_circuit_2025}, but a simplified solution can be made in terms of the macroscopic wavefunction. Assume the superconducting terminals are described by the state
\begin{equation}
    |\Psi(\mathbf{r})\rangle = \sqrt{\rho}e^{i\theta(\mathbf{r})}.
\end{equation}
Here, $\rho$ is the Cooper pair density and $\theta(\mathbf{r})$ is the superconducting phase at a point $\mathbf{r}$. The current density is given by
\begin{align}
        \mathbf{J} = & \, \, \frac{e}{2m_e}\left\{\Psi^*\left(-i\nabla - 2e\mathbf{A}\right)\Psi + \left[ \Psi^*\left(-i\nabla - 2e\mathbf{A}\right)\Psi\right]^*\right\} \\
        = & \, \, \frac{\rho e^2}{m_e}\left(\frac{\Phi_0}{2\pi}\nabla\theta - \mathbf{A}\right),
\end{align}
Where $e$ and $m_e$ are the electron charge and mass. The current is zero in the bulk of the superconductor, so at this point
\begin{equation}
    \label{eq:phase_field_relation}
    \frac{\Phi_0}{2\pi}\nabla\theta = \mathbf{A}.
\end{equation}
The freedom in gauge choice $\mathbf{A} \to \mathbf{A} + \nabla\lambda$ changes the local superconducting phase as $\theta \to \theta + \lambda$, where $\lambda$ can be a function of the position $\mathbf{r}$.

The way to fix the relation between $\theta$ and $\hat{\varphi}$ is by removing the gauge dependence subtracting the contribution from $\mathbf{A}$:
\begin{equation}
    \label{eq:gauge_invariant_phase}
    \hat{\varphi} = \Delta\hat{\theta} -\frac{2\pi}{\Phi_0}\int^R_L \mathbf{A}\cdot\mathbf{dl}.
\end{equation}
Here $\Delta \hat{\theta}_{} = \hat{\theta}_R - \hat{\theta}_L$ measures the difference in ground state phases and the path integral connects both terminals. This leads to a correction to the tunneling Hamiltonian known as the Peierls substitution
\begin{equation}
    \label{eq:JJ_H_charge_Peierls}
    \hat{H}_T = -\frac{1}{2}E_J\sum_n e^{i\frac{2\pi}{\Phi_0}\int \mathbf{A}\cdot\mathbf{dl}}|n\rangle\langle n+1| + e^{-i\frac{2\pi}{\Phi_0}\int \mathbf{A}\cdot\mathbf{dl}}|n+1\rangle\langle n|,
\end{equation}
The extra phase factor is absorbed into the definition of $\hat{\varphi}$. It does not change the dynamics of single junctions, but guarantees that the phase shift of Cooper pairs during tunneling does not depend on the gauge choice.

The importance of $\mathbf{A}$ to the flux sensitivity comes about when we consider a SQUID loop with junctions $J_1$ and $J_2$ (Figure~\ref{fig:JJ_SQUID} b). An external magnetic flux $\Phi_e$ threading the loop is given by an integral of $\mathbf{A}$ along the loop perimeter
\begin{equation}
    \label{eq:external_flux}
    \Phi_e = \int_{\mathcal{L}} \mathbf{A} \cdot \mathbf{dl} = \int_{\textrm{superc.}}\mathbf{A} \cdot \mathbf{dl} + \int_{\textrm{junctions}}\mathbf{A} \cdot \mathbf{dl}.
\end{equation}
The wavefunction phase is continuous along the same path, so
\begin{equation}
    \label{eq:phase_cont}
    \int_{\textrm{superc.}}\nabla \theta \cdot \mathbf{dl} +  \Delta\theta_{J_1} +  \Delta\theta_{J_2} = 2\pi k,
\end{equation}
where $\Delta\theta_{J_i}$ is the phase drop along $J_i$. The phase and the vector potential within the superconductors are related according to Equation~\ref{eq:phase_field_relation}, so we can combine Equations~\ref{eq:external_flux} and~\ref{eq:phase_cont} as
\begin{equation}
    \Phi_e = \Phi_0k + \frac{\Phi_0}{2\pi}\Delta\theta_{J_1} + \frac{\Phi_0}{2\pi}\Delta\theta_{J_2} - \int_{J_1}\mathbf{A} \cdot \mathbf{dl} + \int_{J_2}\mathbf{A} \cdot \mathbf{dl},
\end{equation}
or, using the definition of the junction phases,
\begin{equation}
    \label{eq:fluxoid_quantization}
    2\pi k + \varphi_{1} + \varphi_{2} = \frac{2\pi}{\Phi_0}\Phi_e = \varphi_e.
\end{equation}
This equation is known as the fluxoid quantization, and it summarizes the impact of the reduced magnetic flux $\varphi_e$ over the SQUID loop's dynamics by controlling the relationship of $\varphi_1$ and $\varphi_2$. By incorporating this SQUID into a transmon circuit, we can make its effective Josephson energy $E_{J_{eff}}$ into a function of $\varphi_e$.

\section{Tunable transmons in a time-dependent field}

When the single junction is replaced by a SQUID loop, the transmon Hamiltonian is modified by simply adding two potential terms:
\begin{equation}
    \hat{H} = 4E_C\hat{n}^2 - E_{J_1}\cos\hat{\varphi}_1 - E_{J_2}\cos\hat{\varphi}_2.
\end{equation}
But it is not immediately clear how $\varphi_e$ comes in this equation. We could replace $\varphi_1 \to \varphi_2 - \varphi_e$, or instead $\varphi_2 \to \varphi_1 - \varphi_e$, or in many different combinations. Turns out that this seemingly innocuous choice can lead to physically distinct results for the transmon relaxation rates under flux noise~\cite{you_circuit_2019}.

The correct formulation uses the irrotational constraint~\cite{you_circuit_2019, didier_analytical_2018, lagemann_real_2023}
\begin{equation}
\begin{aligned}
 \varphi_1 &\to \varphi +\frac{1}{2}\varphi_e,  \\
 \varphi_2 &\to -\varphi +\frac{1}{2}\varphi_e,
\end{aligned}
\end{equation}
where the variable $\varphi$ is implicitly defined. This substitution avoids the need to include terms proportional to $\tfrac{d}{dt}\varphi_e$ in the Hamiltonian, significantly simplifying simulations. The Hamiltonian can then be rewritten in terms of a single tunable potential
\begin{equation}
\label{eq:correct_H}
    \hat{H} = 4E_C\hat{n}^2 - E_{J_{eff}}\cos\left(\hat{\varphi} - \varphi_{eff}\right), 
\end{equation}
where 
\begin{align}
    E_{J_{eff}} & = \left(E_{J_1} + E_{J_2}\right)\sqrt{\cos\left(\frac{\varphi_e}{2}\right)^2 + d^2\sin\left(\frac{\varphi_e}{2}\right)^2 },\\
    \varphi_{eff} & = \arctan\left(d\tan\left(\frac{\varphi_e}{2}\right)\right), \, \, \textnormal{and} \\
    \label{eq:junction_asymmetry}
    d & = \frac{E_{J_1} - E_{J_2}}{E_{J_1} + E_{J_2}}.
\end{align}
Simply taking $\varphi_e = \varphi_e(t)$ in these equations accurately represents the time evolution of the transmon.

To numerically simulate the dynamics of a SQUID-transmon, we must be mindful about the basis choice: the zero-point fluctuations in the ladder operator $\hat{b} = \tfrac{1}{2\varphi_{zpf}}\hat{\varphi} - \frac{i}{2n_{zpf}}\hat{n}$ are a function of $\varphi_e$, so the transmon excitation basis $|g\rangle,\, |e\rangle,|f\rangle$ becomes flux-dependent. On the other hand, eigenstates such as $|n\rangle$ can be taken as constant, since they count the number of Cooper pairs regardless of the magnetic field. Therefore, it is convenient to write Eq.~\ref{eq:correct_H} in the charge basis:
\begin{align}
    \label{eq:H_charge_basis}
    \hat{H} = 4E_C\sum_n n^2|n\rangle\langle n| -\frac{1}{2}E_{J_{eff}}\sum_n e^{-i\varphi_{eff}}|n\rangle\langle n+1| + e^{i\varphi_{eff}}|n+1\rangle\langle n |.
\end{align}
A faithful representation of the transmon eigenstates might require a large number of $|n\rangle$, making them computationally costly. For example, Ref.~\cite{hellings_calibrating_2025} uses 31 dimensions by truncating $|n|\leq15$. Another drawback is that the instantaneous eigenenergies of the transmon are not explicit, making it less obvious how to move into a rotating frame.

\begin{figure}
    \centering
    \includegraphics[width=\linewidth]{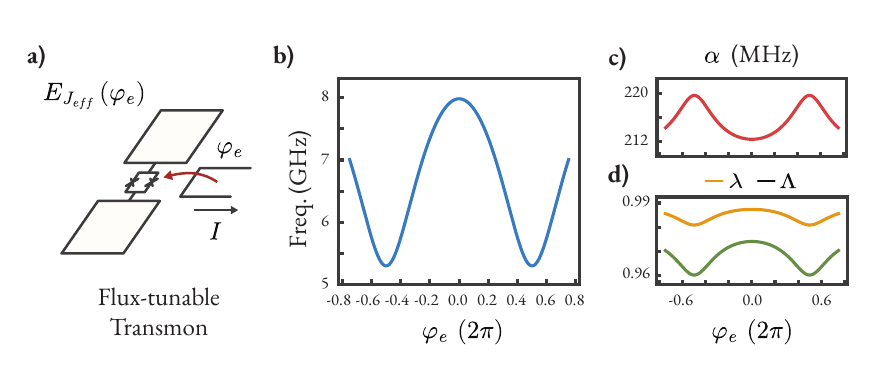}
    \caption[Flux tunability of transmon parameters]{\textbf{Flux tunability of transmon parameters.} (a) A flux-tunable transmon made of two capacitive pads connected by a SQUID loop. The field supplied to the SQUID is generated by a coupled flux line. (b-d) Simulations of transmon parameters as a function of flux for $d=0.454$, maximum $\omega_T/2\pi = 8\,\textnormal{GHz}$, $E_C = 200\,\textnormal{MHz}$. The transmon frequency (b) can be tuned over a large range. The anharmonicity (c) and the coefficients of the charge operator (d) are weakly affected by external flux and are often taken as constant. }
    \label{fig:SQUID_tunability}
\end{figure}

Reformulating the transmon Hamiltonian in the instantaneous eigenbasis avoids these issues. The derivation is quite lengthy, but it is well-explained in~\cite{didier_analytical_2018}. It consists of diagonalizing Equation~\ref{eq:correct_H} perturbatively as a function of the parameter $\xi(t) = \sqrt{2E_C/E_{J_{eff}}(t)}$ to get
\begin{equation}
\label{eq:instant_diag_transmon}
    \hat{H} = \sum_{j\in\{g,e,f\}} E_j\hat{\Pi}_j -\dot{\varphi}_{eff}\hat{n} - \frac{\dot{\xi}}{\xi}v\hat{S}_y. 
\end{equation}
The operator $\hat{\Pi}_j$ is a projector onto the $j$-th instantaneous eigenstate with eigenenergy $E_j$, and the Hamiltonian is truncated to $j\in\{g,e,f\}$. The two last terms are proportional to $\dot{\varphi}_{eff}$ and $\dot{\xi}$, describing non-adiabatic transitions caused by the fast flux dynamics. $\hat{S}_y$ is a Pauli operator in the $\{g, f\}$ subspace, and the charge operator $\hat{n}$ has the form
\begin{equation}
    \hat{n} =  \frac{\lambda}{2\sqrt{\xi}}\hat{\sigma_y} + \frac{\Lambda}{\sqrt{2\xi}}\hat{s}_y,
\end{equation}
with $\hat{\sigma}_y$ and $\hat{s}_y$ the Pauli matrices for transitions $g \to e$ and $e \to f$. Now the situation is reversed: since $\hat{H}$ follows the instantaneous eigenbasis, the time-dependence is pushed back to $\hat{n}$. The series expansion for the parameters $E_i$, $\lambda$, $\Lambda$, and $v$ were calculated in Ref.~\cite{didier_analytical_2018} up to the 25th order, and Figure~\ref{fig:SQUID_tunability} shows how these parameters change as a function of external flux.

Equation~\ref{eq:instant_diag_transmon} is useful for several reasons. First, it makes explicit the flux dependence of the charge operator and the anharmonicity $\alpha = (E_e-E_g)-(E_f-E_g)$, which are often taken as constant for all $\varphi_e$. Second, it allows for the explicit evaluation of non-adiabatic transitions. Third, it makes it simple to move into the rotating frame by applying the unitary transformation 
\begin{equation}
\label{eq:inst_rotatingframe}
    U(t) = \exp\left(-i\int^t_0 dt'\sum_{j\in\{g,e,f\}} E_j\hat{\Pi}_j\right).
\end{equation}
This detailed treatment of time-dependent fields gives us the background necessary for the study of transmons under parametric modulation, as discussed in the next section.

\section{Parametric flux modulation}
\label{sec:param_flux_mod}

As a general formulation for tunable transmons, Equation~\ref{eq:instant_diag_transmon} is useful for any time-resolved numerical simulations with low computational overhead. However, for some specific forms of flux driving, further reformulation can help reveal  hidden dynamics.

This is the case for periodic flux drives. Intuitively, applying a $\varphi_e(t)$ with period $T_m$ induces the transmon parameters to oscillate at the same pattern. This oscillation mixes with the transmon natural evolution given by the average frequency $\bar{\omega}_T = \frac{1}{T_m}\int_0^{T_m}dt\left[E_e(t) - E_g(t)\right]$, creating sidebands at $\bar{\omega}_T +2\pi k/T_m$. The underlying mechanism is similar to how RF mixers combine microwave frequencies, using a nonlinearity to ``multiply'' two incoming signals. These remote sidebands allow the transmon to interact at points far detuned from its average or instantaneous frequencies while adjusting the interaction strength, a resource that has been proven useful for noise spectroscopy~\cite{nir_spectrum_2025}, parametric coupling~\cite{strand_first_2013, naik_random_2017, wu_efficient_2018}, and engineering of flux sweet spots~\cite{didier_ac_2019, didier_flux_2019, valery_dynamical_2022}. In this section, I will derive the parametric interaction between the transmon and a capacitively coupled harmonic oscillator under a bichromatic flux pulse, using similar techniques to those described in Refs.~\cite{didier_analytical_2018, wang_simulating_2019}. The ability to activate and adjust the strength of transmon-resonator coupling will be an important resource for quantum information processing in bosonic circuits.

Consider the external flux
\begin{equation}
\label{eq:flux_modulation}
    \varphi_e(t) = A_1\cos(\omega_mt) + A_2\cos(p\omega_mt + \theta) +\varphi_{dc},
\end{equation}
where $\varphi_{dc}$ is the parking flux, $\omega_m$ is the base angular frequency of the modulation, $p$ is an integer factor, and $\theta$ is the relative phase between the tones. Then consider the transmon is biased by such flux control, leading to a time-dependent Hamiltonian $H_T(t)$. If the transmon is also coupled to a harmonic oscillator, the Hamiltonian of the system is
\begin{equation}
\label{eq:res_transmon_coupling}
    \hat{H} = \omega_a \hat{a}^\dagger \hat{a} + \hat{H}_T(t) + ig\left(\hat{a}^\dagger - \hat{a}\right)\left(\lambda\hat{\sigma}_y + \sqrt{2}\Lambda\hat{s}_y\right).
\end{equation}
I am assuming the fastest modulation rate $p\omega_m$ is slow enough that non-adiabatic transitions are negligible~\footnote{To account for non-adiabatic evolution, this analysis can be done within Floquet theory. }. The coupling modulation becomes apparent in the rotating frame reached with the transformation
\begin{equation}
\label{eq:int_picture_instantaneous}
    \hat{U} = \exp\left(-i\omega_at \hat{a}^\dagger \hat{a}-i\int^t_0 dt'\sum_{j} E_j\hat{\Pi}_j\right).
\end{equation}
The energies $E_j(t)$ are modulated with a fundamental periodicity given by $\omega_m$ (\ref{eq:flux_modulation}), so they can be expressed as a Fourier series
\begin{flalign}
\label{eq:eigen_fourier_exp}
\hspace{0pt}
    & E_j(t) = \sum_{k=0}^\infty E_{jk}\cos\left(k\omega_m t +\theta_k\right) \notag
    \\ & \longrightarrow \int^t_0dt'E_j(t') = \bar{E}_j t + \sum_{k=1}^\infty\frac{E_{jk}}{k\omega_m}\left[\sin\left(k\omega_m t +\theta_k\right) - \sin\theta_k\right],
\end{flalign}
with $\bar{E}_j$ the time-averaged eigenvalue. Moving $\hat{H}$ to the rotating frame using Equations~\ref{eq:int_picture_instantaneous} and~\ref{eq:eigen_fourier_exp}, and using the rotating-wave approximation (RWA) to remove counter-rotating terms, the coupling term becomes:
\begin{align}
    \hat{H}_{int} & = \hat{a}^\dagger  |g\rangle\langle e|\times  g \lambda e^{i\gamma} e^{ 
 i\Delta_{ge}t-i\sum_{k=1}^\infty\frac{\omega_k}{ k \omega_m }\sin\left(k\omega_m t +\theta_k\right)} \notag \\
& + \hat{a}^\dagger|e\rangle\langle f|\times g \sqrt{2}\Lambda e^{i\gamma}   e^{ 
 i\Delta_{ef}t-i\sum_{k=1}^\infty\frac{\omega_k -\alpha_k}{k \omega_m }\sin\left(k\omega_m t +\theta_k\right)} + \textnormal{h.c.},
\end{align}
where $\Delta_{ge} = \omega_a - \bar{\omega}_T$ and $\Delta_{ef} = \omega_a - \left(\bar{\omega}_T - \bar{\alpha}\right)$ are the cavity detuning to the time-averaged transmon transitions, and the Fourier components $\omega_k$, $\alpha_k$ are calculated from the transitions between $E_{jk}$. For simplicity, I denote $\gamma = \sum_{k=1}^\infty\frac{\omega_k}{k\omega_m}\sin\theta_k$. The key step in exposing the modulation sidebands is to rewrite the sinusoidal exponents using the Jacobi-Anger expansion. This results in
\begin{align}
    \hat{H}_{int} & =  \hat{a}^\dagger  |g\rangle\langle e|\times g \lambda e^{i\gamma} e^{ 
 i\Delta_{ge}t}\prod_{k=1}^\infty\sum_{n=-\infty}^\infty J_n\left(\frac{\omega_k}{k\omega_m}\right)e^{-in\left(k\omega_m t + \theta_k\right)} \notag \\
& + \hat{a}^\dagger|e\rangle\langle f|\times g \sqrt{2}\Lambda e^{i\gamma}  e^{ 
 i\Delta_{ef}t}\prod_{k=1}^\infty\sum_{n=-\infty}^\infty J_n \left(\frac{\omega_k-\alpha_k}{k\omega_m}\right) e^{-in\left(k\omega_m t + \theta_k\right)} + \textnormal{h.c.},
\end{align}
where $J_n\left(\cdot\right)$ is the $n$-th order Bessel function of the first kind. This expression has multiple phase factors with frequencies $nk\omega_m$. They can be reorganized in a sum of terms oscillating at $N\omega_m$, $N\in\mathbb{Z}$:
\begin{equation}
    \prod_{k=1}^\infty\sum_{n = -\infty}^\infty J_n\left(\cdot\right)e^{-in\left(k\omega_m t + \theta_k\right)} = \\\sum_{N=-\infty}^\infty \ e^{-iN\omega_m t}\sum_{\{n_k\}} \prod_{k = 1}^\infty J_{n_k}\left(\cdot\right)e^{-in_k\theta_k}.
\end{equation}
The set $\{n_k\}$ represents all possible solutions to the diophantine equation $\sum_{k=1}^\infty n_kk = N$ ($n_k\in\mathbb{Z}$) for a given $N$. Each solution contributes to the coefficient of the $N$-th sideband with frequency $N\omega_m$. In numerical simulations, I limit the number of solutions by truncating $k, |n_k|\leq 6$. Factoring out the sideband exponentials:
\begin{align}
\label{eq:interaction_coupling}
    \hat{H}_{int} & = \sum_{N=-\infty}^\infty   \left[\lambda g^{(N)}_{ge} e^{ 
 i\Delta_{ge}t}\hat{a}^\dagger|g\rangle\langle e| +  \sqrt{2}\Lambda g^{(N)}_{ef} e^{ 
 i\Delta_{ef}t}\hat{a}^\dagger|e\rangle\langle f|\right]e^{-iN\omega_mt} + \textnormal{h.c.},
\end{align}
where
\begin{equation}
\label{eq:geff}
    g^{(N)}_{ge} = g e^{i\gamma}\sum_{\{n_k\}}\prod_{k=1}^\infty J_{n_k}\left(\frac{\omega_k}{k\omega_m}\right)e^{-in_k\theta_k},
\end{equation}
\begin{equation}
    g^{(N)}_{ef} = g e^{i\gamma}\sum_{\{n_k\}}\prod_{k=1}^\infty J_{n_k}\left(\frac{\omega_k-\alpha_k}{k\omega_m}\right)e^{-in_k\theta_k}.
\end{equation}

\begin{figure}
    \centering
    \includegraphics[width=\linewidth]{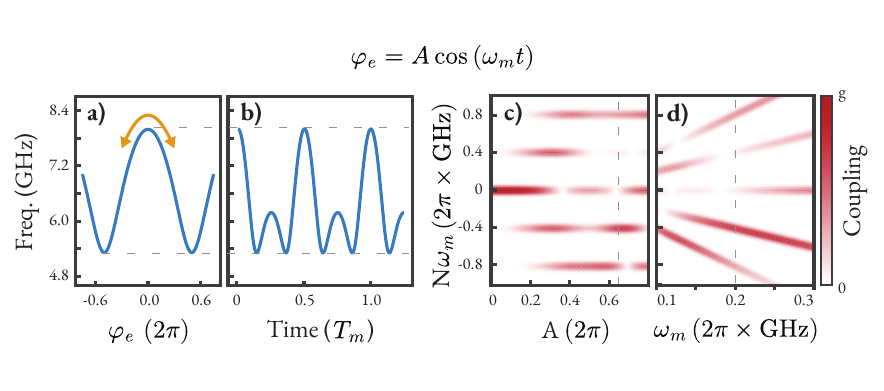}
    \caption[Parametric flux modulation]{\textbf{Parametric flux modulation.} (a) Transmon frequency spectrum $\omega_T(\varphi_e)/2\pi$ for $d=0.454$, maximum $\omega_T/2\pi = 8\,\textnormal{GHz}$, $E_C/2\pi = 200\,\textnormal{MHz}$. The arrow indicates the oscillation induced by a drive $\varphi_e = A\cos\left(\omega_m t\right)$. b) Transmon frequency over time. Notice how parking at $\varphi_{dc} = 0$ makes $\omega_T\left(t\right)$ oscillate with $2\omega_m$. c) Calculated $g_{eff}^{(N)}$ for $\omega_m/2\pi = 200\,\textnormal{MHz}$ as a function of $A$ and $N\omega_m$, with dashed line at $A/2\pi = 0.65$. Only the even $N$ sidebands show nonzero coupling. The bands are broadened in frequency by a Gaussian function to enhance visibility. d) $g_{eff}^{(N)}$ as a function of $\omega_m$ for $A/2\pi = 0.65$, demonstrating the control over the sideband frequency. The dashed line marks the point where $\omega_m/2\pi =  200\,\textnormal{MHz}$.}
    \label{fig:modulated_transmon}
\end{figure}

Following the discussions in Ref.~\cite{didier_analytical_2018}, the factors $g$, $\lambda$, $\Lambda$ and the anharmonicity $\alpha$ are well approximated as constants, in which case $\alpha_k \approx0$ for $k\geq1$, and $g^{\left(N\right)}_{ef} \approx g^{\left(N\right)}_{ge} = g^{\left(N\right)}_{eff}$. Then, the Hamiltonian simplifies further to
\begin{align}
\label{eq:interaction_coupling_final}
    \hat{H}_{int} & = \sum_{N=-\infty}^\infty   g^{(N)}_{eff}\hat{a}^\dagger\left[e^{ 
 i\Delta_{ge}t}\lambda |g\rangle\langle e| +  \sqrt{2}\Lambda e^{ 
 i\Delta_{ef}t}|e\rangle\langle f|\right]e^{-iN\omega_mt} + \textnormal{h.c.}
\end{align}
This is the main result of this section. It shows how the transmon can couple to the cavity through any of its sidebands located at frequencies $\bar{\omega}_T-N\omega_m$, each with adjustable coupling $g_{eff}^{(N)}$.

This formulation is rather general and useful for numerical simulations. But let us look at it in the simpler case of a single modulating tone ($A_2 = 0$). This control has only two free parameters: $\omega_m$ and $A_1$. If the transmon is parked at the point of maximum frequency ($\varphi_{dc} = 0$), the function $\omega_T(\varphi_e)$ oscillates at double the frequency of $\varphi_e$, leading to $\omega_k = 0$ for odd $k$. This symmetry results in only even $N$ sidebands having nonzero $g_{eff}^{(N)}$. Figure~\ref{fig:modulated_transmon} shows the resulting dynamics for such a case, with the calculated coupling factors as functions of modulation amplitude and frequency. If, however, $\varphi_{dc}$ is such that the frequency response with external flux is approximately linear, then the coefficients $\omega_k = \frac{d\omega_T}{d\varphi_e}A_1\delta_{k1}$ (for $k\geq 1$). This leads to the simple relation $g^{(N)}_{eff}=gJ_N\left(\frac{d\omega_T}{d\varphi_e}\frac{A_1}{\omega_m}\right)$, up to a phase factor. In both cases explored above, the flux drive degrees of freedom can be simultaneously calibrated to make the transmon and cavity parametrically resonant, i.e., $\Delta_{eg} = N\omega_m$ for a specific $N$, while at the same time picking a specific value for the coupling strength $g_{eff}^{(N)}<g$. 

We might also want to constrain $\bar{\omega}_T$ to a specific value, or to simultaneously adjust $g_{eff}^{(N)}$ for two different sidebands. In this case, the extra degrees of freedom offered by the bichromatic flux drive becomes useful, since it has four independent variables $A_1, A_2, \omega_m$ and $\theta$, while $p$ is desirably $2$ or $3$ (so $p\omega_m$ is not too large). In the next section, I will show how this flexibility is useful to engineer AC sweet spots at different values of $\bar{\omega}_T$.

There are two last practical observations that are important when implementing flux modulation in practice. First, we can use the same derivation to describe the interaction of the modulated transmon with a charged drive. Simply replace the cavity operator with a drive amplitude using the stiff pump approximation
\begin{equation}
    \hat{a}e^{-i\omega_a t} \longrightarrow \alpha(t)e^{-i\omega_d t},
\end{equation}
with $\alpha(t)$ proportional to the charge drive envelope with carrier frequency $\omega_d$. This means it is possible to access and control the transmon through any of its sidebands, and that this control will consider the transmon to be motionally averaged as long as the modulation frequency is much faster than the Rabi frequency of the drive~\cite{li_motional_2013}. Conversely, it means that a suppressed sideband $g^{(N)}_{eff} = 0$ is inaccessible to the external drive, which may lead to reduced Purcell decay. Second, $\omega_m$ should be chosen carefully. If $N\omega_m = \bar{\omega}_T$ for some $N$, the flux drive will trigger subharmonic driving that excites the transmon with a multi-photon process (see Ref.~\cite{nir_spectrum_2025}, Supplemental Material), affecting the dynamics. 

This discussion establishes the link between flux tunability and the control over the dynamics of light-matter coupling. It shows that SQUID-transmons can control the detuning $\Delta = \omega_a - \omega_T$ by means of changing $\omega_T$, and that they can modulate their effective coupling $g$ with light fields of microwave resonators by means of parametric flux drives. The analytical formulation also sets clear quantitative requirements for the hardware: the flux source must be strong and broadband enough to modulate the transmon with the target amplitudes $A$ and frequencies $\omega_m$. 

So far, I focused on the benefits of flux tunability. However, SQUIDs also inevitably introduce vulnerability to flux noise, which impacts the transmon coherences. In the next section, I will talk about how the flux modulation can be leveraged to construct sweet spots and mitigate the impact of noise.

\section{Noise and sweet spots}


The extra degree of control granted by flux tunability comes with a caveat: the SQUID also makes the transmon more prone to the flux noise from the environment or carried by the control lines. The noise makes the circuit frequency jitter, severely impacting the coherence time $T_2$ and posing a relevant limiting factor for quantum information processing. Thankfully, flux tunability provides tools to attenuate the impact of noise in the form of flux sweet spots. In this section, I will briefly discuss the nature of the noise and show techniques that guide the design of protected circuits.

The noise is modeled by modifying Eq.~\ref{eq:flux_modulation} to include the additive term $\delta \varphi_{dc}(t)$ to the flux offset, and a multiplicative term $\delta\varphi_{ac}(t)$ to the overall modulation amplitude:
\begin{align}
    \label{eq:flux_modulation_noise}
    \varphi(t) = & \, (\varphi_{ac} + \delta\varphi_{ac}(t))\left(a_1\cos(\omega_mt) + a_2\cos(p\omega_mt + \theta)\right) \\ & + \varphi_{dc} + \delta\varphi_{dc}(t).
\end{align}
Here I defined $a_{1,2} = A_{1,2}/\varphi_{ac}$ and $\varphi_{ac} = \sqrt{A_1^2 + A_2^2}$ so the noise acts proportionally to the amplitude of each tone. Both $\delta\varphi_{dc}$ and $\delta\varphi_{ac}$ can be assumed to have a power spectral density of the form 
\begin{equation}
    S(\omega) = \frac{2\pi}{|w|}  A_{1/f}^2+ A_{w}^2.
\end{equation}
$S(\omega)$ is defined as the Fourier transform of the correlation function $\langle\delta\varphi(0)\delta\varphi(t)\rangle$ with units of $1\textnormal{/Hz}$ and, intuitively, it probes the amplitude of the noise that jitters at a frequency $\omega$. 

The noise spectral density has two distinct frequency regions with different effects over the transmon. The factor $A_{1/f}^2$ describes the low-frequency $1/f$ noise, which is a known feature of electronic equipments~\cite{fleetwood_origins_2020}. It causes ``slow jittering'' errors that impact measurements over larger timescales such as the average frequency of a transmon. High-frequency errors, on the other hand, are mainly caused by the white noise term $A_{w}^2$. This noise is generated, for example, by microwave devices thermalized to a temperature $T$. The white noise level is proportional to the temperature as $A_{w}^2 = \left(\frac{2\pi}{\Phi_0}L_m\right)^2S_I(T)$, where $L_m$ is the mutual inductance between the line and the SQUID loop, and $S_I(T)$ is the Johnson-Nyquist thermal current noise. At a frequency range much lower than thermal excitation level $\omega\ll k_BT$ ($k_B$ the Boltzmann constant), the noise can be taken as a constant level $S_I(T) \approx 4 k_BT/R$ for an environment with impedance $R = 50\,\Omega$.

Let us first look at how the $1/f$ noise induces transmon decoherence, and how to suppress this effect. The decoherence rate $\Gamma_{\phi, 1/f}$ due to the $1/f$ noise is proportional to the sensitivity of the average frequency~\cite{didier_ac_2019, flux_didier_2019}:
\begin{equation}
\label{eq:1fnoise}
    \Gamma_{\phi, 1/f} \propto \sqrt{A_{dc,1/f}^2\left(\frac{\partial\bar{\omega}_T}{\partial\varphi_{dc}}\right)^2 + A_{ac,1/f}^2\left(\frac{\partial\bar{\omega}_T}{\partial\varphi_{ac}}\right)^2}.
\end{equation}
This equation points at the overall strategy for reducing the sensitivity to $1/f$ noise: the average transmon frequency must be made insensitive to variations in $\varphi_{dc}$ and $\varphi_{ac}$.

\begin{figure}
    \centering
    \includegraphics[width=\linewidth]{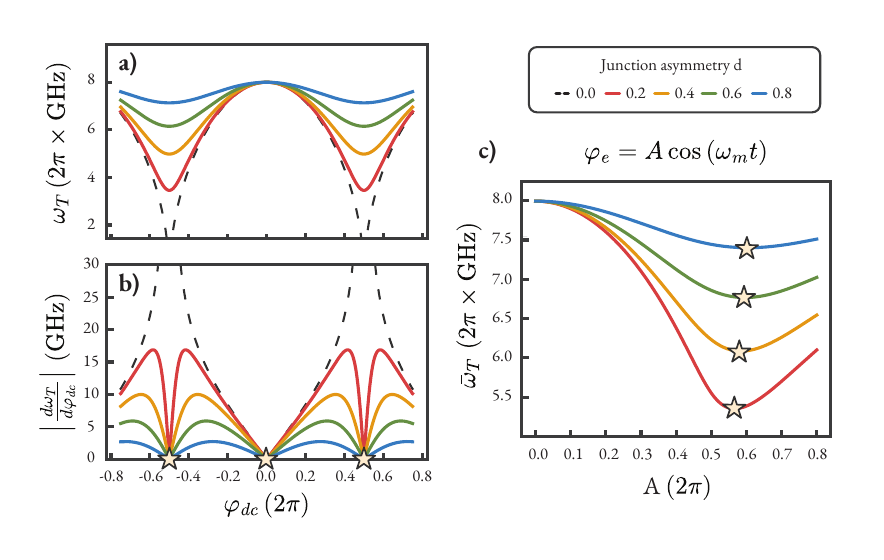}
    \caption[DC and AC sweet spots]{\textbf{DC and AC sweet spots.} (a) Transmon frequency $\omega_T/2\pi$ for a maximum frequency of $8\,\textnormal{GHz}$, $E_C/2\pi = 200\,\textnormal{MHz}$ and variable asymmetry factor $d$. b) Susceptibility to DC noise measured by the derivative $\frac{d\omega_T}{d\varphi_{dc}}$. All values of $d$ have an upper sweet spot at $\varphi_{dc} = 0$, while transmons with $d>0$ have a second sweet spot at the point of lowest frequency $\varphi_{dc} = \pm0.5\times 2\pi$. The higher the value of $d$, the lower is the susceptibility to noise outside of the sweet spots. c) Average frequency $\bar{\omega}_T$ for single-tone flux drive as a function of amplitude. AC sweet spots are indicated by stars around $\varphi_{ac}\approx 0.6\times 2\pi$ where $\frac{d\bar{\omega}_T}{d\varphi_{ac}} = 0$.}
    \label{fig:acdc_sweetspots}
\end{figure}

To make a non-modulated transmon ($a_1 = a_2 =0$) insensitive to $\delta\varphi_{dc}$, it can simply be parked at a point of maximum or minimum frequency, for which $\partial\bar{\omega}_T/\partial\varphi_{dc}=0$. The caveat is that asymmetrical SQUID transmons only have two such sweet spots (see Fig.~\ref{fig:acdc_sweetspots}a-b), severely limiting the operational points and reducing the benefits of having a flux-tunable circuit. The overall sensitivity can also be reduced outside of the sweet spot by making highly asymmetric transmons~\cite{hutchings_tunable_2017}. But this strategy has limited effects in reducing decoherence, and it requires the ability to fabricate junctions in a large range of tunneling energies $E_J$ to reach a high asymmetry $d$ (see Equation~\ref{eq:junction_asymmetry}).

When the transmon is parametrically modulated, it also becomes sensitive to $\delta\varphi_{ac}$. Still, AC sweet spots can be found under the condition $\partial\bar{\omega}_T/\partial\varphi_{dc}=\partial\bar{\omega}_T/\partial\varphi_{ac}=0$. Consider first the case of a single-tone modulation, i.e., $a_1 = 1$ and $a_2 = 0$. Figure~\ref{fig:acdc_sweetspots}c shows $\bar{\omega}_T$ as a function of $\varphi_{ac}$ for $\varphi_{dc} =0$, showing there is a local minimum at $\varphi_{ac}\approx 0.6$. However, this type of modulation only adds two more sweet spots, one for each $\varphi_{dc} = 0$ and $\varphi_{dc} = 0.5$. 

The results are more interesting when considering both modulation tones, which increases the degrees of freedom from one ($\varphi_{ac}$) to three ($A_1$, $A_2$, and the relative angle $\theta$). This allows the existence of a continuum of sweet spots over a large range of $\bar{\omega}_T$, increasing the tunability of the protected transmon. To show that, I first rewrite the frequency sensitivity to multiplicative noise as
\begin{equation}
    \frac{\partial\bar{\omega}_T}{\partial \varphi_{ac}} = \nabla \bar{\omega}_T{(A_1, A_2)}\cdot \hat{n}, 
\end{equation}
where $\hat{n} = \left(\cos\alpha, \sin\alpha\right)$ for $\alpha = \arctan\left(A_2/A_1\right)$. The sweet spots are the points at which the gradient $\nabla\bar{\omega}_T$ is orthogonal to $\hat{n}$, which offers an intuitive geometrical interpretation to the problem.

\begin{figure}
    \centering
    \includegraphics[width=\linewidth]{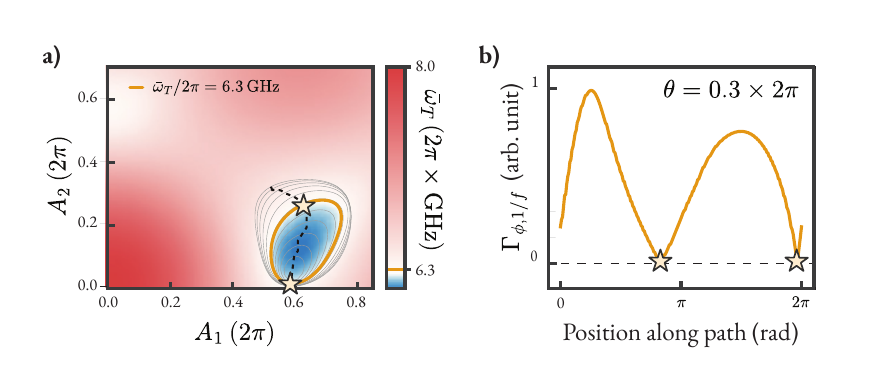}
    \caption[Engineering AC sweet spots with bichromatic flux pulses]{\textbf{Engineering AC sweet spots with bichromatic flux pulses.} (a) Plot of $\bar{\omega}_T(A_1, A_2)$ for a bichromatic pulse modulation with $\theta/2\pi = 0.3$ and $\varphi_{dc} = 0$. The thick gray line indicates the $\bar{\omega}_T=6.3\,\textnormal{GHz}$ path, with two $\partial\bar{\omega}_T/\partial\varphi_{ac}=0$ sweet spots indicated with stars. The thinner gray lines indicate the same frequency paths for different values of $\theta/2\pi$ between $0$ and $0.5$, and the dashed black line indicates the collection of sweet spots. b) Values of $\partial\bar{\omega}_T/\partial\varphi_{ac}\propto \Gamma_{\phi, 1/f}$ along the constant frequency path for $\theta/2\pi = 0.3$. The stars indicate the points at which the susceptibility to noise reaches zero.}
    \label{fig:acbic_sweetspot}
\end{figure}

A geometrical argument guarantees the existence of sweet spots in certain conditions. To illustrate this, Figure~\ref{fig:acbic_sweetspot} plots $\bar{\omega}_T{(A_1, A_2)}$ for the same transmon as in the last example for $d = 0.454$ and $\theta = 0.35\times 2\pi$, again parked at the upper sweet spot ($\varphi_{dc} = 0$). The target frequency $\bar{\omega}_T/2\pi = 6.3\,\textnormal{GHz}$ leads to a closed path in the first quadrant of the $A_1\times A_2$ plane, plotted as a golden line. The gradient $\nabla\bar{\omega}_T$ is normal to this path, and therefore its polar angle sweeps the range $\left[0, 2\pi\right]$; meanwhile, the polar angle of $\hat{n}$ is restricted to a smaller region $\left[\alpha_{min}, \alpha_{max}\right]$. This indicates that there are always two points along this path for which these vectors are orthogonal, leading to dynamical sweet spots. Eq.~\ref{eq:1fnoise} is calculated along the path to find the points at which $\Gamma_{\phi, 1/f} = 0$ (Figure~\ref{fig:acbic_sweetspot}b), indicated with stars. Varying $\theta/2\pi$ between $0$ and $0.5$ leads to a collection of closed paths shown as thinner gray lines, each with their own pair of sweet spots. The set of all available sweet spots for $\bar{\omega}_T/2\pi = 6.3\,\textnormal{GHz}$ is shown with a dashed black line. Note that this argument depends on the path not encircling the origin $(A_1, A_2) = (0,0)$ as happens with $\bar{\omega}_T/2\pi = 7.5\,\textnormal{GHz}$, for example. In this case, the existence of sweet spots is not guaranteed.

The impact of white noise, on the other hand, can usually be neglected in the case of non-modulated transmons ($a_1=a_2=0$). Attenuating the line by $20\,\textnormal{dB}$ below room-temperature noise should be enough to make thermal white noise negligible~\cite{krinner_engineering_2019}. Modulated transmons, on the other hand, are sensitive to noise at higher frequencies $k\omega_m$, $k\in \mathbb{N}$, so white noise can become a dominant~\cite{hong_demonstration_2020}. In this case, the best strategy is to add a low-pass filter to the line with a cutoff frequency between $\omega_m$ and $2\omega_m$, keeping the noise effect to a minimum.

So mitigating transmon dephasing takes engineering sweet spots as well as designing a clean environment for the circuit. The next section talks about the flux line wiring schemes used in this thesis, which take into consideration the flux noise as well as other concerns such as heating and signal amplitude.

\section{Wiring and predistortion}
\label{sec:wiringpredistortion}

Flux control lines come with several practical challenges. There can be extra heating from solder points or resistive microwave components. DC connections might short the cryostat with the room-temperature electronics, creating damaging current loops. The attenuation of the white noise on the line might also compromise the dynamic range of the signal sent to the flux source. Designing experiments involving flux-tunable circuits requires the mindful consideration of all of these aspects. 

As just discussed, the flux line can carry harmful white noise generated from higher-temperature stages until reaching the circuit. The spectral power density of the noise is given by the Johnson-Nyquist formula
\begin{equation}
    S_w = 10\log_{10}\left(\frac{k_B T}{1\,\textnormal{mW}}\right) \, \, \left[\textnormal{dBm/Hz}\right].
\end{equation}
This formula also sets the minimum thermal noise that is achievable at each cooling stage of the refrigerator. That corresponds to $-174$, $-192$, $-200$ and $-219~\textnormal{dBm}/\textnormal{Hz}$ for the room-temperature, $4\,\textnormal{K}$, Still ($800\,\textnormal{mK}$) and mixing chamber ($10\,\textnormal{mK}$) stages, respectively~\footnote{These values are a classical approximation for frequencies $\omega\ll k_BT$. The noise generated at lower temperature stages must be dealt with more care when considering broad-bandwidth noise.}.

Ref.~\cite{krinner_engineering_2019} calculates the minimum white noise level to mitigate transmon decoherence caused by flux noise. Assuming a mutual inductance of $0.5\,\Phi_0/\textnormal{mA}$ between the flux line and the SQUID, the ideal noise carried by the line is between $S_w = -184$ and $-194\,\textnormal{dBm/Hz}$. One way to reach this requirement is to attenuate the line at room temperature until the noise floor of $-174\,\textnormal{dBm/Hz}$, and then add extra $10$ to $20\,\textnormal{dB}$ attenuation at $4\,\textnormal{K}$. With this strategy, most of the active heat load generated inside the cryostat is dissipated on the attenuators at the $4\,\textnormal{K}$ level, which has a higher cooling power than the mixing chamber plate. 

\begin{figure}
    \centering
    \includegraphics[width=\linewidth]{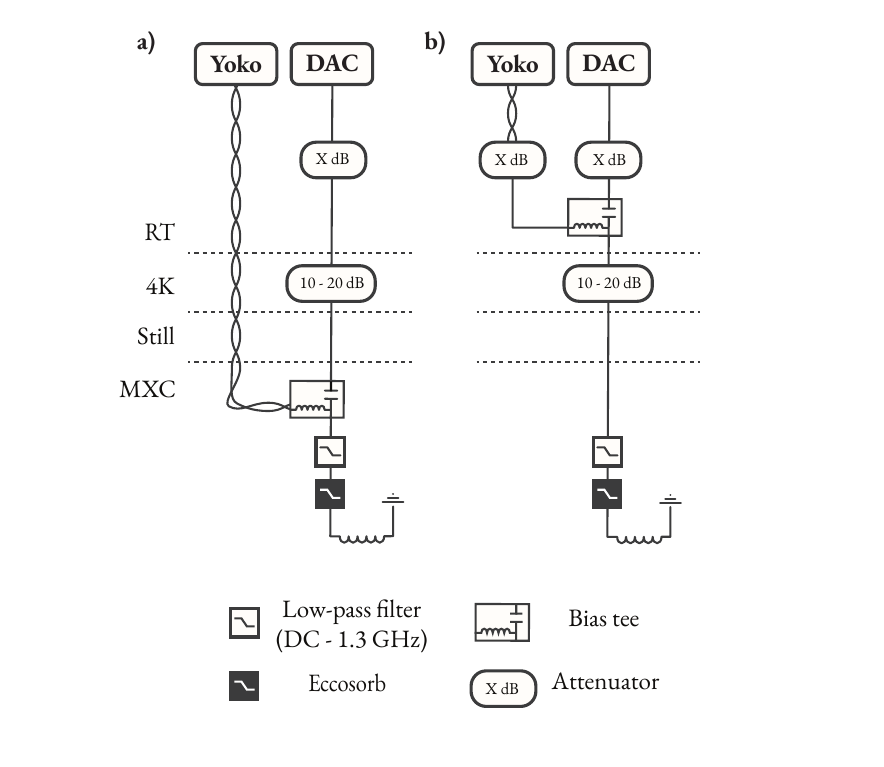}
    \caption[Flux line wiring diagram]{\textbf{Flux line wiring diagram.} The diagrams show an $\textnormal{LD-}250$ dilution refrigerator with two different flux wiring configurations. The DC bias (shown as a twisted pair of wires) is delivered by an Yokogawa GS200 voltage source (Yokogawa Electric Corp.), and the AC control is provided by the DAC of the OPX. a) The DC and AC controls are joined at the mixing chamber (MXC) flange using a soldered bias-tee, which might lead to heating of the cryogenic stage. However, the large current bias from the DC source can be sent through superconducting wires, reducing the total heat load. b) In this configuration, the bias-tee is instead placed outside the fridge, leading to no solder points inside the cryostat.}
    \label{fig:flux_wiring}
\end{figure}

This attenuation can significantly affect the flux control signal. Taking the flux source as a short-circuit, a total flux line attenuation of $X~\textnormal{dB}$ reduces an input current $I_{in}$ by a factor 
\begin{equation}
    \label{eq:current_atten}
    \frac{I_{out}}{I_{in}} = \frac{2K }{K^2 +1}, \, \, \,K = 10^{\frac{X}{20}},
\end{equation}
where $I_{out}$ is the current that reaches the flux source. This formula can be derived from the general circuit of a T-pad attenuator. This thesis' experiments use the DAC of an OPX to generate fast-flux pulses with a dynamic range of $I_{in} =\pm0.5\,\textnormal{V}/50\Omega = \pm10\,\textnormal{mA}$ and a noise level of $-150\,\textnormal{dBm/Hz}$. To take this noise down to $-184\,\textnormal{dBm}/\textnormal{Hz}$, it is necessary to add $34\,\textnormal{dB}$ attenuation, which reduces the dynamic range to $I_{out} = \pm 0.4\,\textnormal{mA}$. Assuming the mutual inductance of $0.5\,\Phi_0/\textnormal{mA}$ leads to a tunability of $\varphi_e/2\pi =\pm 0.2$. This is a very restricted range, which cannot access the full frequency spectrum of the SQUID-transmon.

The general strategies adopted to balance the white noise level and the flux tunability are summarized in Figure~\ref{fig:flux_wiring}. In practice, I place $10-20\,\textnormal{dB}$ attenuation at the $4\,\textnormal{K}$ stage, while the room-temperature attenuation is chosen as the minimum value that maximizes the measured transmon $T_2$. To help reduce the white noise without attenuating the current bias in the relevant frequency ranges, I add a $\textnormal{VLF-}1300+$ low-pass filter (Minicircuits) on the mixing chamber flange.

To circumvent the restricted dynamic range of the OPX, I use a DC source to provide the parking flux $\varphi_{dc}$ while using the DAC only to generate fast-flux pulses. The two sources can be combined using a bias-tee either inside the cryostat (Figure~\ref{fig:flux_wiring}a) or at room temperature (Figure~\ref{fig:flux_wiring}b). The choice of placement depends on several factors. For example, commercial bias-tees such as the Marki Microwave $\text{BT-}0040$ are not designed to operate at cryogenic temperatures, so their characteristics can vary from the manufacturer's specifications when mounted inside the cryostat. The choice also depends on heating: if the bias-tee requires soldering to DC line or has any lossy connections, it might incur excessive heat load in the mixing chamber, and should be placed outside the refrigerator instead. Even so, placing the bias-tee outside the cryostat means the DC current will go through the $4\,\textnormal{K}$ attenuator and lossy SMA cables inside the fridge, instead of using the built-in superconducting DC lines. This also leads to extra heat load. Whichever of these two configurations leads to less heating depends on the current bias required in the experiment.

Another advantage of the bias-tee is that the AC control port comes with a high-pass filter with a cutoff frequency in the $\textnormal{kHz}$ range. This helps reducing the overall noise sent to the flux source. However, this high-pass filter distorts the flux control. It attenuates constant current offsets sent by the OPX, which decay to zero amplitude in a timescale set by the filter cutoff frequency. 

The bias-tee is not the only source of distortion. The flux source itself (e.g. a coil) is an inductive load on the flux line, acting as a low-pass filter on incoming signals. Precise frequency control of a tunable transmon requires taking these distortions into account.

To digitally correct for the line distortions, we assume a linear relation between the DAC voltage signal $V_{in}$ and the current reaching the flux source $I_{out}$
\begin{equation}
I_{out}(s) = G_{line}(s)V_{in}(s).
\end{equation}
The current and voltage are expressed in the $s$-domain of the Laplace transform, and $G_{line}(s)$ is the transfer function that describes the line distortions. In principle, $G_{line}$ is uniquely defined by its impulse response: if $V_{in}(t) = \delta(t)$, then $V_{in}(s) = 1$, and $G_{line}(s) = I_{out,\delta}(s)$. However, producing a Dirac delta impulse is not experimentally feasible. Instead, we can generate a step pulse response, for which $V_{in}(s) = 1/s$. This generates a measurable current at the coil of $I_{out, step}(s) = \frac{1}{s}G_{line}(s)$, which can be used to construct an approximately inverse distortion $\tilde{G}^{-1}_{line}(s)$ such that
\begin{equation}
    V_{predist}(s) = \tilde{G}^{-1}_{line}(s)I_{target}(s).
\end{equation}
That is, knowing $\tilde{G}^{-1}_{line}(s)$, it is possible to generate a target current $I_{target}(s)$ at the source by applying a digitally predistorted voltage signal $V_{predist}(s)$. The construction of $\tilde{G}^{-1}_{line}$ uses a combination of infinite impulse response (IIR) and finite impulse response (FIR) filters, which are obtained from analyzing $I_{step}(t)$. The method for calibrating these filters, as well as other aspects of flux pulse correction are explained thoroughly in References~\cite{hellings_calibrating_2025, valadares_ondemand_2024, butscher_shaping_2018}. 

In this section, I have summarized some practical wiring and control considerations that were critical to setup experiments with tunable transmons. And so far in this chapter, I have shown how these flux tunability introduces interesting dynamics between the transmon and the cavity. Similar discussions are mostly present in the literature, and draw from circuit QED techniques developed in the past couple decades. This raises a question: if these techniques are useful and well-known, why are they underutilized in the specific context of bosonic cQED? In the next section, I will give plausible explanations to the difficulties of adding flux tunability to circuits containing high-$Q$ cavities, and discuss possible solutions to this problem.

\section{The challenge of flux-tunable bosonic cQED}
\label{sec:flux_tunability_challenge}

At the time the work described in this thesis began, flux-tunable superconducting circuits had already gone a long way. In 2002, Martinis et al.~\cite{martinis_rabi_2002} introduced the idea of biasing a qubit with magnetic flux, which provided an electrically isolated control that is more robust than galvanically-connected current-biased junctions~\cite{ramos_design_2001, martinis_superconducting_2009}. A few years later, following the growing interest in coupling qubits to microwave resonators~\cite{wallraff_strong_2004,johansson_vacuum_2006, schuster_resolving_2007}, Ref.~\cite{hofheinz_generation_2008} coupled a phase qubit to a resonator and used Rabi oscillations to create high Fock states. This work was soon followed by a second publication~\cite{hofheinz_synthesizing_2009} demonstrating arbitrary bosonic state preparation with the Law and Eberly protocol~\cite{law_arbitrary_1996}. These early results demonstrated the potential of flux-tunable circuits for manipulating bosonic modes. The invention of the SQUID-based transmon qubit provided a new standard for such systems, which were later used for interesting applications such as the creation of quantum random-access memory~\cite{naik_random_2017} and the digital simulation of the quantum Rabi and Dicke models~\cite{mezzacapo_digital_2014, langford_experimentally_2017}.

This trajectory can mislead one into thinking that combining flux-tunable bosonic cQED is a trivial matter. But in fact, the experimental implementations of these systems lag far behind their full potential, indicating the existence of practical challenges. The reason becomes evident upon examining those early results and recognizing a critical limitation: the bosonic modes were implemented in planar resonators. In such architectures, the field is often strongly concentrated on the oxide surfaces and interfaces between the superconductor and other materials, leading to lifetimes of only a few tens of microseconds. Since the effects of photon loss increase at high populations, these systems lose the main advantage associated with the large Hilbert space of harmonic oscillators, hindering their use for bosonic QIP.

Naively, this problem could be overcome by using three-dimensional microwave cavities. As I discussed in Chapter~\ref{ch:bosonic}, these cavities host most of their field energy in vacuum and can thus achieve lifetimes in the millisecond range when coupled to fixed-frequency transmons~\cite{reagor_quantum_2016, milul_superconducting_2023}. However, a reported attempt at using flux-tunable transmons capped the cavity $T_1$ to values of about $31\,\mu\textnormal{s}$~\cite{reed_entanglement_2013}, a value comparable to planar resonators. The relation between flux tunability and low cavity lifetimes is confirmed by later results, which attempt to partially circumvent the problem by using DC flux trapping~\cite{stammeier_applying_2018} or hybrid aluminum-copper packages~\cite{reshitnyk_microwave_2016}. These results ultimately show that achieving high-quality flux control remains a major challenge, and highlight the need to consider this problem from a new perspective.

Turns out that the same feature that makes $3\textnormal{D}$ cavities long-lived -- their spread-out fields -- also enhances their leakage through flux lines. Here, I will give the high-level understanding of this relation, and explain why solving this contradiction requires looking for solutions outside the established design rules of the field.

\begin{figure}
    \centering
    \includegraphics[width=\linewidth]{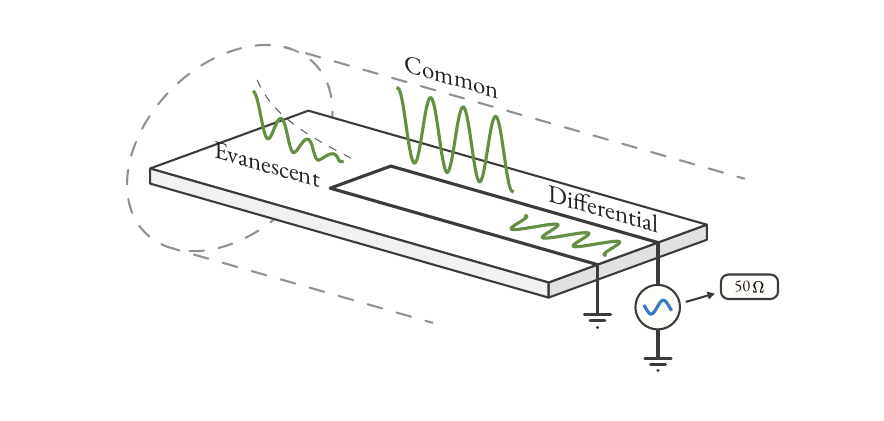}
    \caption[Flux line propagation modes]{\textbf{Flux line propagation modes.} For a flux line designed as two parallel wires inside a waveguide, there are three possible propagation modes: the evanescent mode, which is not supported by the line nor the waveguide and is exponentially suppressed, the common mode between the lines and the waveguide, and the differential mode traveling between the wires. }
    \label{fig:prop_modes}
\end{figure}

Assume the typical coaxline architecture from Figure~\ref{fig:cavity_transmon_coupling}. But now, suppose the transmon is biased by a superconducting flux line printed on the chip (Figure~\ref{fig:prop_modes}). Both ends of the line are terminated on the back of the waveguide, one connected to a short to ground and the other to a voltage source. The source is connected to the external environment and introduces a leakage channel. 

The circuit modes residing within the waveguide have three leakage pathways. In one path, they can directly travel through the waveguide until reaching the source while ignoring the presence of the flux line. This is not usually a concern, because the waveguide is designed to act as a high-pass filter with a cutoff frequency $\gtrsim   15\,\textnormal{GHz}$. This makes the waveguide modes evanescent, i.e., attenuated~\cite{pozar_microwave_2011}. Since the waveguide also usually $20$ to $30\,\textnormal{mm}$ long, the cavity and the transmon are significantly distanced from the leakage point, preventing their interaction through the evanescent modes. Most concerning are the leakage paths mediated by the line, which can be divided into a propagating common mode, and a differential mode.

The common mode is that which considers both arms of the flux line as having the same potential. This mode can be identified by the field being small midways between both arms due to the symmetry. The electric field connects the line to the waveguide, similarly to a coaxial transmission line mode. Due to this ``open'' field profile, the transmon and cavity fields will strongly interact with the common mode. 

The differential mode, on the other hand, considers each arm of the flux line as having opposite voltages. The electric field profile resembles that of an electric dipole, being stronger in-between the arms and quickly decaying the more distant the field is from the flux line. The differential mode field is thus more confined. It overlaps less with the waveguide and is less prone to coupling with other circuits, effects which are enhanced the closer the arms are to each other. Moreover, the flux line section closer to the transmon acts as a short-circuit between both arms. This further reduces the direct coupling with the differential mode.

\begin{figure}
    \centering
    \includegraphics[width=\linewidth]{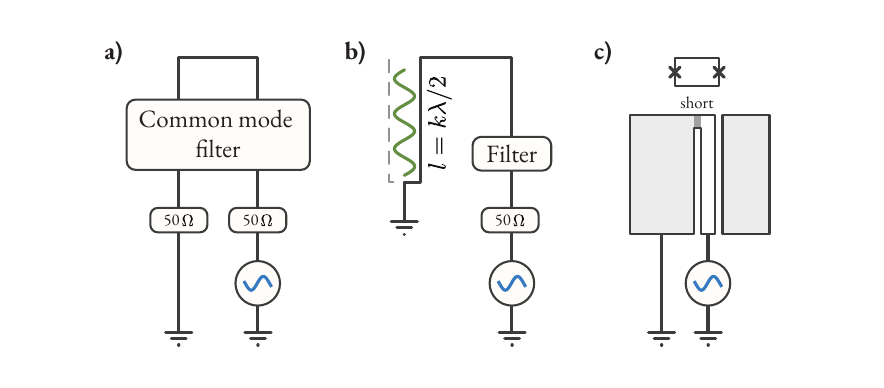}
    \caption[Filtering strategies]{\textbf{Filtering strategies.} (a) A flux line that is symmetric with respect to incoming fields will suppress their propagation through differential modes. The remaining leakage can be mitigated by implementing a common mode filter. b) Asymmetric flux lines can be designed to destructively interfere with incoming fields. The short-circuited arm can be adjusted to reflect the traveling field at a distance multiple of $\lambda/2$, effectively implementing a notch filter on the field with wavelength $\lambda$. c) Typical setup in planar circuit QED systems. This is identical to the previous case for $k=0$, trivially implementing the filter for all $\lambda$.}
    \label{fig:flux_line_strategies}
\end{figure}

In a realistic circuit, leaking fields will travel through a combination of all propagating modes. The key to design a protected flux line is to take all of them into account, designing strategies which involve explicitly adding filtering and leveraging symmetries to reduce the coupling with sensitive circuit fields. Figure~\ref{fig:flux_line_strategies}a-b shows two possible strategies. The first idea is to make both arms of the flux line symmetric from the perspective of the leaking fields. Both paths have to be terminated to similar $50\,\Omega$ environments so as not to induce any uneven reflections. In such geometry, incoming fields would excite both arms equally and transfer all their energy to the common mode, while the differential mode is not excited due to the symmetry. The common mode leakage can then be mitigated by using a stripline filter between the line and the waveguide. This seems to be the strategy followed by the flux delivery design described in Ref.~\cite{maiti_controlling_2025} (Chapter 5), although the implementation details have not yet been publicly disclosed.

The second idea (Figure~\ref{fig:flux_line_strategies}b) leverages the interference between the incoming fields and their reflections from the short-circuited arm. Consider the propagation of a wave between one of the arms and the waveguide, which defines the ground plane. At a distance $l$, the arm and the waveguide are shorted, grounding the incoming signal. The input impedance $Z_{in}$ of the propagation from the perspective of an incoming signal with frequency $\omega_{in}$ is~\cite{pozar_microwave_2011}
\begin{equation}
    Z_{\text{in}} = Z_0 \frac{Z_L + i Z_0 \tan(\frac{2\pi}{\lambda_{in}} l)}{Z_0 + i Z_L \tan(\frac{2\pi}{\lambda_{in}} l)},
\end{equation}
where $Z_{L} = 0$ is the load (a short), $Z_0$ is the characteristic impedance of this propagation, and $\lambda_{in}$ the wavelength corresponding to $\omega_{in}$. By choosing a length $l=k\lambda_{in}/2$ ($k\in\mathbb{N}$), the field will experience an input impedance $Z_{in} = Z_L = 0$. This effectively creates a reflection that destructively interferes with the incoming signal, suppressing the propagation of energy to the second arm. To ensure there is no residual leakage, a filter can be added to the second arm before its terminal. 

Flux lines based on the coplanar waveguide (CPW) architecture, common in $2\textnormal{D}$ superconducting circuits, naturally use this interference strategy (Figure~\ref{fig:flux_line_strategies}c). The ground plane is printed on the chip in close proximity to the flux line, so it is trivial to implement a short circuit nearby the transmon ($l\approx0$). It is not immediately clear how to translate this idea to $3\textnormal{D}$ circuits, where the ground plane is the waveguide surrounding the chip. And this added complexity partially explains why flux control is more challenging in $3\textnormal{D}$ architectures than in $2\textnormal{D}$ circuits, hindering its integration to high-$Q$ microwave cavities.

In this thesis, I will explore three different solutions to this problem. In the first one (Chapter~\ref{ch:somerset}), I try to avoid the problem described here altogether by using a magnetic flux hose. This is a structure that allows the flux to be delivered from a remote source, so there is only the evanescent propagation mode of the waveguide. The second approach (Appendix~\ref{append:pogopin}) attempts to implement the flux line completely in a stripline architecture, and use a superconducting pogo (spring-loaded) pin to establish an out-of-plane short circuit between the chip and the waveguide. This solution was prematurely abandoned due to two critical issues: first, the mechanical instability of the pogo pins complicates device assembly and may lead to scratching of the superconducting film. Second, the open field of the stripline architectures would cause significant crosstalk between different parts of the circuit, complicating the design

The third solution (Chapter~\ref{ch:lakeside}) combines the lessons learned from the previous two. The understanding of the issue allowed challenging the established design conventions of bosonic circuit QED devices. Specifically, I use a hybrid $2\textnormal{D}$--$3\textnormal{D}$ structure that forces the transmon and cavity fields into the differential mode of the flux line that is then heavily filtered. As I will demonstrate experimentally, the presented solutions allow the implementation of the flux control explained in this chapter, offering the tunability of the light-matter interactions between the transmon and the cavity.

\SetPicSubDir{ch-Noodle}
\SetExpSubDir{ch-Noodle}

\chapter{On-demand tunable interactions with a magnetic hose}
\label{ch:somerset}
\vspace{2em}

At the moment I started investigating fast-flux tunability in high-$Q$ bosonic systems, there were many useful results from the literature that helped narrow down the possible solutions. Matthew Reed's 2013 thesis~\cite{reed_entanglement_2013} described the use of an on-chip flux line with a fast response, but that limited the cavity lifetime to $\approx31\,\mu\textnormal{s}$ even with the use of filters. Majumder et al.~\cite{majumder_fast_2022} also reported fast control of a transmon coupled to a $3\textnormal{D}$ cavity. But without filtering or optimization for high quality factors, the cavity showed a single-photon lifetime of about $0.7\,\mu\textnormal{s}$. Perhaps the best first step should be to avoid on-chip flux lines altogether, eliminating the spurious propagating modes of these structures. 

That was the approach adopted in a 2021 publication by Gargiulo et al.~\cite{gargiulo_fast_2021}. They adapted the magnetic hose~\cite{navau_long_2014} into a compact design integrated into a $3\textnormal{D}$ superconducting package to deliver out-of-plane magnetic fields to transmons. The hose routed magnetic fields with high efficiency, allowing the use of low-inductance, broadband flux sources with response times on the order of tens of nanoseconds. The hose also features a built-in filter, providing protection against leakage. Due to this protection, the cavity mode of the hose-integrated device reached a lifetime of $\approx 40\,\mu\textnormal{s}$ -- a value comparable to previous results from the literature, but which presented clear opportunities for improvement. 

In this chapter, I will describe how to optimize a bosonic circuit QED device to incorporate a magnetic hose. Several factors will be considered, including photon loss, coupling strength to the transmon, flux delivery, and active heat load. The resulting device will demonstrate the dynamical coupling between a transmon and a high-$Q$ cavity. This tunability is employed to mix different light-matter interaction regimes within the same experiment for quantum information processing tasks. Specifically, I will prepare Fock states by resonantly loading single photons into the cavity, and then characterize the final state in phase space with both Wigner and characteristic function tomographies. In a subsequent experiment, the dynamical detuning of the transmon is used to suppress the nonlinear interactions, protecting the cavity evolution from self-Kerr and dispersive shifts.

Before presenting the circuit design, I will focus on the magnetic hose itself. The next section will briefly discuss its design, and the features that enable both flux control and leakage suppression.

\section{The hose anatomy}
\label{sec:anatomy}

\begin{figure}
    \centering
    \includegraphics[width=\linewidth]{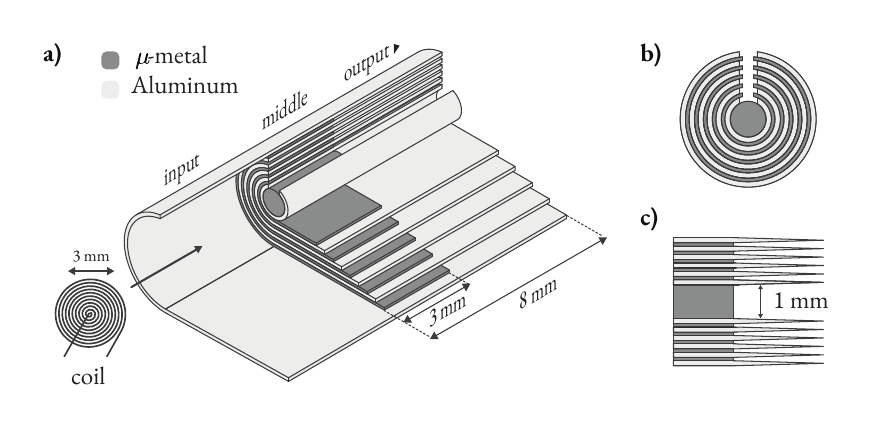}
    \caption[The magnetic flux hose]{\textbf{The magnetic flux hose.} (a) Half-open view of the hose, showing 11 concentric layers wrapped around a cylindrical $\mu$-metal core. The light (dark) gray layers are made of aluminum ($\mu$-metal). The structure is divided in three sections: the input, housing a $3\,\textnormal{mm}$-diameter coil, the middle section, responsible for efficient field transfer, and the output, which delivers the field while working as a lossless filter to prevent circuit leakage. b) Cross-section of the hose, showing a $0.5\,\textnormal{mm}$-wide cut along the structure. The $\mu$-metal layers are made a bit longer to prevent the aluminum layers from touching each other. c) To further prevent contact between the aluminum layers, they are thinned with sandpaper at the output end, increasing their distance.}
    \label{fig:hose}
\end{figure}

The hose's capability to deliver flux is a consequence of its sophisticated electromagnetic design. It consists essentially of a multi-layer cylindrical structure, as shown in Figure~\ref{fig:hose}a in a half-open view, which is split into three main sections.


The middle section provides efficient propagation of the magnetic field along the hose axis. It is made of 11 alternating layers of two different materials, superconducting aluminum and $\mu$-metal. The materials have complementary properties: while the superconductor completely screens the magnetic field via the Meissner effect (with effective magnetic permeability $\mu_{r} = 0$), the $\mu$-metal has an exceptionally high $\mu_r$ on the order of $10^4$. Such a combination creates a metamaterial where the magnetic field cannot travel transversely, since it is blocked by the aluminum layers, but easily propagates in the axial direction due to the $\mu$-metal. In other words, the field lines originating at one end of the cylinder do not curl back to close the magnetic dipole until they reach the opposite end.

The denser the layering, the more efficient is the field transmission. Here I use $0.1\,\textnormal{mm}$-thick $\mu$-metal sheets, and the aluminum layers are etched down from a thickness of $0.4\,\textnormal{mm}$ to $0.07$--$0.10\,\textnormal{mm}$ using aluminum etchant type A. A $1\,\textnormal{mm}$-diameter $\mu$-metal wire is used as a core around which the other layers are rolled. The final structure has a nominal diameter of $3.2\,\textnormal{mm}$, but in practice it reaches up to $3.4\,\textnormal{mm}$.

The input section is an extension of the outermost aluminum layer, which houses the source coil. The coil is made by winding a copper-stabilized superconducting NbTi wire in a disk shape. The wire has a $0.066\,\textnormal{mm}$ diameter, which limits the coil to about $23$ turns. The coil is glued to the bottom of a Teflon cylinder using Eccosorb CR-110, creating a holder. The holder and coil are placed in the input section, which is designed to be long enough to cover both pieces. Additional Eccosorb is used to fill in the gaps. This last step helps secure the coil and improves its thermalization. 

The third section is the output end of the hose. The output is the closest part to the transmon circuit and the cavity, so it does not have any lossy elements such as the $\mu$-metal. In this section, the aluminum layers extend from the middle section by $5\,\textnormal{mm}$, creating a fully superconducting region. Conveniently, this aluminum structure also works as a waveguide with a high cutoff frequency, which attenuates fields propagating up the hose from within the device. This built-in filter prevents the transmon and cavity fields from overlapping with the $\mu$-metal layers, helping mitigate losses.

A critical requirement for flux delivery is that the aluminum and $\mu$-metal layers do not form closed loops around the hose axis. Closed paths allow eddy currents in the resistive $\mu$-metal that heat the system, while supercurrent loops in the aluminum screen the magnetic field flowing through the hose. To prevent loops, all layers are made a bit shorter than the outer perimeter of the previous layer, so they cannot close on themselves. This results in a $0.5\,\textnormal{mm}$ gap along the hose, as shown in the cross-section in Figure~\ref{fig:hose}b. Additional fine adjustments are made to ensure there are no current loops. For example, the $\mu$-metal is cut a bit wider than the aluminum to create a buffer between two adjacent superconducting layers (Figure~\ref{fig:hose}c). Moreover, the sandpaper is used to make the aluminum layers in the output section as thin as possible, increasing the spacing between them.

Several hoses are made following this recipe and assembled onto a testbed containing a tunable transmon chip inserted into a coaxial waveguide. The transmon is $2\,\textnormal{mm}$ away from the end of the hose, which is initially fabricated with a $10\,\textnormal{mm}$-long middle section. Tests with two nominally similar hoses yielded field values of $54\,\textnormal{nT}/\textnormal{mA}$ and $86\,\textnormal{nT}/\textnormal{mA}$. This difference may be influenced by sub-millimeter variations in the position of the hose relative to the transmon, which is difficult to control. To enhance the delivered field, the length of the middle section is reduced to $3\,\textnormal{mm}$. Additionally, the coil is made with two layers stacked on top of each other, and the transmon-hose distance is reduced to $1.5\,\textnormal{mm}$. Combined, these adjustments increased the delivered field strength to about $140\,\textnormal{nT}/\textnormal{mA}$. All SQUIDs used in this chapter have an area of $1000\,\mu\textnormal{m}^2$ leading to a tunability of $0.067\,\Phi_0/\textnormal{mA}$. These numbers compare well with those from Ref.~\cite{gargiulo_fast_2021}. The authors simulated the magnetic field delivery of a similar structure as $22.4\,\textnormal{nT}/\textnormal{mA}$ to a transmon $1.0\,\textnormal{mm}$ away from the hose end. Surprisingly, their experiments resulted in a much stronger field of $240\,\textnormal{nT}/\textnormal{mA}$. Their hose design differs in the number and thickness of aluminum and $\mu$-metal layers, but is otherwise the same. A possible explanation to the relatively low delivered field and its high dependence on the middle section length could be the $\mu$-metal degradation during the manufacturing process, which can greatly reduce the material's magnetic permeability.

Having validated the flux delivery, the next step is to optimize the coupling of a hose-biased transmon to a high-$Q$ cavity.

\section{Somerset: a fast-flux tunable bosonic circuit}

The tunable device is designed to be as similar as possible to a conventional bosonic circuit QED architecture. There are two reasons for this choice. First, the hose will ideally complement the existing hardware toolkit without imposing more physical constraints than necessary. Second, our laboratory reliably fabricates circuits such as the one in Figure~\ref{fig:cavity_transmon_coupling}a with cavity single-photon lifetimes in the $0.8$ to $1.2\,\textnormal{ms}$ range. Therefore, any additional losses can be attributed to the presence of the hose.

\begin{figure}
    \centering
    \includegraphics[width=\linewidth]{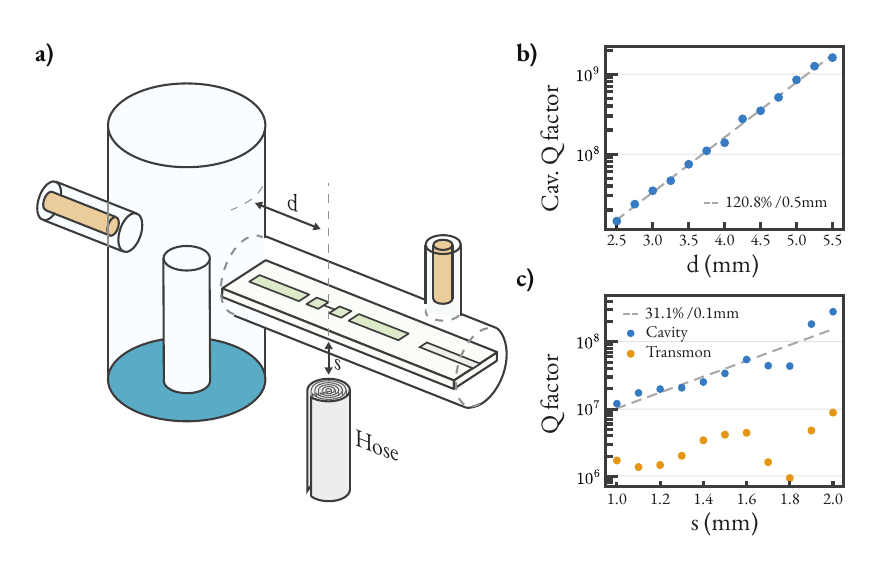}
    \caption[Integrating the hose to a high-$Q$ bosonic device]{\textbf{Integrating the hose to a high-$Q$ bosonic device.} (a) Schematics of the Somerset design, highlighting the hose-cavity distance $d$ and the hose-transmon vertical distance $s$. The final design uses two coupler lines connecting the transmon to the cavity and to the readout resonator.  b) Simulated cavity $Q$ factor (without the transmon) as a function of $d$. The values of $\log{Q}$ fit a linear function with a slope of $120.8\%/0.5\,\textnormal{mm}$. c) Simulated transmon and cavity $Q$ factors as a function of $s$, showing a cavity $Q$ sensitivity of $31.1\%/0.1\,\textnormal{mm}$. }
    \label{fig:somerset}
\end{figure}

The circuit losses are estimated using finite-element simulations of the electromagnetic environment. The simulated setup is shown in Figure~\ref{fig:somerset}a, and it consists of a $\lambda/4$ cavity at $5.70\,\textnormal{GHz}$ and a $21\,\textnormal{mm}$ chip placed in the transverse waveguide. The hose is directed to the surface of the chip, coming from the flip side to avoid accidental scratching while mounting. The $\mu$-metal layers are modeled with ideal values of $\mu_r = 8\times 10^4$ and conductivity of $1.67\times 10^6 \, \textnormal{S}/\textnormal{m}$, and are the only source of loss in the simulation. 

I use this model to test different geometric configurations of the system and identify the critical parameters that impact the cavity $Q$ factor. The most important parameter, as expected, is the distance between the cavity and the hose, with the $Q$ factor increasing by $120$\% for every $0.5\,\textnormal{mm}$ of extra spacing (see Figure~\ref{fig:somerset}b). This means the cavity losses can be greatly reduced by just increasing this distance, but such a strategy leads to other problems. Since the SQUID loop has to follow the hose position, the transmon also has to be distanced from the cavity to increase the $Q$ factor of the latter. This significantly reduces their coupling factor $g$. One way to counter the decrease in $g$ is to elongate the transmon pads to cover the extra distance and pick up the cavity fields. However, the simulations show that stretching the capacitor pads also spreads out the transmon field, leading to a larger overlap with the $\mu$-metal and lower transmon quality factors in the order of $10^5$. Therefore, instead of directly streching the pads, I keep the transmon short so its fields are concentrated in-between the capacitor pads (Figure~\ref{fig:somerset})a. Then, a $2.85\,\textnormal{mm}$ transmission line is placed between the transmon and the cavity~\footnote{This design was inspired by the vertical transmon from Ref.~\cite{vlastakis_controlling_2015}.}, recovering their coupling. The optimal configuration is when the center of the hose is $d=3.5\,\textnormal{mm}$ away from the edge of the cavity, at which point a coupling factor of $g= 7.4\,\textnormal{MHz}$ is achievable. The transmon has $Q\approx4\times10^6$, which establishes an upper bound of $T_1< 120\,\mu \textnormal{s}$. The $Q$ factor of the cavity is $7\times10^7$ when the transmon is not added to the simulation, and reduces to $2\times10^7$ when it is included. The latter value corresponds to a cavity single-photon upper bound of $560\,\mu\textnormal{s}$, indicating that the hose will likely be a limiting factor.

The second critical parameter is the vertical distance between the transmon and the hose, denoted as $s$ in Figure~\ref{fig:somerset}. When this distance is swept between $1\,\textnormal{mm}$ and $2\,\textnormal{mm}$, the cavity $Q$ factor increases by $31\%$ for every additional $0.1\,\textnormal{mm}$, ranging from $1\times10^7$ to $3\times10^8$. This means that even sub-millimeter variations to this parameter have significant impact on the cavity lifetime. Surprisingly, the transmon $Q$ is less sensitive, ranging between $2\times 10^6$ and $8\times 10^6$. Additionally, the distance $s$ impacts the transmon frequency, which varies over $150\,\textnormal{MHz}$ in the same parameter range. This shows that displacing the hose changes the electromagnetic environment of the transmon, and may also affects its coupling with the cavity and readout resonator. Since the hose assembled manually, the circuit parameters will be vulnerable to inaccurate hose insertion. To improve the mounting and reduce these variations, I adapted an $\textnormal{M}8$ screw by drilling a tunnel through its axis that could securely accommodate the hose. The screw was assembled into the device so the hose insertion could be finely adjusted by counting the number of turns.

The final design was named Somerset after the Singaporean Mass Rapid Transit station. As discussed in Chapter~\ref{ch:bosonic}, it was machined out of $4\textnormal{N}6$ aluminum and treated with etchant A to remove machining defects on the top $150\,\mu\textnormal{m}$-thick layer. After inserting the transmon chip and the hose into the package, the device was loaded into the dilution refrigerator for characterization.

\subsection{First tests \& mitigation of flux line heating}

Somerset was mounted on the mixing chamber flange of the dilution refrigerator and wired according to Figure~\ref{fig:wiring_diagram}. In some experiments, the resonator output was connected to a SNAIL parametric amplifier~\cite{frattini_optimizing_2018} for single-shot readout of the transmon state. The flux line followed the wiring configuration in Fig.~\ref{fig:flux_wiring}a. The coil was connected to a bias-tee with a $4\,\textnormal{kHz}$-cutoff high-pass filter mounted inside the refrigerator, with $10\,\textnormal{dB}$ attenuation at the $4\,\textnormal{K}$ stage of the AC line. The system had two solder points: one at the DC port of the bias-tee, and another connecting the coil to the SMA line routing the signal from the AC+DC port. Both solder points were thermally anchored to a region close to the mixing chamber using copper braids, helping dissipate the heat generated by the current bias.

\begin{figure}
    \centering
    \includegraphics[width=\linewidth]{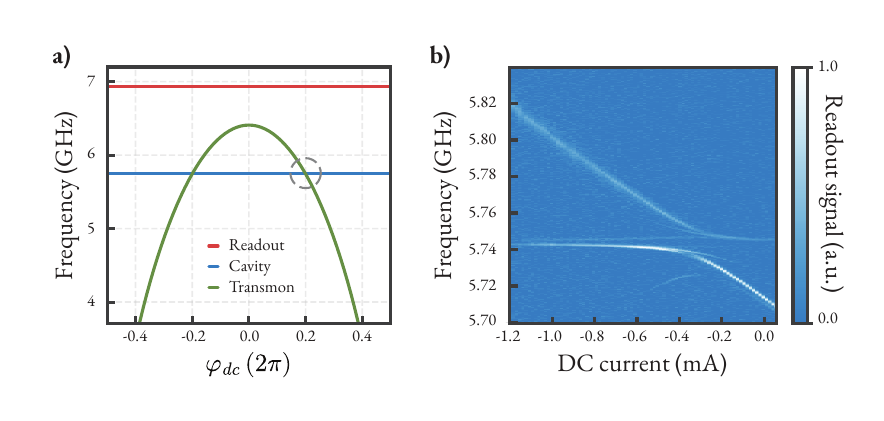}
    \caption[System's frequency spectrum]{\textbf{System's frequency spectrum.} (a) Frequency diagram showing The cavity at $\omega_a/2\pi =5.744\,\textnormal{GHz}$, readout resonator at $\omega_c/2\pi = 6.932\,\textnormal{GHz}$, and symmetric SQUID-transmon frequency as a function of flux bias for maximum $\omega_T/2\pi = 6.409\,\textnormal{GHz}$. The dashed circle indicates the point of avoided crossing between cavity and transmon. b) Transmon spectroscopy experiment versus the DC current bias around the avoided crossing. The minimum frequency split between the transmon and the cavity feature indicates a coupling $g/2\pi = 7.3\,\textnormal{MHz}$. }
    \label{fig:freq_charact}
\end{figure}

The frequency of the readout resonator was measured as $6.932\,\textnormal{GHz}$, a negligible $6\,\textnormal{MHz}$ error from the target, and its drive line was overcoupled with a total $Q=13\times10^3$. The transmon anharmonicity was measured as $\alpha/2\pi = 202\,\textnormal{MHz}$ (compared to simulated $210\,\textnormal{MHz}$), and its maximum frequency was $6.409\,\textnormal{GHz}$, about $100\,\textnormal{MHz}$ lower than the target. This frequency imprecision is expected and comes from known variations in the Josephson junction fabrication~\cite{kreikebaum_improving_2020}. Still, since the transmon is imbued with a symmetric SQUID, it is possible to accurately set is frequency to any value lower than the upper sweet spot by applying flux bias through the hose. By sweeping the transmon frequency, the avoided crossing with the cavity mode was found at $5.744\,\textnormal{GHz}$, with a minimum frequency gap that indicates $g/2\pi = 7.3\,\textnormal{MHz}$ (Fig.~\ref{fig:freq_charact}b). All measured values are close to design and validate the quality of the simulations.

\begin{figure}
    \centering
    \includegraphics[width=\linewidth]{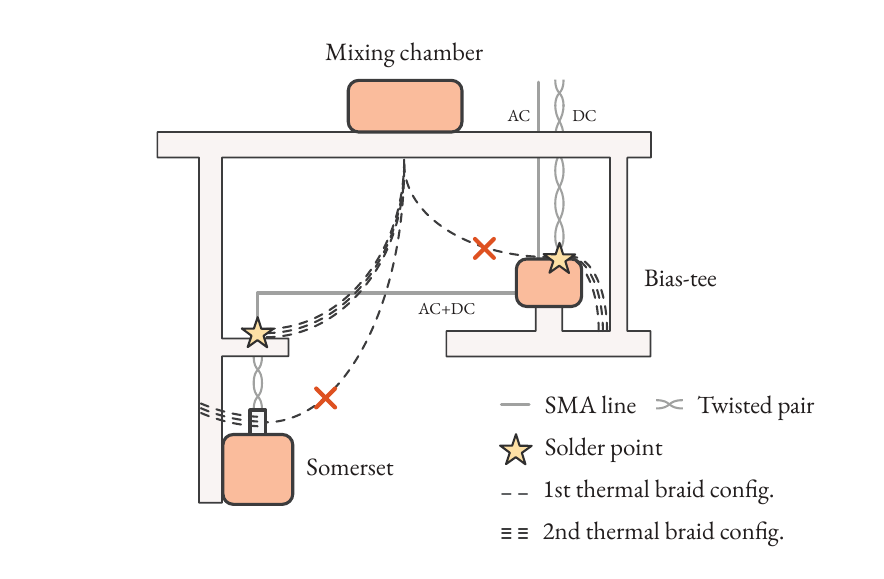}
    \caption[Managing the flux line heat load]{\textbf{Managing the flux line heat load.} The schematics shows Somerset, the bias-tee and the mixing chamber inside the dilution refrigerator. There are three main sources of heat: the two solder points along the flux line, indicated with stars, and the $\mu$-metal in the hose. The first strategy was to use copper braids to connect all three heat sources to the mixing chamber to pump the heat out of the refrigerator. This configuration inadvertently connected Somerset to the solder points, increasing the temperature of the transmon. The second strategy was to connect each copper braid to a different point, thoughtfully redistributing the heating away from the device.}
    \label{fig:braids}
\end{figure}

However, it was not yet possible to perform meaningful experiments due to the excessive heating of the flux control. At a relatively low current of $2.7\,\textnormal{mA}$ through the coil, the baseline $|e\rangle$ population of the transmon was measured to be $17\%$ using the method described in Ref.~\cite{geerlings_demonstrating_2013}. This excitation level is much higher than expected non-equilibrium transmon populations~\cite{jin_thermal_2015}, typically less than $2\%$. Still, the refrigerator thermometer did not indicate any increase in temperature. Another peculiar effect was observed when step flux pulses were applied to the AC line: an intense transmon excitation could be observed from the multi-resonance response in resonator spectroscopy, indicating excitations up to the $|f\rangle$ state. The transmon excitations from ground state were proportional to the step amplitude and dissipated quickly with a characteristic timescale of $\approx 15\,\mu\textnormal{s}$. However, the same result was obtained when using flux pulses with slower ramp-up times, indicating the transmon excitation is not due to non-adiabatic transitions. These observations indicate that the flux line generated heat that was small enough to not be detectable by the thermometer of the refrigerator, whereas the heat source was close enough to the chip to directly affect the transmon. Therefore, the heat source was assumed to be the $\mu$-metal layers of the hose. 

To redirect the heat generated at the $\mu$-metal layers away from the device, I detached the hose from the $\textnormal{M}8$ screw and thermally anchored it near the mixing chamber using a copper braid (see 1st thermal braid configuration in Figure~\ref{fig:braids}). Surprisingly, this made the problem worse. The same heating effects were observed at currents four times lower. The cause of the worsening was that, since all heat sources were then anchored to the same location near the mixing chamber, Somerset became indirectly connected to the two solder points of the flux line.

The solution was to rearrange the copper braids and redistribute the heat. The solder point closest to Somerset was kept connected near the mixing chamber, redirecting the heat away from the device. The hose was thermally anchored to the copper bracket immediately close to the device, providing a fast sink for the heat generated at the $\mu$-metal. The DC line solder point was anchored to the large plate where the bias-tee was mounted, allowing the heat to diffuse into a high-thermal-capacity environment. This careful copper braiding scheme (Fig.~\ref{fig:braids}, 2nd thermal braid configuration) led to excited-state populations consistently below $2.5\%$, measured for coil currents up to $3.1\,\textnormal{mA}$. This strategy also mitigated the transmon excitation following fast step flux pulses. Step pulses could tune the transmon frequency over a range of $160\,\textnormal{MHz}$ without increasing the $|e\rangle$ population to a detectable amount beyond $\pm0.1\%$ error in the measurement.

After overcoming the major issues with the active heat load of the flux line, the tunability of Somerset could be fully realized to control the light-matter interactions between the transmon and the cavity.

\section{Protected flux control}

As discussed in Chapter~\ref{ch:bosonic}, with fine control over $\Delta = \omega_a-\omega_T$, the cavity-transmon dispersive shift $\chi_{ab}~$\footnote{I will most often be referring to the dispersive shift between the transmon and the cavity, so for now on I will drop the subscript from $\chi_{ab}$.} can be tailored for specific applications. The analytical formula for $\chi$ is obtained with the Schrieffer-Wolff transformation as~\cite{koch_charge_2007, blais_circuit_2021}
\begin{equation}
\label{eq:dispersive_SW}
    \chi = g^2\left( \frac{1}{\Delta}-\frac{1}{\Delta + \alpha}\right),
\end{equation}
showing how these two parameters are related.

\begin{figure}
    \centering
    \includegraphics[width=\linewidth]{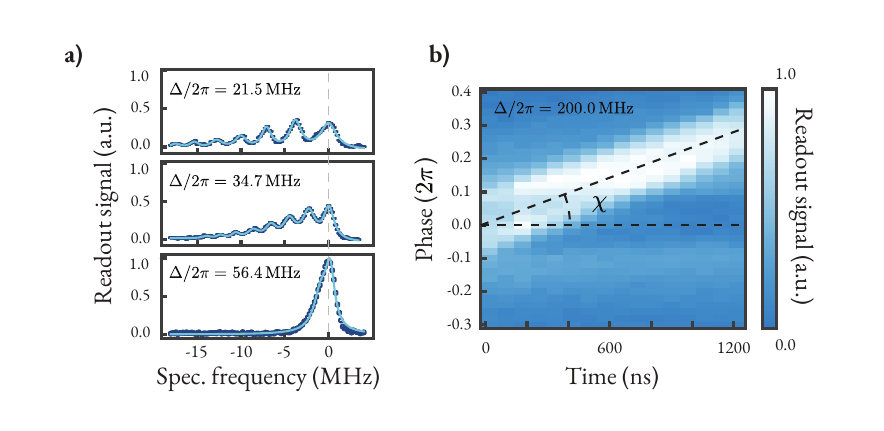}
    \caption[Dispersive shift measurement methods]{\textbf{Dispersive shift measurement methods.} (a) Number-splitting spectroscopy at different detunings $\Delta = \omega_a - \omega_T$, showing smaller peak-to-peak distance at higher $\Delta$. The $x$ axis is measured relative to the first peak. The values of $\chi$ are slightly different from those in Table~\ref{tab:chi_values} due to cooldown variations. b) Out-and-back measurement for large detuning $\Delta/2\pi = 200.0\,\textnormal{MHz}$, showing $\chi/2\pi = 0.18\,\textnormal{MHz}$ from the linear phase evolution of the return displacement. The readout signal is proportional to the transmon population after applying a $\pi$ pulse with frequency $\omega_T$.}
    \label{fig:chimethods}
\end{figure}

To illustrate how Somerset tunes $\chi$, I performed photon-number-splitting spectroscopy at different flux biases. In this experiment, the cavity was prepared in a coherent state $|\alpha\approx 1.5\rangle$, followed by a measurement of the transmon frequency spectrum. The results are shown in Figure~\ref{fig:chimethods}a. For a small detuning $\Delta/2\pi \approx-21.5\,\textnormal{MHz}$, the peaks were well-resolved and separated by a large $\chi\approx3.5\,\textnormal{MHz}$~\footnote{At this point $\Delta/g\approx 3.5$, so the dispersive shift approximation is not valid and requires higher-order corrections. Still, the number-splitting spectroscopy illustrates well the tunability of the energy levels.}. The amplitudes of each peak probe the Fock state population $|\langle z|\alpha\rangle|^2$ according to Equation~\ref{eq:cohstate_population}. As the transmon was detuned to $\Delta/2\pi = -34.7\,\textnormal{MHz}$, $\chi$ was reduced, deteriorating the peak resolution. At larger detunings, as shown for $\Delta/2\pi = -56.4\,\textnormal{MHz}$, the peaks were no longer resolvable.

To measure small values of $\chi$, the peak resolution can be enhanced by using longer, narrowband qubit pulses. However, this strategy is limited by the minimum linewidth determined by the transmon $2/T_2$. For $\chi < 2/T_2$, the system is in the weak dispersive regime, and a different method is necessary to measure $\chi$. One option is to use out-and-back measurements ~\cite{eickbusch_quantum_2024}. In this experiment, a cavity coherent state $|\alpha\rangle$ is created and let evolve under the dispersive shift interaction. In the rotating frame of the cavity $\omega_a$,
\begin{equation}
    e^{i\chi t\frac{1}{2}\left(1-\hat{\sigma}_z\right)\hat{a}^\dagger\hat{a}}|\alpha\rangle = |\alpha e^{i\chi t\frac{1}{2}\left(1-\hat{\sigma}_z\right)|\alpha|^2} \rangle ,
\end{equation}
i.e., the dispersive shift causes the phase-space rotation by an angle $\chi t$ when the qubit is in $|e\rangle$. To measure this rotation, after a time $t$ a displacement $D\left(\alpha e^{i\theta}\right)$ is applied with a variable angle $\theta$, leading to the final state
\begin{equation}
    |\psi_f\rangle = D(-\alpha e^{i\theta})e^{i\chi t\hat{a}^\dagger\hat{a}}D(\alpha)\, |0\rangle\otimes|e\rangle = |\alpha e^{i\chi t } - \alpha e^{i\theta}\rangle\otimes|e\rangle,
\end{equation}
up to a global phase. When $\theta = \chi t$, the cavity goes back to vacuum. The vacuum state can be identified by measuring $|\langle0|\psi_f\rangle|^2$ with a long qubit $\pi$ pulse, which works reasonably well even if the peaks are not individually resolved. 

The out-and-back experiment is shown in Figure~\ref{fig:chimethods}b for $\Delta/2\pi = 200\,\textnormal{MHz}$. The angular evolution fits a linear function with $\chi/2\pi = 0.18\,\textnormal{MHz}$. The same measurement was repeated for $\Delta/2\pi = 150\,\textnormal{MHz}$ and $600\,\textnormal{MHz}$. Table~\ref{tab:chi_values} shows the values of $\chi$ for six different points, which fit Eq.~\ref{eq:dispersive_SW} for $g/2\pi = 8.6\pm0.2\,\textnormal{MHz}$. I assume this coupling factor value to be less accurate than that obtained from the avoided crossing, since the latter provides a more direct estimation of $g$.

\vspace{0.5cm}
\begin{table}[htp]
\centering
\begin{tabular}{c c c c}
\toprule
\addlinespace
$\omega_T/2\pi$ (GHz) & $\Delta/2\pi$ (MHz) & $\chi/2\pi$ ($\textnormal{MHz}$) & $K/2\pi$ ($\textnormal{kHz}$)   \\
\addlinespace
\midrule
\addlinespace	
5.709 & $35$ & $1.71$ & $44$ \\
\addlinespace
5.677 & $67$ & $0.94$ & $6$  \\
\addlinespace
5.639 & $105$ & $0.57$ & $1.9$ \\
\addlinespace
5.594 & $150 $ & $0.29$ & $0.19$ \\
\addlinespace
5.544 & $200$ & $0.18$  & $0.09$ \\
\addlinespace
5.144 & $600$ & $0.05$ & $0.005$ \\
\addlinespace
\bottomrule
\addlinespace
\end{tabular}
\caption[System parameters at different detunings from the cavity]{\textbf{System parameters at different detunings from the cavity.} The dispersive shift $\chi$ was measured either with number-splitting spectroscopy or out-and-back experiment. Assuming $\alpha/2\pi = 202\,\textnormal{MHz}$, the values of $\chi$ fit Eq.~\ref{eq:dispersive_SW} for $g/2\pi = 8.6\pm0.2\,\textnormal{MHz}$. For consistency, all of these values were obtained in the same cooldown. The values of cavity self-Kerr $K$ are estimated from numerical simulations that match the corresponding values of $\chi$.}
\label{tab:chi_values}
\end{table}

The transmon relaxation and coherence times were also characterized at different flux points. The $T_1$ was measured by tracking the decay of the excited-state population as $\textnormal{P}_{|e\rangle}(t) \propto e^{-\frac{t}{T_1}}$, while the transmon Ramsey $T_2^*$ was extracted from Ramsey interferometry~\cite{krantz_quantum_2019}. The results are summarized in Table~\ref{tab:qubit_coherence}. The transmon $T_1$ decreased from $18.0\,\mu\textnormal{s}$ to $9.2\,\mu\textnormal{s}$ as the detuning from the readout resonator was reduced, indicating that the Purcell decay is a dominant factor. The highest $T_2^*$ of $2.9\,\mu\textnormal{s}$ was measured at the sweet spot, which decreased to $T_2^*=0.86\,\mu\textnormal{s}$ after tuning the transmon to lower frequencies. The Hahn echo $T_2=3.6\,\mu\textnormal{s}$ was measured at $\omega_T/2\pi = 5.807\,\textnormal{GHz}$ (not shown).

\vspace{0.5cm}
\begin{table}[htp]
\centering
\begin{tabular}{c c c c}
\toprule
\addlinespace
$\omega_T/2\pi$ (GHz) & $\Delta_\text{RR}$ (GHz) & $T_1$ ($\mu$s) & $T_2^*$ ($\mu$s) \\
\addlinespace
\midrule
\addlinespace
5.607 & $1.325$ & $18.0$ & --- \\
\addlinespace
5.667 & $1.265$ & $15.9$ & $0.86$ \\
\addlinespace
5.714 & $1.218$ & $15.5$ & $0.86$ \\
\addlinespace
5.874 & $1.058$ & $13.7$ & $1.00$ \\
\addlinespace
6.413 & $0.519$ & $9.2$ & $2.92$ \\
\addlinespace
\bottomrule
\addlinespace
\end{tabular}
\caption[Qubit relaxation and coherence times at different flux points]{\textbf{Qubit relaxation and coherence times at different flux points.} The detuning $\Delta_{RR} = \omega_c - \omega_T$ is relative to the readout resonator. The relaxation times $T_1$ decrease as the transmon becomes closer to the readout, which indicates Purcell leakage. The $T_2$ increases at high frequencies because of the lower susceptibility to noise closer to the upper sweet spot at $\omega_T/2\pi \approx 6.413\,\textnormal{GHz}$.}
\label{tab:qubit_coherence}
\end{table}

\begin{figure}
    \centering
    \includegraphics[width=\linewidth]{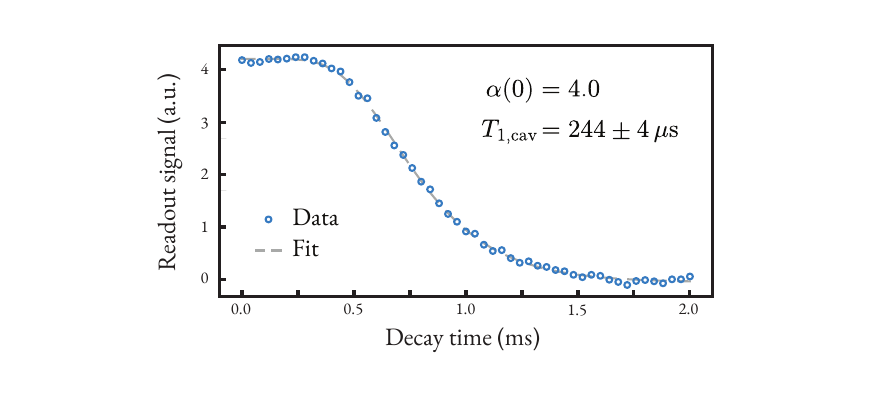}
    \caption[Measurement of the cavity single-photon lifetime.]{\textbf{Measurement of the cavity single-photon lifetime.} A coherent state was prepared in the cavity and let decay for a variable time, at the end of which the vacuum population was measured with a selective transmon pulse. The data fit Eq.~\ref{eq:cohstate_decay} for $\alpha(0) = 4.0$ and $T_{1, \text{cav}} = 244\pm4\,\mu\textnormal{s}$. }
    \label{fig:cavt1}
\end{figure}

Next, I measured the cavity single-photon lifetime $T_{1,\text{cav}}$ at a detuning $\Delta/2\pi = 77\,\textnormal{MHz}$. At this point, the dispersive shift $\chi/2\pi\approx1\,\textnormal{MHz}$ allows using number-selective measurements to track the decaying population of a coherent state $|\alpha\rangle$. Since $\langle\hat{z}(t)\rangle = |\alpha(t)|^2 \propto e^{-\frac{t}{T_{1,\text{cav}}}}$, the coherent state decays as
\begin{equation}
    \alpha(t) = \alpha(0)e^{-t/\left(2T_{1,\text{cav}}\right)},
\end{equation}
and the vacuum state population follows the double-exponential function
\begin{equation}
\label{eq:cohstate_decay}
    |\langle 0 |\alpha(t)\rangle|^2 = e^{-|\alpha|^2} = \exp\left(-|\alpha(0)|^2e^{-t/T_{1,\text{cav}}}\right).
\end{equation}
Figure~\ref{fig:cavt1} shows the decay of a coherent state over time for an initial $\alpha = 4.0$, together with a fit to Equation~\ref{eq:cohstate_decay}. 

The measurement indicates a cavity lifetime of $T_{1,\text{cav}} = 244\pm4\,\mu\textnormal{s}$, where the error is the standard deviation estimated from the covariance matrix of the least-squares fit. This value corresponds to a total $Q=9\times 10^6$, an improvement of almost one order of magnitude over previously reported results using the magnetic hose~\cite{reed_entanglement_2013, gargiulo_fast_2021}. Assuming that other loss channels limit the photon decay to $1\,\textnormal{ms}$ ($Q_{\text{other}} = 3.6\times10^7$) based on typical lifetimes of similar cavities coupled to fixed-frequency transmons, the hose-induced losses can be estimated to correspond to a $Q_{\text{hose}} = \left(Q^{-1} - Q_{\text{other}}^{-1}\right)^{-1} = 1.2\times10^7$. This value approaches the simulated $Q_{\text{hose}} = 2\times10^7$, while the difference is possibly due to deviations from the nominal design parameters.

These results achieve the first goal of Somerset: to demonstrate tunable light-matter interactions while protecting the cavity mode from decay. I will move on to the second objective of this chapter. I will use the transmon tunability to move between different interaction regimes on demand, and perform QIP tasks in ways that are not typically possible in fixed-frequency bosonic circuits.

\section{Quantum information processing with tunable interactions}

\begin{figure}
    \centering
    \includegraphics[width=\linewidth]{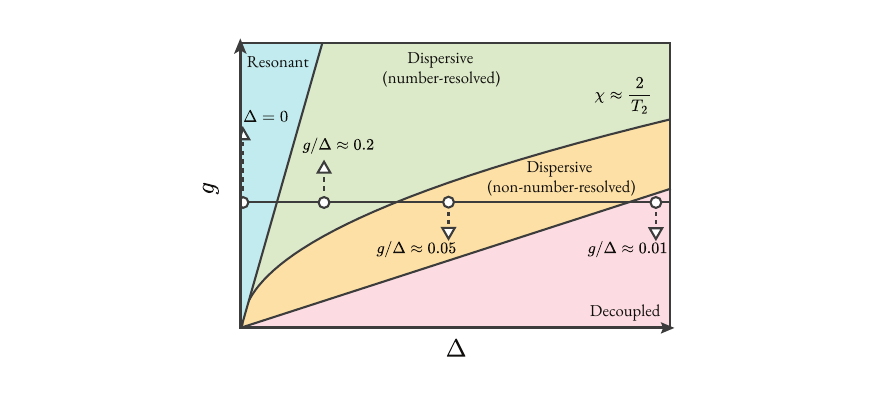}
    \caption[Light-matter interaction regimes]{\textbf{Light-matter interaction regimes.} The diagram qualitatively shows the interaction regimes as a function of $g$ and $\Delta$. Somerset can switch between the indicated points and use techniques from different regimes. }
    \label{fig:regimes}
\end{figure}

The interplay between the transmon linewidth and $\chi$ leads to a  distinction between the strong and weak dispersive regimes, leading to different approaches for manipulating information in the cavity. Moreover, the control over $\Delta$ allows access to the resonant regime, where the nonlinearities are the strongest. And in the limit $\Delta\to\infty$, the interactions of the system are suppressed in the decoupled regime. Figure~\ref{fig:regimes} shows a qualitative diagram of these regimes as a function of $g$ and $\Delta$, illustrating operational points accessible by Somerset.

To harness different interaction regimes on demand, the frequency control must be fast and precise. As discussed in Section~\ref{sec:wiringpredistortion}, this is not trivial. Microwave components along the flux line distort the current pulses that reach the coil. The finite inductance of the coil itself slows down sharp pulse edges, while the high-pass filter of the bias-tee causes the current to decay over time. 

\begin{figure}
    \centering
    \includegraphics[width=\linewidth]{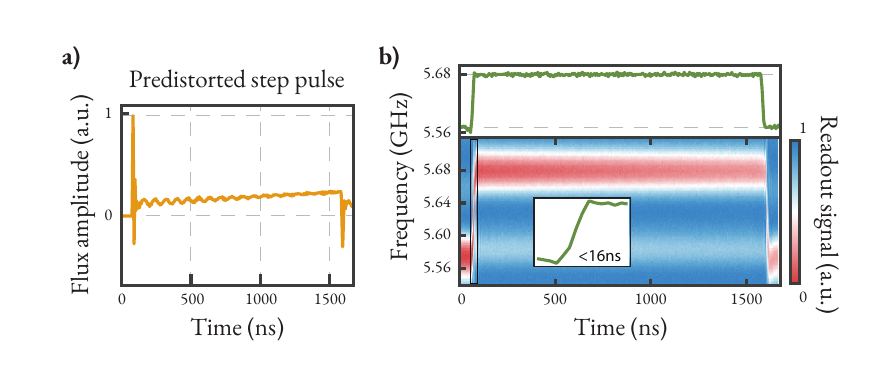}
    \caption[Fast flux control]{\textbf{Fast flux control.} a) Current control necessary to generate a step flux pulse at the SQUID. The predistortion is done by 11 IIR filters and 2 FIR filters calibrated with the methodology described in Section~\ref{sec:wiringpredistortion}. The large initial voltage speeds up the charging time of the coil. The slow increasing edge counters the attenuation from the $4\,\textnormal{kHz}$ high-pass filter of the bias-tee. b) $\pi$-scope results showing the step frequency response of the transmon using the predistorted pulse. The dynamical voltage range allows detunings of over $100\,\textnormal{MHz}$ within less than $16\,\textnormal{ns}$. The upper panel shows the fitted frequency at each instant. }
    \label{fig:predistortion}
\end{figure}

The first step in correcting these distortions is to characterize them. To measure the frequency trajectory of the transmon during a step flux pulse, I carried out a time-dependent transmon spectroscopy experiment known as $\pi$-scope. The transmon was initially parked at a flux-sensitive point out of the sweet spot, $\omega_T(0)/2\pi \approx 5.60\,\textnormal{GHz}$, and a step pulse was applied to the coil. Then a short $\pi$ pulse ($16\,\textnormal{ns}$) was applied to the transmon with a variable frequency while also sweeping the relative delay with respect to the flux pulse, effectively probing the trajectory of $\omega_T(t)$. From the transmon frequency $\omega_T(t)$, it was possible to obtain $\varphi_e(t)$ and compare it to an ideal step pulse. 

The distortions from an ideal step pulse were digitally inverted by calibrating $11$ IIR and $2$ FIR filters. Using these filters, a predistorted $1.5\,\mu\textnormal{s}$ step pulse shows an $\omega_T(t)$ trajectory within $\pm0.2\%$ of the target frequency (Figure~\ref{fig:predistortion}). The pulse predistortion sped up the rising edge to a duration lower than the time resolution of $16\,\textnormal{ns}$, enabling frequency control in timescales much shorter than the decoherence rates in Somerset. But such a fast control comes with a caveat: the predistorted pulses must compensate for the slow charging of the coil by applying a large initial amplitude. This starting excitation saturates the dynamical range of the DAC, imposing a trade-off between the speed and the amplitude of the flux control. Regardless, the fastest step pulses with $16\,\textnormal{ns}$ rising edge have enough amplitude for a detuning of more than $100\,\textnormal{MHz}$, which is suitable for the following experiments.

\begin{figure}
    \centering
    \includegraphics[width=\linewidth]{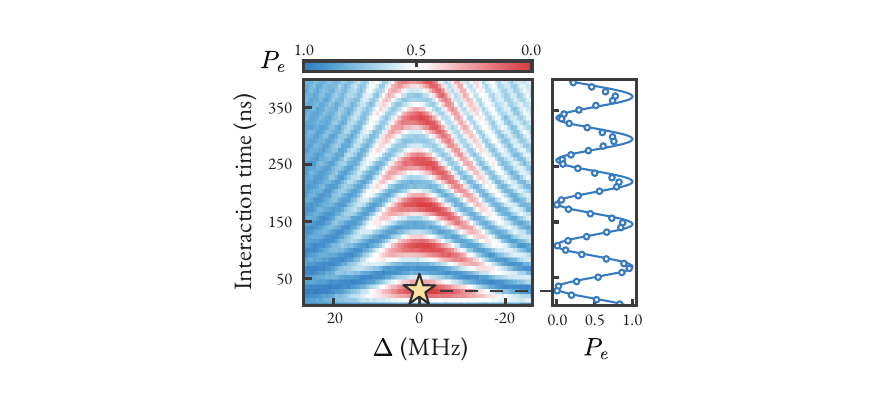}
    \caption[Vacuum Rabi oscillations]{\textbf{Vacuum Rabi oscillations.} The transmon is excited and quickly tuned into resonance with the cavity, allowing the photon to coherently oscillates between the two circuits. The plot shows the transmon state after a variable interaction time and $\Delta = \omega_a - \omega_T$. The side panel shows the $1\textnormal{D}$ cut at $\Delta =0$. Fock state $|1\rangle$ can be prepared by interrupting the experiment at the point of minimum transmon population, at $32\,\textnormal{ns}$ (indicated by a star). The transmon population was measured using single-shot readout and rescaled to account for readout errors.}
    \label{fig:vacuum_rabi}
\end{figure}

The fast step pulses were employed to prepare non-Gaussian Fock states in the cavity. The transmon was initially biased at $\Delta/2\pi = 35\,\textnormal{MHz}$ and then excited to $|e\rangle$ with an $80\,\textnormal{ns}$-long $\pi$ pulse. A step pulse then quickly tuned the transmon into resonance with the cavity ($\Delta = 0$). This caused the system to non-adiabatically\footnote{This non-adiabaticity refers to a switching time much faster than $2\pi/g \approx 150\,\textnormal{ns}$. However, the system still evolves adiabatically with respect to the eigenfrequencies $\omega_a$, $\omega_T$.} evolve to the Hamiltonian
\begin{equation}
\label{eq:beamsplitting}
    H = g\left(\hat{a}^\dagger\hat{\sigma}_- + \hat{a}\hat{\sigma}_+\right),
\end{equation}
which caused the excitation to swap between the transmon and the cavity as 
\begin{equation}
    |0\rangle\otimes|e\rangle \leftrightarrow |1\rangle\otimes|g\rangle
\end{equation}
with period $\pi/g$. These are known as vacuum Rabi oscillations. I performed this experiment while sweeping the amplitude of the step pulse (which corresponds to a sweep of $\Delta$ during the interaction), and its duration, which defines the total oscillation time. After the interaction, the transmon returned to the initial flux bias and its population was measured. Figure~\ref{fig:vacuum_rabi} shows the result of this experiment, with the characteristic chevron pattern depicting the photon oscillating between the two modes. The side panel shows the $1\textnormal{D}$ cut at $\Delta = 0$, which matches well with simulations for $g/2\pi = 6.65\,\textnormal{MHz}$. I attribute the change in the value of $g$ relative to the previously reported value of $g/2\pi = 7.3\,\textnormal{MHz}$ to cooldown-to-cooldown variations in the device parameters. The oscillations decay with a time $\tau = 1.5\,\mu\textnormal{s}$, which should correspond to the transmon $T_2^*$ enhanced by the cavity hybridization. The Fock state $|1\rangle$ was then prepared by evolving under resonant oscillations with a flux pulse with nominal duration $t_{|1\rangle}= 32\,\textnormal{ns}\approx \pi/2g$. The same procedure was repeated a second time to create a Fock state $|2\rangle$ with an additional flux pulse of nominal duration $t_{|2\rangle}= 23\,\textnormal{ns} \approx t_{|1\rangle}/\sqrt{2}$. The shorter time of the second oscillation is due to the bosonic enhancement of the coupling.

The created states were characterized in two different ways. In the first method, the transmon was detuned to $\Delta/2\pi = 35\,\textnormal{MHz}$, and then used to perform a full Wigner tomography. The parity mapping sequence consisted of two $16\,\textnormal{ns}$-long transmon pulses $R_x\left(\pi/2\right)$ separated by a wait time of $ 296\,\textnormal{ns}\approx\pi/\chi$ (total protocol duration of $328\,\textnormal{ns}$). The measurement was repeated twice while varying the second pulse between $R_x\left(\pi/2\right)$ and $R_x\left(-\pi/2\right)$, mapping the parity to different transmon states~\cite{krisnanda_demonstrating_2025}. The two measurement results were subtracted to cancel out coherent errors such as the cross-Kerr $K_{ac}$, leading to a more accurate parity measurement. The parity results were then rescaled according to the contrast of the Wigner function of the vacuum state $|0\rangle$. This step removed loss of contrast due to tomography imperfections, highlighting the state preparation errors. 

\begin{figure}
    \centering
    \includegraphics[width=\linewidth]{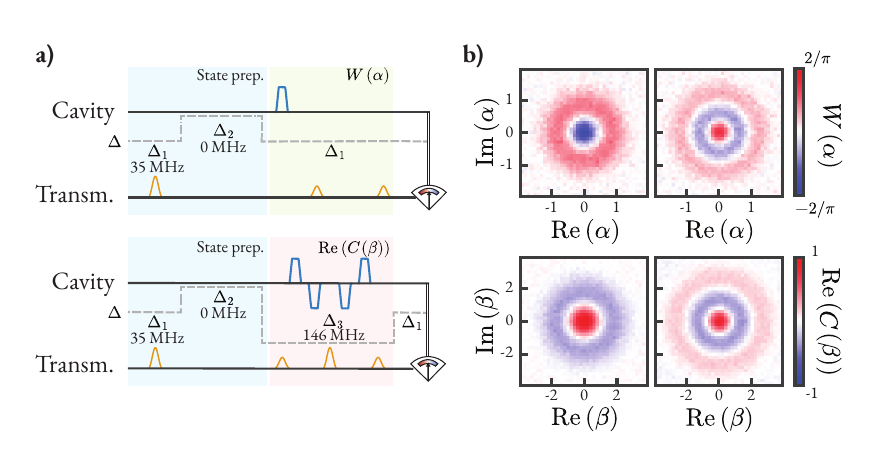}
    \caption[On-demand access to Wigner and characteristic function tomographies]{\textbf{On-demand access to Wigner and characteristic function tomographies.} (a) Measurement sequence for both tomographies, showing the evolution of $\Delta$ at different points in time. b) Respective results. The signals are rescaled with respect to the tomography of vacuum, which removes measurement imperfections to highlight the quality of the state preparation. }
    \label{fig:tomo}
\end{figure}

In the second method, the real part of the characteristic function was measured using the $\textnormal{ECD}$ gate at $\Delta = 146\,\textnormal{MHz}$. The gate parameters were calibrated to generate a vacuum characteristic function $C\left(\beta\right) = e^{-\frac{1}{2}x^2}$, whose amplitude and offset were used to rescale the tomography results, similar to the rescaling used for the Wigner tomography. The displacement amplitudes in the $\textnormal{ECD}$ gate sequence were maximized, allowing the displacement durations and the free $\chi$ evolution times to be as short as possible ($20\,\textnormal{ns}$ and $116\,\textnormal{ns}$ for each displacement and $\chi$ evolution interval, respectively). Including two $16\,\textnormal{ns}$-long $\pi/2$ pulses and one $\pi$ pulse of same length, the characteristic function tomography had a total length of $360\,\textnormal{ns}$. Figure~\ref{fig:tomo} shows the results for both Wigner and characteristic function tomographies of the $|1\rangle$ and $|2\rangle$ Fock states. 

The state creation quality was quantified by the state fidelity calculated from the overlap of the tomography function with that of the ideal state. For the Wigner (characteristic function) tomography, the fidelities are $91.1\%$ ($82.8\%$) for $|1\rangle$ and $66.6\%$ ($62.8\%$) for $|2\rangle$. The imperfections in state preparation are attributed to a low $T_2^*\approx 500\,\textnormal{ns}$ during the transmon excitation pulses. Again, I attribute the reduction in transmon coherence times compared to Table~\ref{tab:qubit_coherence} to cooldown-to-cooldown variations. 

Overall, the characteristic function was the most efficient tomography method in terms requiring less measurement repetitions for the same signal-to-noise ratio. This higher efficiency comes from its echo sequence, which is resilient to low frequency noise on the transmon. So, while Wigner is affected by the Ramsey $T_2^* = 500\,\textnormal{ns}$, $C(\left(\beta\right)$ is affected by the longer Hahn echo $T_2 = 800\,\textnormal{ns}$. This shows that the fast flux control is useful not only to access observables pertaining to different interaction regimes, but also to tailor the tomography to the specificities of the system. 

These experiments achieve the second objective of this chapter: to use on-demand control of light-matter interactions to mix control and tomography techniques from different regimes. They demonstrate the use of the strong nonlinearities of the resonant regime to create non-Gaussian $|1\rangle$ and $|2\rangle$ states, a technique that can be expanded for arbitrary state preparation~\cite{hofheinz_synthesizing_2009}. They also demonstrate Somerset is able to access on demand the three main observables analytically available for bosonic mode tomography in circuit QED, namely the photon number distribution $P(z) = \textnormal{Tr}\left[\boldsymbol{\rho}|z\rangle\langle z|\right]$, the Wigner function $W(\alpha) = \frac{2}{\pi}\textnormal{Tr}\left[\hat{D}(\alpha)\boldsymbol{\rho}\hat{D}(-\alpha)\hat{\Pi}\right]$, and the characteristic function $C(\beta) = \textnormal{Tr}\left[\boldsymbol{\rho}\hat{D}(\beta)\right]$.

In the next section, I will exploit the fast frequency tunability of Somerset to access larger detunings and suppress both the $\chi$ and $K$ terms. This dynamical decoupling is useful for protecting the cavity state from nonlinear distortions, addressing a key challenge of bosonic circuit QED.

\section{Dynamical suppression of nonlinearities}

\begin{figure}
    \centering
    \includegraphics[width=\linewidth]{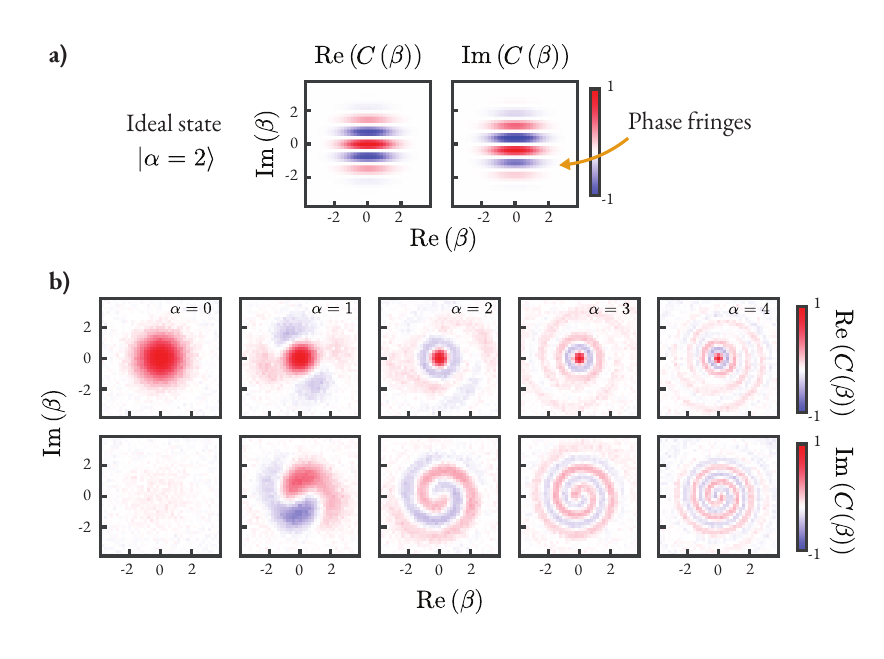}
    \caption[Effect of Kerr on coherent states of the cavity]{\textbf{Effect of Kerr on coherent states of the cavity.} a) Characteristic function of an ideal coherent state with $\alpha=2$. The fringes are present in both the real and imaginary components and indicate the phase of the state. b) Measured characteristic functions for diverse coherent states $|\alpha\rangle$ ($\alpha=0$ to $\alpha = 4$) after $5\,\mu\textnormal{s}$ of free evolution at $\Delta/2\pi = 35\,\textnormal{MHz}$. The high value of cavity self-Kerr (estimated $44\,\textnormal{kHz}$) causes the fringes to spiral around the origin, significantly distorting the state. }
    \label{fig:alpha_dist}
\end{figure}

The main advantage of using the weak dispersive regime for QIP is to mitigate nonlinearities such as the self-Kerr term $-\frac{K_a}{2}\hat{a}^\dagger\hat{a}^\dagger\hat{a}\hat{a}$. This term adds a quadratic phase shift to the cavity state as a function of the photon number, that is
\begin{equation}
\label{eq:selfkerr}
    e^{i\frac{K_a t}{2}\hat{a}^\dagger\hat{a}^\dagger\hat{a}\hat{a}}\sum_{z=0}^\infty c_z|z\rangle = \sum_{z=0}^\infty c_z e^{i\frac{K_a t}{2}\left(z-1\right)z} |z\rangle, 
\end{equation}
where $c_z$ are the Fock-basis expansion coefficients of an arbitrary state. The Kerr evolution reshapes the phase-space distribution of the cavity state beyond a simple rotation, and for many applications it is an unwanted distortion that obscures quantum information~\cite{kirchmair_observation_2013, pietikinen_strategies_2024}. While Kerr is a coherent evolution that can in principle be reversed with SNAP gates~\cite{heeres_cavity_2015}, its combination with single-photon loss leads to uncorrectable dephasing of the cavity~\cite{puri_engineering_2017}. Considering that Kerr becomes a limiting factor in longer and more complex QIP protocols, there has been significant effort to investigate methods to mitigate this term, such as drive cancellation techniques~\cite{wang_photon_2021, zhang_drive_2022} and Kerr-free nonlinear circuits~\cite{pietikinen_strategies_2024, frattini_wave_2017}.

Figure~\ref{fig:alpha_dist} shows the effect of Kerr on the characteristic function of the cavity state. Coherent states with different values of $\alpha$ were prepared in the cavity and left to evolve at a transmon detuning of $\Delta/2\pi = 35\,\textnormal{MHz}$ for $5\,\mu s$. At the end of the evolution, both the real and imaginary parts of $C\left(\beta\right)$ are measured to show the distortion of the initial state. Ideally (Fig.~\ref{fig:alpha_dist}a), the coherent state exhibits fringes that indicate its phase. Due to the self-Kerr, the fringes were distorted, spiraling around the center of the reciprocal phase space (Fig.~\ref{fig:alpha_dist}b). This distortion is more pronounced for coherent states with larger average photon number $|\alpha|^2$, as expected from Eq.~\ref{eq:selfkerr}. Such a scaling is unfortunate, since accessing higher dimensions of the cavity state is generally desirable for quantum information processing.

\begin{figure}
    \centering
    \includegraphics[width=\linewidth]{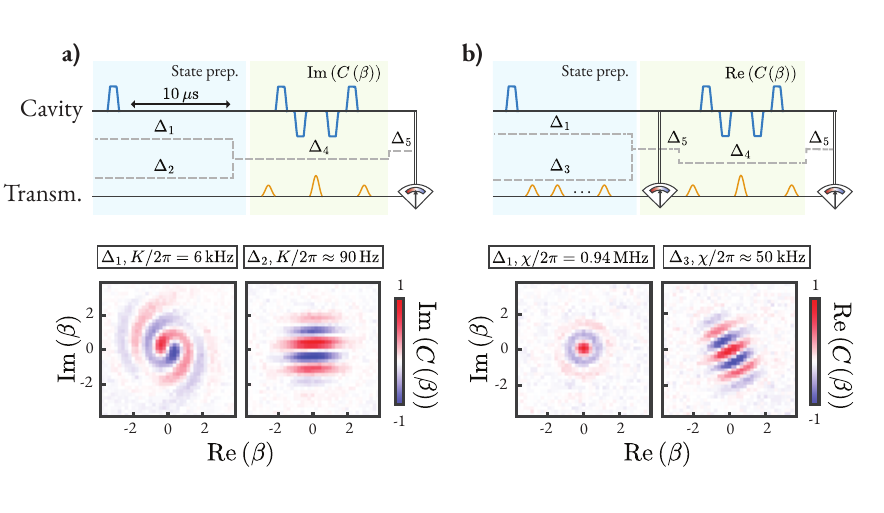}
    \caption[Suppression of nonlinear distortions]{\textbf{Suppression of nonlinear distortions.} (a) Evolution of a coherent state with $|\alpha|\approx6.25$ under the effect of Kerr. The experiment was executed at two different detunings: $\Delta_1/2\pi = 67\,\textnormal{MHz}$ ($K/2\pi= 6\,\textnormal{kHz}$), which induced significant Kerr distortion, and $\Delta_2/2\pi = 200\,\textnormal{MHz}$ ($K/2\pi\approx90\,\textnormal{Hz}$), in which the distortions were significantly suppressed. b) Evolution during a sequence of transmon excitations. The transmon $\pi/2$ pulses were used to induce periodic dispersive shifts, which cause the loss of coherence of the cavity state when the transmon is at $\Delta_1$. If the transmon is detuned to $\Delta_3/2\pi = 600\,\textnormal{MHz}$, the real part of the characteristic function shows the cavity state still preserves its coherence fringes. In both experiments, the tomography and the transmon readout are done at $\Delta_4/2\pi = 150\,\textnormal{MHz}$ and $\Delta_5/2\pi = 105\,\textnormal{MHz}$, respectively.}
    \label{fig:nonlinear_supp}
\end{figure}

The fast-flux tunability of Somerset can dynamically suppress the self-Kerr of the cavity. To show this, a coherent state $|\alpha|^2\approx 6.25$ is prepared and left to evolve $10\,\mu s$ under different flux points. In the first point, $\Delta_1/2\pi = 67\,\textnormal{MHz} $ with an estimated self-Kerr $K/2\pi \approx 6\,\textnormal{kHz}$. The characteristic function (Fig.~\ref{fig:nonlinear_supp}a) of the final state shows spiraling fringes, as expected. However, when the state evolved at a more detuned point $\Delta_2/2\pi =200\,\textnormal{MHz} $ (estimated $K/2\pi\approx 90\,\textnormal{Hz}$.), the distortion was strongly suppressed, with the final state closely resembling the expected shape of $|\alpha\rangle$.

The dynamical decoupling can also suppress the dispersive interaction $-\frac{\chi}{2}\left(1-\hat{\sigma}_z\right)\hat{a}^\dagger \hat{a}$. As discussed previously, the dispersive interaction introduces a transmon-state dependent frequency shift in the cavity, which can entangle the two modes. This prevents the independent operation of the transmon, since it cannot be operated without affecting the state of the cavity. To show the dynamical suppression of $\chi$, again a coherent state $|\alpha|^2\approx 6.25$ is created and left to evolve at two different flux bias: $\Delta_1 = 67\,\textnormal{MHz}$ ($\chi = 0.94\,\textnormal{MHz}$) and $\Delta_3 = 600\,\textnormal{MHz}$ ($\chi = 0.05\,\textnormal{MHz}$). To induce dispersive shifts, a sequence of ten $\pi/2$ pulses was applied to the transmon, each followed by an interval of $\tau = 400\,\textnormal{ns}$. The action of $\chi$ caused evolutions of the form
\begin{equation}
    |\alpha\rangle\otimes\frac{1}{\sqrt{2}}\left(|g\rangle + |e\rangle\right) \longrightarrow  \frac{1}{\sqrt{2}}\left(|\alpha\rangle|g\rangle + |\alpha e^{i\chi\tau}\rangle|e\rangle\right),
\end{equation}
which, given $T_2^* = 0.5\,\mu\textnormal{s}$, further evolved toward the mixed state
\begin{equation}
    \boldsymbol{\rho} = \frac{1}{2}|\alpha\rangle\langle \alpha|\otimes|g\rangle\langle g| + \frac{1}{2}|\alpha e^{i\chi\tau}\rangle\langle \alpha e^{i\chi\tau}|\otimes|e\rangle\langle e|.
\end{equation}
By repeating this evolution multiple times, the coherent state was split into many branches with phase factors $k\chi\tau$, $k = 0, 1,..., 10$, similarly to a random-walk evolution in phase. Then, the transmon was projected onto the ground state using single-shot readout and the real part of the characteristic function was measured (Fig.~\ref{fig:nonlinear_supp}b). When this protocol was executed at $\Delta_1$ ($\chi/2\pi = 0.94\,\textnormal{MHz}$), the state fringes were completely washed away, showing a rotational symmetry that indicates a complete absence of a coherent phase. On the other hand, executing this evolution at a larger detuning $\Delta_3$ where the dispersive shift is reduced to $\chi/2\pi \approx 50\,\textnormal{kHz}$, the resulting tomography is much closer to that of a coherent state. The final state still conserves the phase fringes, despite a residual overall rotation given by $5\chi\tau$ and the narrowing of the features caused by residual dephasing. 

This experiment concludes the third and last objective of this chapter, which is to use dynamical tuning of the transmon-cavity coupling to suppress distortions caused by nonlinear terms. Before closing chapter, in the next section will explore in more depth what place these results have in the current state of the art of bosonic circuit QED, and what is ahead for the development of flux-tunable architectures.

\section{Discussion}

This chapter has provided and answer to the research question of this thesis, leading to a peer-reviewed journal publication~\cite{valadares_ondemand_2024}. By adapting a magnetic hose design and optimizing its integration into a bosonic superconducting circuit, I demonstrated real-time control of light-matter interactions between a transmon and a high-$Q$ cavity. I used the capability of moving between interaction regimes for different quantum information processing tasks: fast creation of Fock states, on-demand access to different cavity observables for tailored tomography, and suppression of harmful cavity nonlinearities during idle times. These results provide a platform for the experimental implementation of several protocols that rely on such capabilities~\cite{hofheinz_synthesizing_2009, mezzacapo_digital_2014, park_efficient_2024}, encouraging further work on flux-tunable bosonic circuit QED.

During this project, the field of bosonic circuit QED saw significant progress in the topic of flux integration. In 2023, Chapman et al.~\cite{chapman_high_2023} used a superconducting loop to transfer the DC field of a coil to a SNAIL loop without direct electrical connection between them, mitigating energy leakage from the circuit. The SNAIL could then be used as a Kerr-free coupler between two cavities without compromising their lifetimes. In another work published in the same year, Lu et al.~\cite{lu_high_2023} biased a SQUID using the magnetic field of a buffer resonator, which oscillates at the correct frequency to activate a parametric beamsplitter between two bosonic modes. Compared to these works, Somerset remains advantageous for implementing broadband transmon tunability, since the flux provided by the hose ranges from DC to fast-switching step pulses. Soon after this chapter's results were published, another work by Gerhard Kirchmair's group successfully coupled the cavity to a fluxonium biased with a magnetic hose~\cite{atanasova_insitu_2025}. These recent developments demonstrate the interest of the community in developing flux-tunable architectures, as well as the timeliness of this thesis' topic.

Still, many aspects of Somerset require improvements, and there are still open problems. For one thing, the difficulty in manufacturing the hose poses a high entry barrier to other researchers. It is desirable to create a more reproducible solution to speed up development. A future solution should also prioritize solving the mechanical instabilities that lead to significant cooldown-to-cooldown variations. Ideally, the $\mu$-metal should also be removed. Its presence still imposes a limit on the cavity Q, and is possibly the cause of excessive flux line heating. In Section~\ref{sec:anatomy}, I observed that reducing the $\mu$-metal length was beneficial for flux delivery, which suggests the possibility of a $\mu$-metal-free hose.

Future work will also benefit from improved flux strength and bandwidth. In Somerset, the flux strength is relatively low. The mutual inductance between the coil and the SQUID is $0.067\,\Phi_0/\textnormal{mA}$, one order of magnitude lower than values routinely achieved on planar devices~\cite{krinner_engineering_2019}. This requires larger SQUID loops, which increase the transmon's susceptibility to environmental noise, or larger currents, which incur heating. As for the bandwidth, Somerset is capable of nanosecond scale control. But that comes at the cost of applying a very large initial voltage to the coil that reduces the control's dynamical range. Moreover, the limited bandwidth does not allow using parametric flux modulation. Thus, to improve the tunability over light-matter interactions by modulating the coupling factor $g$, a significant improvement in hardware is necessary.  

At the moment of writing this thesis, there are ongoing investigations in the field to use an updated, all-superconducting version of the hose, which should tackle some of these issues. I decided, however, to investigate a completely new direction, which is the topic of the next chapter.
\chapter{Broadband on-chip flux control for qudit operations in the light field}
\label{ch:lakeside}

The experiments discussed so far provide a partial answer to the topic of this thesis. I have shown that the magnetic hose can tune the light-matter dynamics by controlling $\Delta$ even as the cavity losses remain suppressed, and that this feature can be used to manipulate the bosonic mode across different interaction regimes. 

However, there is still significant unexplored potential in flux-tunable circuits. One important goal is to expand the space of accessible dynamics by providing control over the coupling factor $g$. Moreover, improving the coherences of the circuit would grant access to more sophisticated bosonic features of the cavity beyond Fock and coherent states, improving its usefulness for quantum information processing.

Somerset provides a proof-of-principle demonstration for flux tunability, and a benchmark for these improvements. But it is not immediately clear how its design can be modified to support parametric flux modulation. Flux control in the range of hundreds of $\textnormal{MHz}$ would require reducing the load inductance of the coil by decreasing its size. Using smaller coils would further weaken the flux control, with no obvious way to compensate for this loss. Increasing the transmon coherence times is also not trivial, since naively it requires decreasing the SQUID loop or increasing the hose distance -- two modifications that would compromise the flux tunability. Considering other issues such as heating, mechanical stability, and reproducibility, the design described in the previous chapter does not offer a clear pathway for improvement.

Given this challenge, it is necessary to consider a pathway that has been avoided until now: using on-chip flux lines. In principle, they provide the means to overcome all of these issues. On-chip lines can be placed in close proximity to the SQUID, which enhances the transmon tunability even when using low-inductance, broader-bandwidth sources. They are printed using standard circuit fabrication techniques, i.e., at no additional cost and with high reproducibility. However, once again, the critical question is whether such a structure can be implemented without compromising the quality factor of the cavity.

In the course of this thesis, I implemented on-chip flux lines in bosonic circuits using two different methods. In the first attempt, the aim was to suppress the cavity leakage using a strategy similar to Figure~\ref{fig:flux_line_strategies}b. That is, the flux line had an asymmetric design in which propagating cavity fields destructively interfered with their reflection, suppressing the leakage to the terminal of the line connected to the current source. The reflections were created by connecting one end of the flux line to the ground at a distance that was an integer multiple of $\lambda/2$ from the cavity, where $\lambda$ is the cavity's electrical length. This effectively implements a band-reject Purcell filter around the frequency corresponding to $\lambda$. To further reduce the leakage, the line was equipped with a lithographically defined $4\,\textnormal{GHz}$-cutoff low-pass filter. With this double filtering, the cavity losses through the flux line could be suppressed to reach $Q$ factors of up to $10^{12}$ in finite-element simulations. These values indicate that filtered on-chip lines can be implemented while protecting the cavity from leakage. At least in theory.

In practice, connecting to the flux line terminals was difficult. To make contact with a flux line inside the waveguide without introducing any seams, I used spring-loaded pins. These pins could connect to SMA ports outside the aluminum package and make contact with the chip inside the waveguide with an adjustable force. One pin grounded the line by directly shorting it to the waveguide, while a second pin was connected to the current source. However, the contact between the pins and the flux line was unstable. The electrical continuity strongly relied on manual mounting skill, and the pin could easily scratch the aluminum circuit during the process. The bad contact caused excessive heating when a DC current was applied, quickly destroying the superconducting state of the line. For these reasons, this attempt, described in detail in Appendix~\ref{append:pogopin}, was abandoned. Moving forward and overcoming these issues would require rethinking the on-chip approach from the ground up.

In this chapter, I will present the second approach used to implement an on-chip flux line. The off-plane connector pins are replaced with wirebonding, eliminating mechanical instabilities. The design of the flux line was changed from a stripline to a coplanar waveguide architecture, allowing strong filtering with lower crosstalk between circuit elements. The wirebonding required introducing seams to the waveguide, but the seam losses were mitigated by reducing their coupling with the cavity field as much as possible. This solution reimagined the typical bosonic circuit QED device, creating a hybrid $2\textnormal{D}$--$3\textnormal{D}$ circuit that leveraged the advantages of both architectures. This made it possible to achieve the two main goals of this chapter: introducing the parametric modulation of light-matter coupling, and effectively manipulating more complex, higher-dimensional states of the cavity mode, as demonstrated with qudit control protocols.

\section{Lakeside: a hybrid 2D-3D superconducting circuit}

\begin{figure}
    \centering
    \includegraphics[width=1\linewidth]{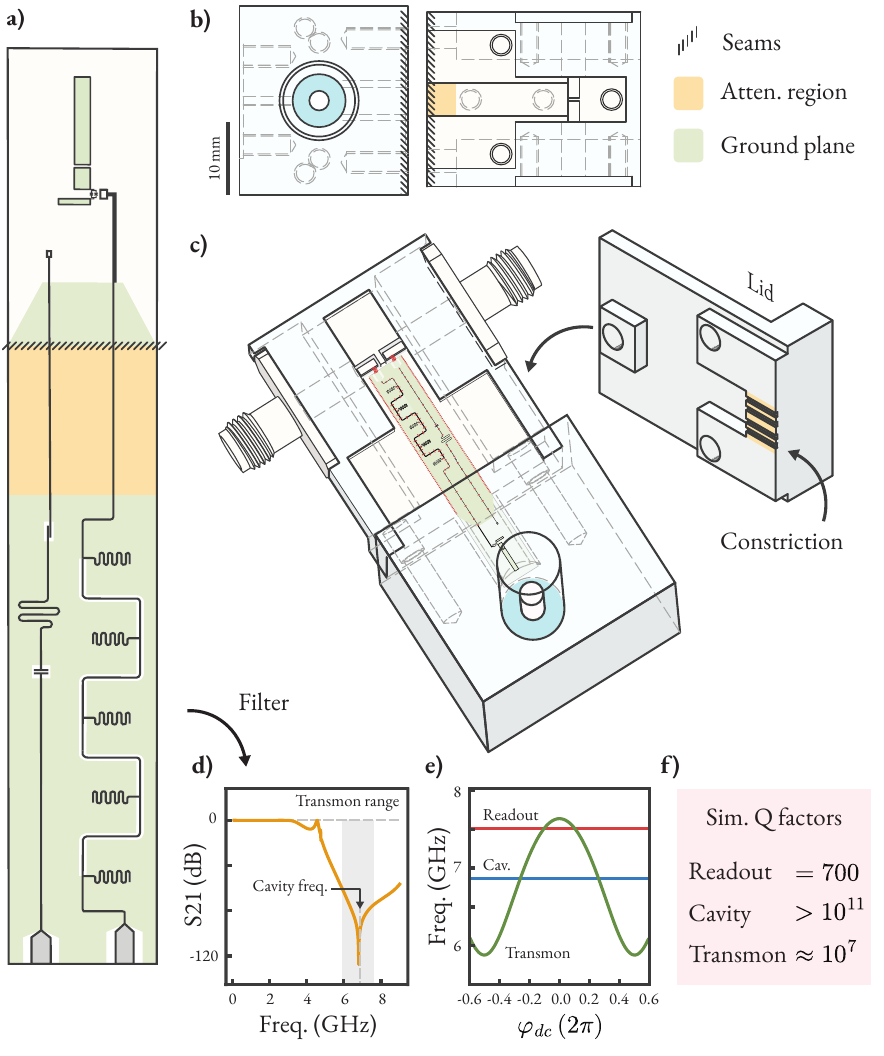}
    \caption[Bosonic circuit with protected on-chip flux line]{\textbf{Bosonic circuit with protected on-chip flux line} (a) Circuit printed on the chip, showing the location of the package seams and the attenuation region. (b) Main body (left), and bottom piece of the clamp (right). The connection between these two parts forms the seam. (c) Device assembly. The ground plane of the chip is wirebonded to the bottom clamp piece, and the line terminals are connected to the SMA ports. The lid closes the clamp, creating the attenuation region from a constriction of the waveguide. d) Simulated flux line filter $\textnormal{S}21$, with $98\,\textnormal{dB}$ attenuation at the cavity frequency and $>50\,\textnormal{dB}$ attenuation for frequencies larger than $5.62\,\textnormal{GHz}$. e) Frequency diagram of the three main elements as obtained in experiment later. f) Total simulated $Q$ factors. }
    \label{fig:lakeside}
\end{figure}

The new device, named \textit{Lakeside}, retained many familiar traits from typical bosonic cQED devices (see Fig.~\ref{fig:lakeside}). The cavity was made from a $\lambda/4$ coaxial stub resonator, and the coupled transmon was placed inside a transverse waveguide forming the ground plane of the stripline geometry. The cavity and the transmon coupled through a superconducting line, similar to Somerset, designed for $g/2\pi\approx 12\,\textnormal{MHz}$. This target coupling factor is almost twice as that of Somerset, which contributes to increase the speed and the fidelity of cavity operations.

The redesign focused on the flux line and the readout resonator. The current loop of the flux line that produces the field was placed $58\,\mu\textnormal{m}$ from the center of the SQUID loop (area of $470\,\mu\textnormal{m}^2$), leading to a simulated mutual inductance of $0.360\,\Phi_0/\textnormal{mA}$. The readout was designed to couple to the transmon with a simulated $g/2\pi\approx30\,\textnormal{MHz}$, a value that can be easily adjusted by changing their physical separation. At a distance $2.55\,\textnormal{mm}$ from the transmon, both the readout and the flux line were designed to merge into a ground plane printed on the chip, where they were reshaped into $50\,\Omega$-matched coplanar waveguide (CPW) lines (see Figure~\ref{fig:lakeside}a). From this point on, these circuit elements were subject to two stages of filtering. The first stage consists of $4\,\textnormal{mm}$ of a constrained waveguide section, which creates an attenuation region. In this section, the waveguide is as close as $0.07\,\textnormal{mm}$ from the surface of the ground plane, reducing the volume of the electromagnetic environment. This constriction greatly increases the cutoff frequency of the waveguide, enhancing the suppression of all free fields traveling outside the ground plane. This structure is conceptually similar to the high-pass filter at the output side of the magnetic hose (Fig.~\ref{fig:hose}a), with the difference that, in Lakeside, signals can still propagate across the filter through the CPW line without loss.

The second stage of filtering acts on the signals traveling on the CPW lines. The readout is connected to a Purcell bandpass filter, designed to be close in frequency to the readout and strongly coupled to the feed line. Because of this filter, the readout resonator is strongly coupled with $Q \approx 700$ at a frequency of $\omega_c/2\pi=7.51\,\textnormal{GHz}$ without significantly affecting the transmon $Q$ factor. In fact, the readout was conveniently designed to also carry transmon drives, with coupling $>4\times10^6$ to the transmon for frequencies lower than $7.27\,\textnormal{GHz}$. This corresponds to an upper bound of about $90\,\mu\textnormal{s}$ for the transmon $T_1$. The flux line, on the other hand, goes through two $3$rd-order $4\,\textnormal{GHz}$-cutoff Butterworth low-pass filters in series~\cite{pozar_microwave_2011}. The simulated $\textnormal{S}21$ of this section has $98\,\textnormal{dB}$ attenuation at the cavity frequency of $6.868\,\textnormal{GHz}$ and at least $50\,\textnormal{dB}$ attenuation for frequencies larger than $5.62\,\textnormal{GHz}$ (Figure~\ref{fig:lakeside}d). 

Because of this strong filtering, the flux line does not introduce significant leakage to the transmon and the cavity. The total simulated $Q$ factor of the cavity, including the readout and the flux line, is between $5\times10^{11}$ and $2\times10^{12}$, depending on the transmon frequency ($Q$s of other elements summarized in Figure~\ref{fig:lakeside}f). The transmon $Q$ factor is limited by the intentional coupling with the readout, which carries the charge drive. The design of Lakeside is thus in a favorable position in which the control channels do not impose a relevant limit to photon loss, despite their high coupling strengths.

These simulated conditions cannot be realized in practice without a carefully designed superconducting package. The package creates the attenuation region responsible for the first filtering step. Moreover, the on-chip ground plane is directly wirebonded to the package, creating a monolithic ground structure that enhances the microwave hygiene and eliminates package modes. To understand how this works, we must examine Lakeside's package in more detail.

The package can be divided into two main parts (Fig.~\ref{fig:lakeside}b). The main body is made of $4\textnormal{N}6$ aluminum and contains the cavity and the circular waveguide housing the transmon. This circular waveguide is relatively short, with only $9.7\,\textnormal{mm}$. This is a relevant detail because the seam at the end of the waveguide is closer to the cavity than usual, possibly introducing seam losses. The second part is called the clamp\footnote{It is called ``clamp'' for legacy reasons. In earlier designs, the lid was supposed to touch the ground plane of the chip in the attenuation region. But this step is risky and unnecessary. So instead the lid and the chip are just designed to be very close to each other without contact.}, which hosts the CPW section of the circuit.

The clamp consists of a bottom and a top piece. The bottom piece is the chip holder. It contains a $0.5\,\textnormal{mm}$-deep pocket where the chip is mounted with cryogenically compatible vacuum grease. After placing the chip, the on-chip ground is wirebonded to the surface of the clamp surrounding it (see Figure~\ref{fig:lakeside}c). On the back of the chip, the flux line and readout terminals are wirebonded to two flat Beryllium Copper pins connected to SMA ports, which provide the control signals. As for the top piece, it is essentially the lid that closes the structure and blocks radiated fields. Additionally, it has a $4\,\textnormal{mm}$-long extrusion that gets in close proximity with the top surface of the chip, creating the constriction that forms the attenuation region. This region separates the electromagnetic environments of the main body and of the clamp, and prevents the cavity and transmon fields from traveling outside the CPW lines. The two pieces of the clamp have rather small features which would be challenging to machine with high-purity $4\textnormal{N}6$ aluminum. Instead, I use 6061 aluminum alloy, which is more lossy but should not be a limiting factor to the quality of the cavity since its fields are mostly contained in the main body of the device.

This description covers the basic structure of Lakeside. This device was the result of an extensive, multivariable optimization which addressed the shortcomings of the previous research. There are many aspects of this optimization that are worth exploring in more depth, making clear what the advantages and limitations of this device are. Before moving on to the experimental characterization, I will discuss three key aspects of Lakeside.

\subsection*{The advantages of a hybrid circuit}

The most distinct feature of Lakeside is certainly the way in which the stripline $3\textnormal{D}$ circuits and the CPW $2\textnormal{D}$ lines coexist on the same chip. The reason for this hybrid approach is to extract the best of both architectures. $3\textnormal{D}$ circuits can reach high lifetimes, since their cavity and transmon fields are mostly stored in vacuum or in bulk sapphire, with reduced overlap with surface losses. Meanwhile, $2\textnormal{D}$ CPW lines offer low crosstalk, since their fields are well-shielded by the surrounding ground plane. This shielding isolates circuit elements and eliminates propagation pathways that bypass the filtering stages, which improves the suppression of leakage. It also makes circuits more compact, with multiple filters packed in a small chip area.

\begin{figure}
    \centering
    \includegraphics[width=\linewidth]{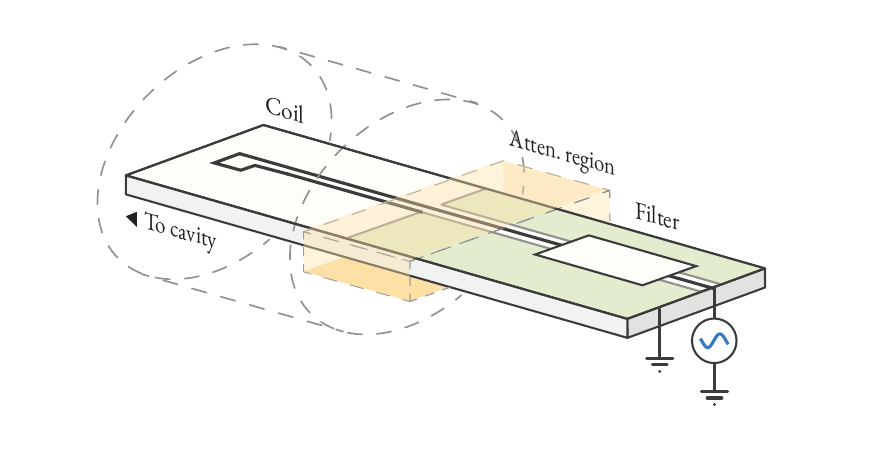}
    \caption[Lakeside flux line filtering strategy]{\textbf{Lakeside flux line filtering strategy} The circular waveguide carries the transmon and cavity energy through three modes: differential, common and evanescent. The differential mode becomes the CPW line mode at the point the coil connects to the ground plane, while the common mode becomes evanescent. Then, all evanescent modes are heavily filtered by the attenuation region. Only the CPW line mode is left, which is then filtered by two stages of $3$rd order Butterworth low-pass filters. The filter cutoff is chosen at $4\,\textnormal{GHz}$ to block the circuit leakage while allowing broadband flux control.}
    \label{fig:multifilter}
\end{figure}

Beyond low crosstalk and compactness, the stripline-to-CPW transition offers an elegant solution to the multimode propagation problem discussed in Section~\ref{sec:flux_tunability_challenge} (see Fig.~\ref{fig:multifilter}). In the stripline section, the flux loop can pick up the cavity and transmon fields in both common and differential modes. However, when these fields reach the ground plane, the differential mode transforms into the native CPW traveling mode, while the common mode becomes evanescent. The constriction of the attenuation area then accelerates the decay of the evanescent modes. Therefore, the sensitive fields only reach the back of the clamp through the CPW mode, which is heavily filtered to prevent leakage.

This mode filtering mechanism only works if the ground plane is shorted to the clamp, forming a single ground. If the ground plane is not shorted, it will instead act as the central pin of a coaxial line mode. This structure would support the propagation of sensitive fields into the clamp, a region that hosts many leakage channels such as the flux line and readout ports.

\subsection*{Heat load and thermalization}

One major limitation of the magnetic hose was the heating of the $\mu$-metal layers and the solder points along the flux line wiring. In Lakeside there is no $\mu$-metal, and all solder points are replaced by wirebonded connections. Therefore, the active heat load comes mostly from the SMA cable and is not expected to be significantly harmful when compared with the low heat load of planar circuits that use similar flux line designs~\cite{krinner_engineering_2019}.  

Moreover, the clamp should enhance the chip thermalization. In Somerset and in typical bosonic circuit QED devices, the chip is held by the clamp over a very short length of about $5\,\textnormal{mm}$. In Lakeside, $19.4\,\textnormal{mm}$ of the chip rests on the clamp surface, and the contact is improved using vacuum grease. I expect this increased contact to improve thermalization and provide a more effective sink eventual flux-related heating on the chip.

\subsection*{Seam losses}

The strong filtering of the control lines results in a very high upper bound for the $Q$ factor of the cavity. Therefore, in practice the cavity lifetime will be limited by a loss channel that is not featured in the finite-element simulations. Putting aside transmon-mediated cavity leakage, which can be mitigated by dynamically detuning the transmon, the cavity $T_1$ will be most likely affected by seam losses. 

Following this reasoning, much of the device's optimization focused on placing the seam as far as possible from the cavity. The seam losses were estimated by integrating the field over a $0.4\,\textnormal{mm}$-wide ribbon along the seam path. This makes it possible to gauge the relative increase or decrease of losses resulting from modifications on the circuit. As a result of the design optimization, I chose to use the same coupler line that was originally introduced with Somerset. The line is placed in-between the transmon and the cavity to retain some of their coupling while increasing their separation, as well as the length of the waveguide. 

However, this strategy to increase the cavity $T_1$ ultimately comes from a trade-off with the cavity-transmon coupling strength $g$. To overcome this constraint, future devices should invest in increasing the quality of the seam. In Lakeside, the main body and the clamp are connected simply with the use of indium wire. This contact could be improved with welding~\cite{milul_superconducting_2023}, with more advanced surface treatment or by designing Purcell cavities~\cite{krayzman_entanglement_2013}.

\begin{center}
$\ast\ \ast\ \ast$
\end{center}

In this detailed discussion, I explained why it was necessary to make unconventional design choices to solve the challenges presented by previous devices. The final device has unusual features compared to most bosonic circuits, but every change is justified by an expected increase in performance. With this background, we now turn to the characterization results and examine how they validate the design choices.

\section{Protected flux-parametric modulation}

Owing to its multi-factored optimization, Lakeside managed to overcome the current limitations of flux-tunable bosonic circuits. This device not only provided much cleaner light-matter interactions by means of improving transmon and cavity coherence times, but also gave access to the parametric modulation of the coupling strength.

To show this, I will first discuss the system properties. The transmon was found to be in the $\omega_T/2\pi = 5.894-7.634\,\textnormal{GHz}$ frequency range, with an asymmetry factor of $d=0.601$. The measured mutual inductance of the flux line with the SQUID loop was measured to be $0.387\,\Phi_0/\textnormal{mA}$, similar to the simulated value of $0.360\,\Phi_0/\textnormal{mA}$. The cavity was found at $\omega_a/2\pi = 6.868\,\textnormal{GHz}$, with an avoided crossing indicating a coupling of $g/2\pi = 12.07\,\textnormal{MHz}$ with the transmon. At the lower flux sweet spot ($\Delta/2\pi = 974\,\textnormal{MHz}$), the cavity-transmon dispersive shift is $\chi/2\pi = 40\,\textnormal{kHz}$, offering a convenient point for operations in the weak dispersive regime. At $\Delta/2\pi = 126\,\textnormal{MHz} \approx 10\times g/2\pi$, the coupling could attain good photon number selectivity given $\chi = 1.1\,\textnormal{MHz}$ and Ramsey $T_2^* = 1.3\,\mu\textnormal{s}$. 

To enable accurate control of the detuning and the dispersive shift, I calibrated the flux pulse predistortion as described in Section~\ref{sec:wiringpredistortion}. FIR filters were not required to correct the flux step pulse rising edge, which was instantaneous up to the $16\,\textnormal{ns}$ resolution of the $\pi$-scope experiment over a $100\,\textnormal{MHz}$ frequency range. This indicates the flux line bandwidth is much larger than that of the magnetic hose. It also means that it is not necessary to compensate for the inductive charging time of the flux line by using large current pulses, which consumes the dynamical range of the source. With 2 IIR filters, the step pulses were corrected to within $0.3\%$ error for the first $2\,\mu\textnormal{s}$, and $0.7\%$ during the first $6\,\mu\textnormal{s}$. In the following experiments, I used the first $2\,\mu\textnormal{s}$ of the step pulse for operations that are sensitive to the transmon frequency such as selective drives. Other operations such as readout are less sensitive to the transmon frequency and can be safely executed at any point within $6\,\mu\textnormal{s}$ after the flux pulse rising edge.

\begin{figure}
    \centering
    \includegraphics[width=1\linewidth]{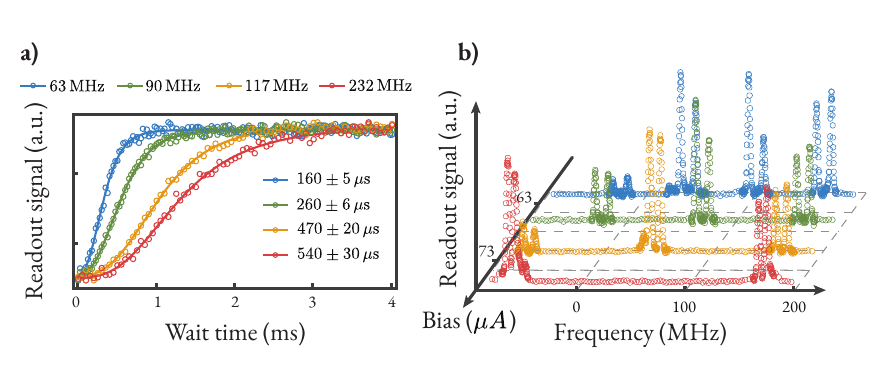}
    \caption[Protected flux modulation]{\textbf{Protected flux modulation.} (a) Cavity coherent state decay for progressively larger values of $\Delta$, showing the increase in $T_1$ up to  $540\pm30\,\mu\textnormal{s}$ at $\Delta/2\pi = 232\,\textnormal{MHz}$. b) Transmon spectroscopy for the same values of $\Delta$ under a flux pump $\varphi_{e}\propto\cos(\Delta t)$. The flux pump activates the parametric resonance between the transmon and the cavity, as shown from the frequency split that appears on every sideband due to the avoided crossing.}
    \label{fig:lakeside_t1}
\end{figure}


The calibrated flux pulses were used to measure the cavity lifetimes at different detunings of the transmon. The cavity was prepared with a coherent state with $\alpha = 2.5$ and left to decay for variable times at different values of $\Delta$. At the end of the decay, the transmon was tuned to $\Delta_{spec}/2\pi = 100\,\textnormal{MHz}$ ($\chi/2\pi \approx 1.8\,\textnormal{MHz}$), where the ground state population $|\langle 0|\alpha(t)\rangle|^2$ of the cavity was measured using a number-selective transmon $\pi$ pulse. Figure~\ref{fig:lakeside_t1}a shows the cavity $T_1$ increasing from $160\pm 5\,\mu\textnormal{s}$ at $\Delta/2\pi = 63\,\textnormal{MHz}$ up to $540\pm 30\,\mu\textnormal{s}$ when the transmon is detuned to $\Delta/2\pi = 232\,\textnormal{MHz}$ (uncertainties correspond to the standard deviation estimated from the covariance matrix of the least-squares fit). The latter value is not limited by transmon-induced decay, and is likely to be due to the seam losses as discussed in the previous section. The maximum cavity $T_1$ obtained across cooldowns was $590\pm40\,\mu\textnormal{s}$ at $\Delta/2\pi=260\,\textnormal{MHz}$. This represents a cavity $Q$ factor improvement of more than a factor of two compared with Somerset, reaching a total $Q = 2.6\times 10^7$.

The same flux points were used to test the activation of parametric resonances between the transmon and the cavity via flux modulation. The modulation frequency was tested up to a value of $232\,\textnormal{MHz}$, limited by the bandwidth of the DAC of $\pm250\,\textnormal{MHz}$. For each value of $\Delta$, I applied a flux pump $\varphi_e(t) \propto \cos(\Delta t)$, and performed a transmon spectroscopy experiment during the duration of the pump. As discussed in Section~\ref{sec:param_flux_mod}, this flux driving creates sidebands at $\bar{\omega}_T \pm N\Delta$, where $\bar{\omega}_T$ is the average transmon frequency and $N$ is the sideband index. Since in this experiment the transmon was parked at a point where the frequency response to flux is approximately linear, $\bar{\omega}_{T}\approx\omega_T(\varphi_{dc})$, and the $N=1$ sideband was resonant with the cavity frequency at $\bar{\omega}_T +\Delta \approx\omega_a$. Figure~\ref{fig:lakeside_t1}b shows the results of the spectroscopy. Multiple sidebands are visible in the frequency spectrum -- even the $N=-2$ sideband, shown as the leftmost feature in the spectroscopy for $\Delta/2\pi = 63\,\textnormal{MHz}$. Moreover, all features show a frequency split. This corresponds to the resolved avoided crossing between the $N=1$ sideband and the cavity, which can be accessed by the spectroscopy drive from any other sideband. These results validate the ability of the flux line to activate strongly-coupled parametric resonances through flux modulation. 

\begin{figure}
    \centering
    \includegraphics[width=1\linewidth]{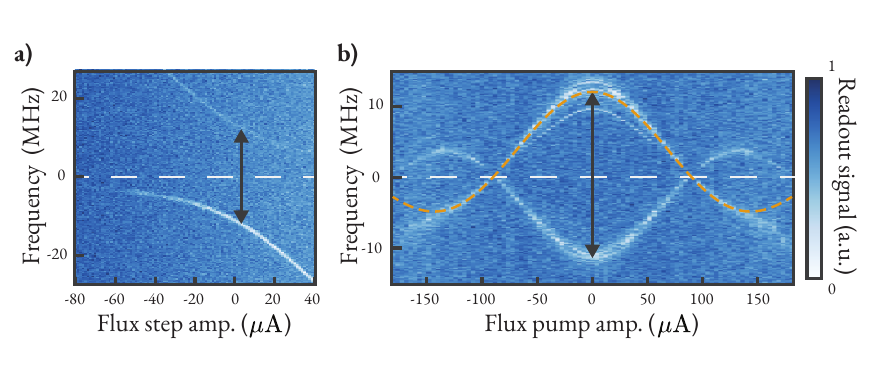}
    \caption[Parametric modulation of the coupling strength]{\textbf{Parametric modulation of the coupling strength.} (a) Spectroscopy around the cavity frequency as a function of DC flux when the transmon is near resonance, showing the avoided crossing with the cavity. The arrow indicates the minimum frequency split that is twice the value of $g/2\pi = 12.07\,\textnormal{MHz}$. b) Spectroscopy with the transmon parked at $\Delta = 0$ during a flux pump $\varphi_e(t) = A\cos(\omega_mt)$ for variable $A$ (converted to current) and $\omega_m/2\pi = 60\,\textnormal{MHz}$. The frequency split is compared with the function $g_{eff}^{(0)}= gJ_0\left(\frac{2.4048}{90\,\mu \textnormal{A}}\times I\right)$, shown as a golden dashed line, where $I$ is the current running through the coil in units of $\mu A$. }
    \label{fig:modulated_crossing}
\end{figure}

The magnitude of the coupling $g_{eff}^{(N)}$ can be calibrated by changing the modulation amplitude. This capability was tested in two ways. First, the transmon was put into resonance with the cavity, i.e., $\Delta =0$. Then, transmon spectroscopy was performed around the cavity frequency during a flux pump $\varphi_e(t) = A\cos(\omega_m t)$ for $\omega_m/2\pi = 60\,\textnormal{MHz}$ and a variable amplitude $A$, as shown in Figure~\ref{fig:modulated_crossing}b. The results show that the avoided crossing, which is twice the value of $g/2\pi= 12.07\,\textnormal{MHz}$ when $A=0$, becomes smaller as the modulation amplitude increases. At $A\approx90\,\mu \textnormal{A}$, the cavity-transmon frequency splitting vanished, indicating  $g_{eff}^{(0)} =0$, at which point the light-matter interactions were effectively switched off. The coupling factor is expected to follow a first-order Bessel function $J_0(x)$ (see Eq.~\ref{eq:geff}) as a function of modulation amplitude; so, from the first zero of this function at $x = 2.4048$, it is possible to infer $g_{eff}^{(0)} \approx gJ_0\left(\frac{2.4048}{90\,\mu \textnormal{A}}\times I\right)$, where $I$ is the current at the coil in units of $\mu\textnormal{A}$. This result is comparable to the expected function $g_{eff}^{(0)} \approx gJ_0\left(\frac{2.4048}{70\,\mu \textnormal{A}}\times I\right)$ from the measured mutual inductance of the line and the transmon frequency curve. For larger $A$, however, the avoided crossing curve deviated from the expected Bessel function shape. This happened at the point the linear approximation of $\omega_T(\varphi_e)$ broke down and the average transmon frequency $\bar{\omega}_T$ was no longer resonant with the cavity. 

\begin{figure}
    \centering
    \includegraphics[width=1\linewidth]{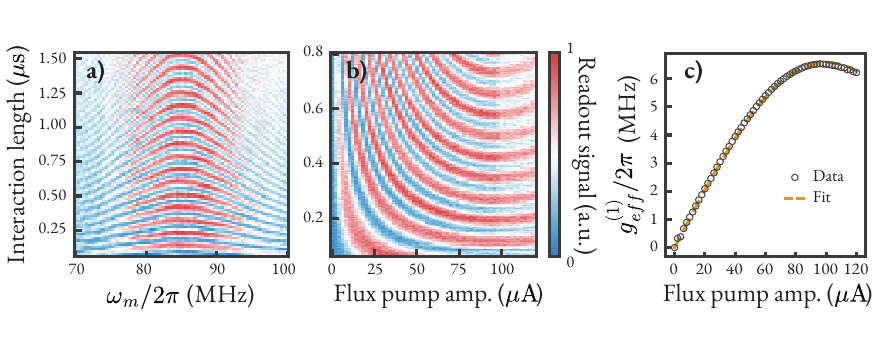}
    \caption[Flux-tunable parametric resonances]{\textbf{Flux-tunable parametric resonances.} (a) Vacuum Rabi oscillations activated by a flux sideband. The transmon is detuned with $\Delta/2\pi= 85\,\textnormal{MHz}$. After being excited with a $\pi$ pulse, a flux pump with $\omega_m\approx\Delta$ creates a sideband that activates the transmon-cavity exchange for a variable interaction time. b) Resonant vacuum Rabi oscillations as a function of the flux pump amplitude. The speed of Rabi oscillations can be continuously tuned in the $g_{eff}^{(1)}/2\pi = 0-6.57\,\textnormal{MHz}$ range. c) The effective coupling factors as a fucntion of the pump amplitude fits the Bessel function $g_{eff}^{(1)}= g J_1\left(I/I_0\right)$ with parameters $g = 11.3\,\textnormal{MHz}$ and $I_0 = 53\,\mu\textnormal{A}$}
    \label{fig:modulated_sideband_coupling}
\end{figure}

Next, I show the modulation of the sideband coupling factor $g_{eff}^{(1)}$ using a parametric vacuum Rabi experiment with adjustable Rabi frequency (Figure~\ref{fig:modulated_sideband_coupling}). The transmon was parked at a detuning $\Delta/2\pi = 85\,\textnormal{MHz}$ and excited to $|e\rangle$. Then the flux pump was activated with a variable frequency $\omega_m$ and variable duration. After the pump was turned off, a readout pulse measured the transmon state. For $\omega_m\approx \Delta$, the final transmon population revealed the familiar chevron pattern of vacuum Rabi oscillations with the cavity (Figure~\ref{fig:modulated_sideband_coupling}a), which characterizes their exchange of energy. The exchange was mediated by virtual photons of the flux pump, which bridged the frequency detuning $\Delta$ between the two modes and guaranteed the conservation of energy. In Figure~\ref{fig:modulated_sideband_coupling}b, the speed of resonant oscillations was adjusted as a function of the flux pump amplitude up to $120\,\mu\textnormal{A}$. Figure~\ref{fig:modulated_sideband_coupling}c shows that the effective coupling strength fits well to a Bessel function $g_{eff}^{(1)}/2\pi=11.3\,\textnormal{MHz}\times J_1(I/53\,\mu\textnormal{A})$, where $I$ is the current in the flux line in units of $\mu\textnormal{A}$.

These experiments show the dynamical tuning of the light-matter coupling through parametric modulation. Moreover, introducing this capability did not require sacrificing the cavity lifetime, which reached quality factors significantly higher than in previous devices. Lakeside has thus achieved two critical objectives for which it was designed. 

Lakeside also improved several other practical aspects that impact the quality of quantum information processing. For example, the baseline excited state population of the transmon was measured at $1.6\%$ until a bias of $0.28\,\Phi_0$, and had similarly low excitations for different values of the DC current. Also, none of the flux pulses used in this chapter were identified to introduce heat to the transmon. These improvements are a direct consequence of removing resistive elements such as solder points, and of increasing the thermal contact between the chip and the clamp.

Transmon coherence times have also improved. The flux line was wired according to Figure~\ref{fig:flux_wiring}b, with $10\,\textnormal{dB}$ attenuation at the $4\,\textnormal{K}$ stage of the refrigerator. The attenuation at room temperature was increased until the transmon $T_2^*$ reached its maximum value, resulting in additional $16\,\textnormal{dB}$ attenuation on both the AC and DC lines. At $\omega_T/2\pi =  7.147\,\textnormal{GHz}$, out of the sweet spot, the transmon has a stable $T_2^*=1.3\,\mu\textnormal{s}$. This improvement over Somerset's $T_2^* =500\,\textnormal{ns}$ is due to a few factors. First, the SQUID loop in the current device is asymmetric with $d=0.601$, reducing the overall vulnerability to flux noise (see Fig.~\ref{fig:acdc_sweetspots}). This vulnerability is further reduced by using a smaller SQUID loop of $470\,\mu\textnormal{m}^2$ (compared to the previous $1000\,\mu\textnormal{m}^2$ ), which is only possible because the mutual inductance to the flux line was significantly increased in this design. The higher mutual inductance also makes it possible to increase the total flux line attenuation from $10\,\textnormal{dB}$ to $26\,\textnormal{dB}$, considerably reducing the temperature of the white noise reaching the transmon. These adjustments improve the overall control over the transmon, thereby increasing the cavity control capabilities.

These hardware improvements significantly increase the usefulness of flux-tunable interactions for quantum information processing with bosonic modes. The enhanced interaction speeds, together with longer coherence and relaxation times, allow Lakeside to manipulate the Jaynes-Cummings Hamiltonian at a higher effective dimension and with greater precision than could be demonstrated with Somerset. From here on, I will apply these capabilities to a task that directly benefits from the large Hilbert space of cavities: creating superconducting qudits.

\section{Resonant qudit state preparation}

As superconducting quantum processors scale, problems with hardware overhead become increasingly prominent. Larger Hilbert spaces require more qubits, increasing the chip footprint. The extra control lines introduce heat load and decay channels, and their perimeter-limited routing makes them more susceptible to crosstalk~\cite{bjanin_dimensional_2016}. These issues motivate research efforts toward leveraging the multi-level structure of transmons as qudits~\cite{liu_performing_2023, nguyen_empowering_2024, wang_high_2025} ($d$-level systems), effectively increasing the computational power per unit cell of the circuit without increasing its footprint.

However, using transmons as qudits is not trivial. Higher levels are more prone to charge dispersion (see Eq.~\ref{eq:qp_dephasing}), which undermines the transmon resilience to charge noise. Coherently accessing up to $d=12$ levels requires fabricating devices with an especially high $E_J/E_C$ ratio~\cite{wang_high_2025}. This comes at the cost of either reducing the anharmonicity $|\alpha| = E_C$ or significantly increasing the frequency $\omega_T = \sqrt{8E_JE_C} - E_C$.

Microwave cavities are not bound by the same constraints. They have a virtually infinite number of discrete levels, with experimental demonstrations reaching Fock states up to 100~\cite{deng_heisenberg_2024} and 1024-photon cat states~\cite{milul_superconducting_2023}. Given the extraordinarily long lifetimes they can achieve~\cite{kim_ultracoherent_2025} and the possibility of using bosonic qudit codes~\cite{brock_quantum_2025}, microwave cavities are a promising platform for hardware-efficient higher-dimensional quantum computation.

Several works have explored qudit control in oscillators coupled to qubits, both theoretically~\cite{santos_universal_2005, strauch_quantum_2011, mischuck_qudit_2013, liu_constructing_2021} and experimentally~\cite{hofheinz_synthesizing_2009, huang_fast_2025}. These works build on the Law and Eberly protocol~\cite{law_arbitrary_1996}, which was initially designed for state preparation in the oscillator, but has since been extended to include the synthesis of arbitrary unitaries~\cite{mischuck_qudit_2013, liu_constructing_2021}. Its efficiency is limited by the leakage of the oscillator state outside of the qudit subspace, that is, to the $(d+1)$-th energy level. This leakage follows from the non-photon-selective nature of the control primitives, specifically the qubit-oscillator energy exchange through $H_{int} = g\left(\hat{a}^\dagger \sigma_- + \hat{a}\sigma_+\right)$. While ref.~\cite{huang_fast_2025} addresses this issue by implementing selective charge-activated sideband drives, the selectiveness is restricted to a small subspace of transitions, which can become a hurdle for general gates and oscillator states.

Lakeside is in an excellent position to explore an alternative approach to qudit control. Exploiting the strong nonlinearities of the resonant Jaynes-Cummings spectrum (Eq.~\ref{eq:jaynes-cummings}), all energy levels of the system can be accessed through photon-number-selective transitions. These operations provide the building blocks for universal control of the system, while leakage issues can be suppressed by adjusting the length of the operations. Moreover, the frequency tunability of the transmon can be used to suppress the coupling to the cavity during idle times, fully leveraging the high quality factor of the bosonic mode. In the following experiments, I will use Lakeside's unique advantages to implement the protocol of Ref.~\cite{strauch_resonant_2012}, demonstrating the state preparation and control of a qudit encoded in the microwave cavity.

The protocol is based on the connection between the resonant Jaynes-Cummings eigenstates, $|N\pm\rangle, \, \forall N\in\mathbb{N}^*$, and the eigenstates of the decoupled system $|N,s\rangle = |N\rangle\otimes|s\rangle$, with $s = g, e$ labeling the transmon state. On resonance ($\omega_a = \omega_T$), these states are related as 
\begin{align}
    |N+\rangle = |N, g\rangle + |N-1,e\rangle, \\
    |N-\rangle = |N, g\rangle - |N-1,e\rangle.
\end{align}
That is, the fully hybridized eigenstates $|N\pm\rangle$ contain $N$ photons, of which $N-1$ are stored in the cavity and one is equally shared with the transmon. The relative phase of the photon-hopping, labeled by $\pm$, determines whether the hybridization is bonding $\left(\textnormal{with eigenenergy } E_{|N-\rangle} = \hbar\omega_TN - \hbar g\sqrt{N}\right)$ or antibonding $\left(E_{|N+\rangle} = \hbar\omega_T+  \hbar g\sqrt{N}\right)$. 

This energy relation has practical consequences. If the transmon frequency is slowly shifted to lower values ($\Delta = \omega_a - \omega_T > 0$), the states of lower energy $|N-\rangle$ will adiabatically evolve into a state where the $N$-th excitation is in the lowest-energy mode, i.e., the transmon~\footnote{See, for example, Ref.~\cite{chamberland_theory_2014} for further explanation on adiabatic control.}. Conversely, the higher energy hybridization $|N+\rangle$ will evolve to a state where all $N$ excitations are stored in the high-energy mode, the cavity. In summary, adiabatically detuning the system to $\Delta > 0$ maps the eigenstates
\begin{align}
    |N+\rangle & \longrightarrow |N,g\rangle, \\
    |N-\rangle & \longrightarrow |N-1,e\rangle,
\end{align}
with the trivial vacuum state mapping $|0\rangle \to |0,g\rangle$. 

Therefore, the qudit control consists in first manipulating the system in the hybridized basis $|\varphi_0\rangle = \sum_0^{d-1}c_N|N+\rangle$, with eigenbasis decomposition given by the coefficients $c_N$, and then adiabatically connecting the $|N+\rangle$ states to the cavity Fock ladder, resulting in the preparation of a state $|\varphi_0\rangle \to |\varphi_\Delta\rangle = \sum_0^{d-1}c_n|N\rangle\otimes|g\rangle$.

\begin{figure}
    \centering
    \includegraphics[width=\linewidth]{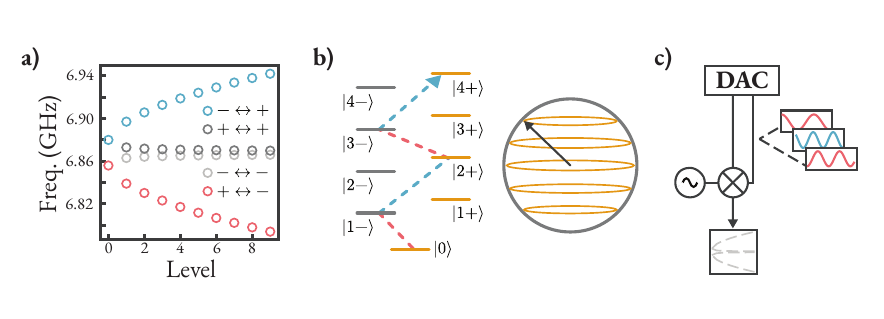}
    \caption[Microwave control of the Jaynes-Cummings spectrum]{\textbf{Microwave control of the Jaynes-Cummings spectrum.} (a) Transition frequencies for a resonant Jaynes-Cummings Hamiltonian with $\omega/2\pi=6.868\,\textnormal{GHz}$ and $g/2\pi = 12.07\,\textnormal{MHz}$. The horizontal axis indicates the starting number of photons. The labels indicate the sign of the starting state on the left, and that of the ending state on the right. b) Path connecting states $|0\rangle$ and $|4+\rangle$ using only diagonal transitions. The $|N+\rangle$ states $\left(0<N<d-1\right)$ are used to encode the $d$ levels of the qudit. c) Schematics of the microwave generation of multi-tone control. The baseband pulses are generated within the $\pm 250\,\textnormal{MHz}$ bandwidth of the DAC and then upconverted to the $\textnormal{GHz}$ range using a microwave mixer, allowing the amplitude and phase calibration of each tone.}
    \label{fig:qudit_control}
\end{figure}

The resonant Jaynes-Cummings spectrum is shown in Fig.~\ref{fig:qudit_control}. A transmon charge drive can manipulate the dressed states either via carrier (vertical) transitions $|N\pm\rangle \leftrightarrow |(N+1)\pm\rangle$, or sideband (diagonal) transitions~\footnote{Here, "sideband" refers to the charge-activated transitions that induce exchange of photons between the cavity and the transmon. Should not be confused with the flux-modulation sidebands discussed previously in this chapter.} $|N\pm\rangle \leftrightarrow |\left(N+1\right)\mp\rangle$. The latter can be divided into red, lower-frequency transitions $|N+\rangle \leftrightarrow |(N+1)-\rangle$, and blue, higher-frequency transitions $|N-\rangle \leftrightarrow |(N+1)+\rangle$. Figure~\ref{fig:qudit_control}a shows how these transitions are ideally distributed in frequency as a function of $N$ in the case of Lakeside. Compared to the carriers, the diagonal transitions are more spaced in frequency relative to each other, which contributes to their photon number selectivity. Moreover, they can connect any two states of the ladder, as shown in the path connecting $|0\rangle$ and $|4+\rangle$ in Figure~\ref{fig:qudit_control}b. Accordingly, I will restrict all control protocols to diagonal transitions and neglect the carrier transitions.

To control multiple transitions simultaneously, a multi-tone pulse is generated digitally and sent through a frequency upconversion chain at the DAC output (Figure~\ref{fig:qudit_control}c). With this scheme, it is possible to generate an arbitrary number of tones within the DAC bandwidth of $\pm250\,\textnormal{MHz}$ while controlling all their relative phases. Each tone is also individually calibrated with phase and gain adjustments to correct for mixer imbalance, creating a clean signal spectrum.

Using this setup, every transition up to a desired photon-number cutoff $d-1$ was individually calibrated in frequency and amplitude through standard spectroscopy and Rabi oscillation experiments. Since every transition entails a parity flip, the excitations were tracked by detuning the transmon to the strong dispersive regime ($\chi/2\pi = 1.6\,\textnormal{MHz}$) and using a standard parity measurement protocol. The ramp duration of the detuning pulse was calibrated by first preparing $|1+\rangle$, applying a flux pulse with variable ramp-up time and then measuring the transmon state $|e\rangle$. The optimal ramp time of $200\,\textnormal{ns}$ was chosen as the shortest time for which no excitations to $|e\rangle$ could be detected.

\begin{figure}
    \centering
    \includegraphics[width=\linewidth]{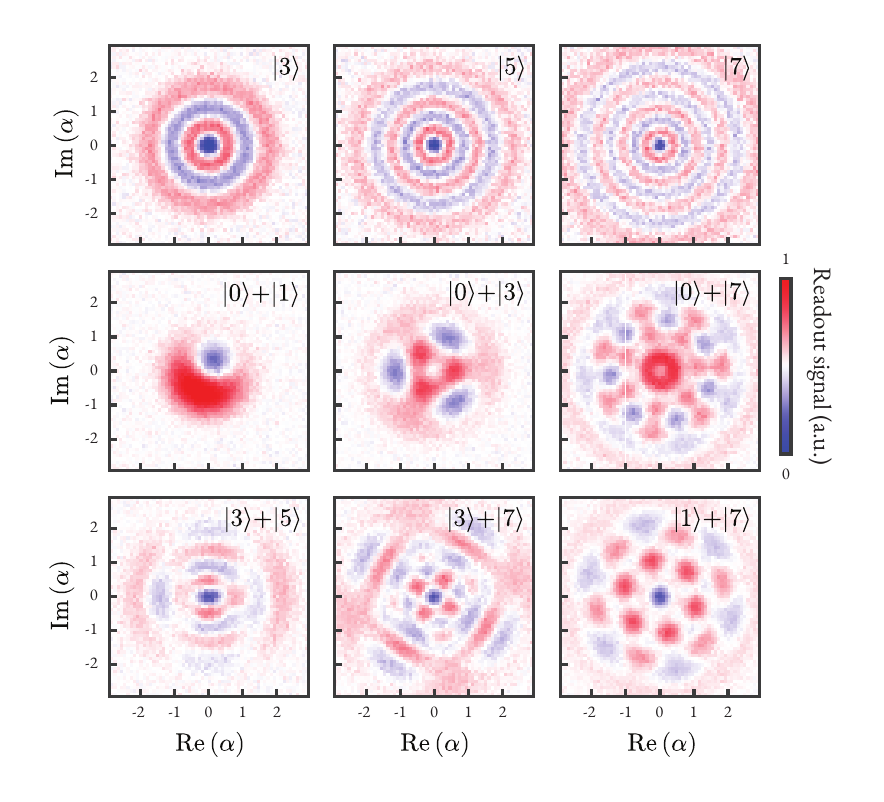}
    \caption[Qudit arbitrary state preparation]{\textbf{Qudit arbitrary state preparation.} Wigner plots showing diverse qudit states prepared by sequentially exciting diagonal transitions in the Jaynes-Cummings ladder.}
    \label{fig:qudit_preparation}
\end{figure}

I used this protocol to demonstrate arbitrary state preparation in the cavity with an analytical control sequence. The sequence consisted of activating the transitions sequentially using variable-amplitude, Gaussian-shaped pulses with $4\sigma$ duration, $\sigma \in \left[64, 96\right]\,\textnormal{ns}$. The pulse sequences were designed to create diverse cavity states, as shown in Figure~\ref{fig:qudit_preparation}, characterized with Wigner function tomography. Since the transmon does not have single-shot readout, in these experiments I calibrated the parity signal by comparing it to a $1\textnormal{D}$ Wigner measurement of cavity vacuum $|0\rangle$ measured beforehand. The Wigner plots closely resemble the target states in shape, though with a reduction in contrast. This indicates that the state preparation is mostly limited by decoherence.

\begin{table}[htp]
\centering
\begin{tabular}{c c c c}
\toprule
\addlinespace
Transition & $\omega_s/2\pi$ (GHz) & $\Omega_s/2\pi$ (MHz) & $\sigma$ (ns)\\
\addlinespace
\midrule
$|0\rangle \leftrightarrow |1+\rangle$ & $6.8809$ & $29.0$ & $68$ \\
$|0\rangle \leftrightarrow |1-\rangle$ & $6.8566$ & $25.5$ & $52$ \\
$|1-\rangle \leftrightarrow |2+\rangle$ & $6.8987$ & $22.3$ & $44$ \\
$|1+\rangle \leftrightarrow |2-\rangle$ & $6.8402$ & $18.9$ & $48$ \\
$|2-\rangle \leftrightarrow |3+\rangle$ & $6.9074$ & $24.1$ & $44$ \\
$|2+\rangle \leftrightarrow |3-\rangle$ & $6.8315$ & $19.4$ & $32$ \\
$|3-\rangle \leftrightarrow |4+\rangle$ & $6.9142$ & $25.9$ & $44$ \\
$|3+\rangle \leftrightarrow |4-\rangle$ & $6.8246$ & $22.0$ & $24$ \\
$|4-\rangle \leftrightarrow |5+\rangle$ & $6.9201$ & $25.5$ & $32$ \\
$|5+\rangle \leftrightarrow |6-\rangle$ & $6.8143$ & $22.7$ & --- \\
$|6-\rangle \leftrightarrow |7+\rangle$ & $6.9291$ & $24.5$ & --- \\
\bottomrule
\addlinespace
\end{tabular}
\caption[Calibration of diagonal transitions]{\textbf{Calibration of diagonal transitions.} The table shows the frequencies of each transition and their respective maximum Rabi frequencies $\Omega_s$. $\Omega_s$ is defined as the coefficient of the $\frac{\Omega_s}{2}|\left(N+1\right)\mp\rangle\langle N\pm|$ matrix element when activated through the transmon drive with maximum DAC output of $0.4\,\textnormal{V}$. The parameter $\sigma$ is the standard deviation of the Gaussian $\pi$ pulse calibrated for each transition, with total duration of $4\sigma$. The length of the $|5+\rangle \leftrightarrow |6-\rangle$ and $|6-\rangle \leftrightarrow |7+\rangle$ transitions were not optimized.}
\label{tab:cal_sidebands}
\end{table}

\begin{figure}
    \centering
    \includegraphics[width=\linewidth]{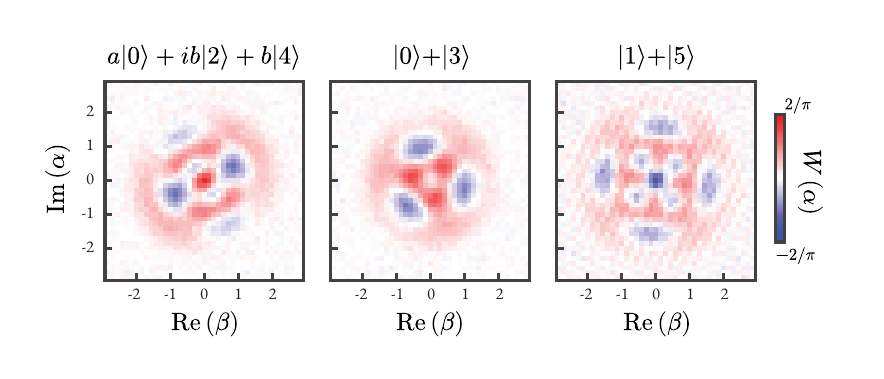}
    \caption[Optimized qudit state preparation]{\textbf{Optimized qudit state preparation.} For these three states, the excitation pulses were optimized to be as short as possible. The parity was rescaled to a $2\textnormal{D}$ Wigner function of vacuum obtained simultaneously, enhancing contrast reliability. For the state shown in the first plot on the left, the coefficients $a = \left(1/4\right)^\frac{1}{2}$ and $b = \left(3/8\right)^\frac{1}{2}$.}
    \label{fig:qudit_preparation_optimized}
\end{figure}

To better quantify these errors, I optimized each pulse to the shortest duration for which the error of a $2\pi$ oscillation is minimum (see Table~\ref{tab:cal_sidebands}). This results in a compromise between decoherence and leakage to adjacent transitions. The parity contrast calibration method was also modified: a $2\textnormal{D}$ Wigner function of vacuum was measured concomitantly with the prepared state and used as a reference. Simultaneous measurements account for low-frequency variations in the parity measurement protocol that can affect flux-tunable devices. The fidelities of each state are then estimated via density matrix reconstruction. Figure~\ref{fig:qudit_preparation_optimized} shows the results for three states: $\frac{1}{\sqrt{2}}\left(|0\rangle + |3\rangle\right)$ (fidelity of $93\pm 3\%$), $\left(\frac{1}{4}\right)^\frac{1}{2}|0\rangle + i\left(\frac{3}{8}\right)^\frac{1}{2}|2\rangle + \left(\frac{3}{8}\right)^\frac{1}{2}|4\rangle$ ($88\pm 3\%$), and $\frac{1}{\sqrt{2}}\left(|1\rangle + |5\rangle\right)$ ($87\pm 2\%$). The state fidelity and error were obtained from density matrix reconstruction using Bayesian inference.

The power of this protocol is in its simplicity. Preparing a different state is as straightforward as choosing the amplitudes and phases that will generate the desired superposition. In comparison, typical dispersive control methods require numerical optimization of the pulse waveform~\cite{deFouquieres_second_2011, ma_quantum_2021} or of the parameters of a predefined set of operations~\cite{eickbusch_fast_2022, krastanov_universal_2015} for every target state. Moreover, the intuitive nature of the protocol and the finite set of control parameters (one set of frequency and amplitude per transition) simplifies troubleshooting and optimization. Lastly, frequency-dependent distortions in the signal chain are compensated for when each tone is independently calibrated. See Table~\ref{tab:cal_sidebands}: the effective Rabi amplitude $\Omega_s$ of a transition depends on its frequency $\omega_s$. This dependence is not trivially taken into account in numerically-optimized waveforms, but is immediately incorporated in the presented method.

I have considered the simplest case in which the resonant control activates only one transition at a time. Multi-tone control, however, enables more complex evolutions that can be used to create arbitrary unitary qudit gates. That will be the topic of the next section.

\section{Multi-tone Hamiltonian engineering and Givens rotations}

\begin{figure}
    \centering
    \includegraphics[width=\linewidth]{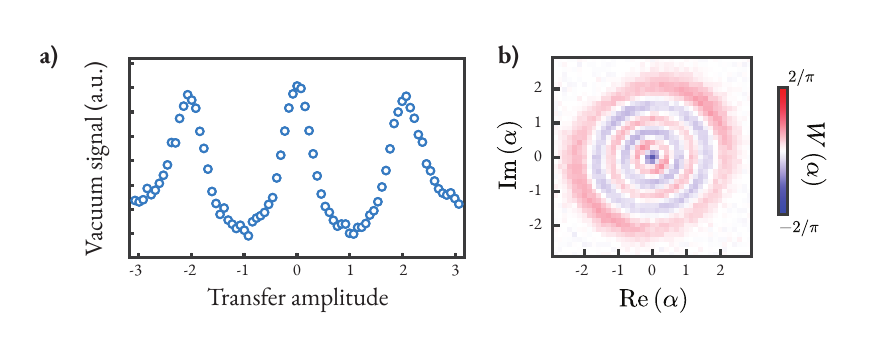}
    \caption[Multi-tone state transfer pulse]{\textbf{Multi-tone state transfer pulse.} a) Oscillations between states $|0\rangle$ and $|3+\rangle$ by activating the drive in Eq.~\ref{eq:jx_matrix}. The plot tracks the vacuum state population through number-splitting spectroscopy, which is minimum when the transfer to $|3+\rangle$ is maximum. b) Wigner tomography of Fock state $|5\rangle$ created from transfer from vacuum.}
    \label{fig:state_transfer}
\end{figure}

The resourcefulness of the resonant qudit control method can be significantly expanded by activating multiple transitions  simultaneously. For example, it is possible to construct optimal transfer operations between two levels. Consider any two Jaynes-Cummings eigenstates such as $|5+ \rangle$ and $|2+\rangle$. The string of intermediate states connecting them following $|5+\rangle \to |4-\rangle \to ... \to|0\rangle\to|1-\rangle\to|2+\rangle$ forms a virtual qudit with $d=7$ levels. Following Ref.~\cite{strauch_resonant_2012}, its angular momentum matrix $J_x$ can be built as
\begin{equation}
\label{eq:jx_matrix}
J_x =
\begin{pmatrix}
0     & A_1   & 0     & 0      & \cdots       & 0    \\
A_1   & 0     & A_2   & 0      & \cdots       & 0    \\
0     & A_2   & 0     & A_3    & \cdots       & 0    \\
0     & 0     & A_3   & 0      & \ddots & \vdots       \\
 \vdots & \vdots  & \vdots  & \ddots & \ddots & A_{d-1} \\
 0    & 0     & 0     & \dots      & A_{d-1} & 0
\end{pmatrix}
\end{equation}
with the weights
\begin{equation}
\label{eq:Jx_weights}
    A_n = 2A\frac{\Omega_1}{\Omega_n}\sqrt{n\left(d-n\right)},
\end{equation}
where the matrix is represented in the basis of the connecting states and $A$ is an overall amplitude. $J_x$ is the generator of displacements within this subspace and can create optimal state transfer between the two end states. To demonstrate this, I program a multi-tone pulse following the parameters in Table~\ref{tab:cal_sidebands} and Equation~\ref{eq:Jx_weights} to build a $J_x$ pulse between $|0\rangle$ and $|3+\rangle$. Figure~\ref{fig:state_transfer}a shows the post-detuning Fock $|0\rangle$ population for a $660\,\textnormal{ns}$ pulse with variable amplitude, showing back-and-forth transfer. At the point of minimal $|0\rangle$ population, the transfer to Fock state $|3\rangle$ is maximized.

The population transfer can be numerically optimized. Instead of using constant pulse envelopes as in Eq.~\ref{eq:Jx_weights}, each drive is programmed with a time-varying complex amplitude $A_n(t) + iB_n(t)$ such that
\begin{equation}
\begin{aligned}
    A_n(t) &= \sum_{k=1}^M a_n^{(k)}\left[ 1- \cos(2\pi k t/T)\right], \\
    B_n(t) &= \sum_{k=1}^M b_n^{(k)}\sin(2\pi k t/T).
\end{aligned}
\end{equation}
I use $M=3$, meaning each transition has $6$ coefficients $a_n^{(k)}$, $b_n^{(k)}$ that are optimized using gradient descent to maximize the fidelity of the final state compared to an ideal transfer. Note that for $t=0,T$, all amplitudes are zero and increase smoothly, which softens the edges of the pulse and restricts its spectral band. This method was applied to generate a state transfer between $|0\rangle$ and $|5\rangle$. Figure~\ref{fig:state_transfer} shows the Fock $|5\rangle$ generated with a $800\,\textnormal{ns}$ pulse with a fidelity of $75\pm2\%$. The slight asymmetry of the state indicates the presence of coherent errors from the optimization.

\begin{figure}
    \centering
    \includegraphics[width=\linewidth]{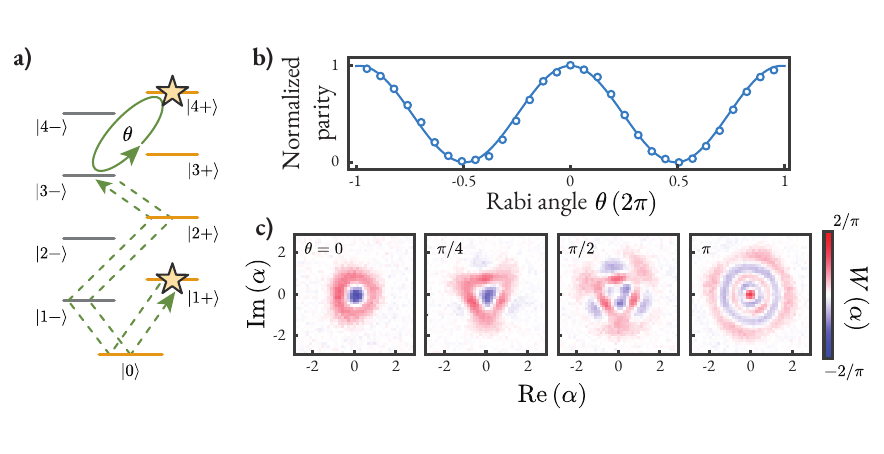}
    \caption[Givens rotations on a $d=5$ qudit.]{\textbf{Givens rotations on a $d=5$ qudit.} a) Schematics of the pulse sequence implementing Givens rotations between $|1+\rangle$ and $|4+\rangle$. This is equivalent to an operation between cavity Fock states $|1\rangle$ and $|4\rangle$ due to the adiabatic connection between these states. b) Normalized parity measurement of the final cavity state initially prepared in $|1\rangle$ after applying $G_{1,4}(\theta)$ for variable rotation angle. c) Wigner tomography plots of the final cavity state after detuning for selected values of $\theta$.}
    \label{fig:givens}
\end{figure}

This optimal state transfer was used to implement rotations in the two-dimensional subspace of two cavity qudit levels $|N_A+\rangle$, $|N_B+\rangle$, known as Givens rotations~\cite{brennen_criteria_2005}. Consider the two qudit levels $|1+\rangle$, $|4+\rangle$, and the negative-sign level that is right below the latter, i.e., $|3-\rangle$. As done in the previous experiment, I optimized a $J_x$ pulse for the virtual qudit between $|1+\rangle$ and $|3-\rangle$:
\begin{equation}
    |1+\rangle \leftrightarrow|0\rangle \leftrightarrow|1-\rangle \leftrightarrow|2+\rangle \leftrightarrow|3-\rangle.
\end{equation}
Denoting the full state transfer as $R^{|1+\rangle}_{|3-\rangle}(\pi)$, an arbitrary-angle Givens rotation can be implemented with the sequence (see Fig.~\ref{fig:givens}a)
\begin{equation}
\label{eq:givens}
   G_{1,4}(\theta,\phi) =  R^{|1+\rangle}_{|3-\rangle}(\pi)R^{|3-\rangle}_{|4+\rangle}(\theta, \phi)\left[R^{|1+\rangle}_{|3-\rangle}(\pi)\right]^\dagger,
\end{equation}
where $R^{|3-\rangle}_{|4+\rangle}(\theta,\phi)$ is the rotation corresponding to the $|3-\rangle \leftrightarrow|4+\rangle$ transition, and $\theta, \phi$ are the rotation angle and the phase of the operation. In the $\{|1+\rangle,|4+\rangle\}$ subspace,
\begin{equation}
    G_{1,4}(\theta,\phi)|1\rangle =  \cos\left(\theta/2\right)|1\rangle
    + e^{i\phi}\sin\left(\theta/2\right)|4\rangle.
\end{equation}
This sequence effectively implements a rotation between two levels of the cavity qudit without disturbing any of the intermediate states, since the second application of $R^{|1+\rangle}_{|3-\rangle}(\pi)$ reverts the changes on the spectating levels. 

The Givens rotations were demonstrated by first preparing state $|1+\rangle$, and then applying the gate sequence of Eq.~\ref{eq:givens} with a total duration of $1536\,\textnormal{ns}$. Figure~\ref{fig:givens}b shows the normalized parity measurement of the cavity as a function of the rotation angle $\theta$, demonstrating continuous, coherent oscillations. Figure~\ref{eq:givens}c shows the Wigner tomography of the cavity state for four points, $\theta=0,\,\frac{\pi}{4},\,\frac{\pi}{2},\,\textnormal{and}\,\pi$, respectively. The quality of the states suffer from the length of protocol relative to the transmon $T_2^*$, leading to fidelities of $79\pm2\%$, $76\pm2\%$, $71\pm2\%$, and $67\pm2\%$. Still, the tomography demonstrates the capability of this protocol of executing arbitrary Givens rotations, which are a sufficient condition for universal qudit control~\cite{brennen_criteria_2005}.

The multi-tone control enabled faster state transfer and Givens rotations in the cavity-encoded qudit. In principle, it can support even more complex operations. Numerically optimizing each tone's coefficients can target gates such as a qudit generalization of a Hadamard gate~\cite{wang_qudits_2020}. However, a detailed implementation of these gates is left for future work. 

\section{Discussion and future directions}


In this chapter, I have identified and solved several problems that limited the on-demand tunability of light-matter interactions. These improvements were made possible by the development of a new bosonic cQED design, Lakeside, which combined familiar hardware elements with an unconventional hybrid $2\textnormal{D}-3\textnormal{D}$ architecture. Specifically, this solution improved on:
\begin{itemize}
    \item \textbf{Lifetime}: The cavity $T_1$ was increased to $590\pm40\,\mu\textnormal{s}$, corresponding to a total $Q=2.6\times 10^7$. This value approaches the lifetimes of similar cavities in non-tunable circuits, and has clear paths for improvement by enhancing the quality of the seam.
    \item \textbf{Tunable interactions}: The broadband on-chip flux line can deliver fast-switching step pulses and flux pumps to the transmon. This flux control provided three new operational capabilities: modulation of the coupling strength $g$, activation of parametric resonances, and access to number-selective transitions between the transmon and the cavity.
    \item \textbf{Heat load management}: The flux line in Lakeside incurs no greater heat load than in typical SQUID-transmons in planar architectures, which can comfortably host several flux-tunable units~\cite{vigneau_quantum_2025, besedin_realizing_2025}. The device did not require any additional thermalization strategies such as copper braids to reach a baseline excited-state population $\leq2\%$, and no heating was detected during the experiments presented in this chapter. 
    \item \textbf{Compactness}: The flux line, readout and filters are integrated in a compact $2\textnormal{D}$ architecture. This results in a footprint comparable to that of non-tunable circuits even with the additional functionalities.
    \item \textbf{Reproducibility}: This design requires no additional steps beyond the nanofabrication of the superconducting circuit and the machining of the aluminum package. There are no manual fabrication step that could introduce significant performance uncertainty, as was the case with the magnetic hose. This upgraded solution is therefore easily reproducible and provides a relatively low entry barrier for other researchers.
\end{itemize}

These results granted access to quantum information processing operations of much higher quality and of novel capabilities compared to the previous devices. I used the flux tunability to drive number-selective transitions between the transmon and the cavity in the resonant regime of the Jaynes-Cummings Hamiltonian. These transitions were employed for preparing arbitrary qudit states to $d=8$ cavity levels, and for applying continuous Givens rotations between Fock states $|1\rangle$ and $|4\rangle$. This number-selective control was proved to be a useful tool for operating qudits in microwave cavities, enabling access to a Hilbert space of higher effective dimension without hardware overhead.

Improvements to the qudit control protocol should focus on the limited transmon coherence time $T_2^*$. This is a known drawback of SQUID-transmons, but fortunately there are ways to mitigate it. For example, flux modulation can be used to engineer dynamical sweet spots, as explained in Chapter~\ref{ch:flux}. Another approach is to use a flux-tunable coupler circuit to mediate the interaction between a cavity and a fixed-frequency transmon. Such couplers are designed to toggle interactions without introducing significant operational errors~\cite{bialczak_fast_2011, cai_multiplexed_2025, an_zz_2025}.

There are more general aspects of this device that could be improved in future iterations. One of them is the flux line mode. The low-pass filters on the flux line create a reflective boundary for frequencies $>4\,\textnormal{GHz}$, which leads to a resonant mode in the coil region. This mode lies relatively close in frequency to the transmon, at about $\gtrsim8\,\textnormal{GHz}$, and they couple strongly with a simulated factor of $g/2\pi\approx200\,\textnormal{MHz}$ due to their close proximity. While at first sight this mode may seem problematic: it crowds the frequency spectrum and might lead to undesired dynamics. However, it can be adapted and repurposed as a useful component of the circuit. Preliminary simulations show this coupling can be reduced to $g/2\pi\approx30\,\textnormal{MHz}$, which is well suited for dispersive readout. It is then possible to remove the readout resonator entirely, couple the readout line to the flux line mode, and use the latter to measure the transmon state. This adaptation would simplify the frequency spectrum of the device and further reduce its footprint.

During the examination process of this thesis, it came to my knowledge a recently available preprint reporting related results by Ziqian Li et al.~\cite{li_cascaded_2025}. In this work, they demonstrate a broad bandwidth on-chip flux transformer coupled to high-$Q$ multimode cavity. This is a similar design to the one described in Ref.~\cite{maiti_controlling_2025}, mentioned in Sec.~\ref{sec:flux_tunability_challenge}. Even though both of our works strive for similar objectives, we have developed independent solutions that leverage different design principles. These solutions are complementary and could be mutually beneficial. For example, the device reported in this Chapter leverages CPW lines to provide a more compact circuit, allowing integration of the readout resonator and cascading filters on the same chip. On the other hand, the device reported in Ref.~\cite{li_cascaded_2025} employs a microwave balun to prevent common-mode driving of the SQUID, which is essential for high-amplitude flux modulation without ionizing the transmon circuit. Future iterations of Lakeside could use similar strategies to further improve the parametric modulation capabilities.

Future research can also focus on applications for which Lakeside offers unique advantages. One such direction is engineered dissipation~\cite{harrington_engineered_2022}. Although this topic was not explored in this thesis, the readout resonator shown in Figure~\ref{fig:lakeside} is specially suitable as a gateway to a lossy environment. The CPW architecture allows the $2\textnormal{D}$ resonator to couple strongly to the external line. Moreover, the density of states of the environment can be reshaped using filters to protect the transmon and cavity modes. The interaction can then be activated with the use of flux-modulation transmon sidebands, whose frequency and strength are adjustable in-situ and in real time. Combined with the long lifetime of the cavity, the highly engineerable losses of the readout resonator are key to harnessing dissipation as a resource for quantum information processing.

There are many ways in which controlled dissipation can be exploited. Refs.\,\cite{kim_fast_2025, magnard_fast_2018, zhou_rapid_2021, vivek_demand_2024} use flux-modulation sidebands to couple cavities and transmons to an engineered bath on demand, which acts as an unconditional reset. Refs.\,\cite{lacroix_fast_2025, kim_fast_2025} use a similar setup but target the enhanced losses at the $|e\rangle\leftrightarrow|f\rangle$ transmon transition to reduce the leakage from fast qubit gates. Taking a different route, Refs.~\cite{reglade_quantum_2024, putterman_hardware_2025} use two-photon dissipation on a bosonic mode to stabilize cat qubits with suppressed bit-flip errors. The single-photon loss of the cavity, which is about $60\,\mu\textnormal{s}$ for \cite{putterman_hardware_2025}, determines the dominant phase-flip errors. One can therefore naively expect that a Lakeside-like device, with its $\approx 600\,\mu\textnormal{s}$ cavity lifetime, could suppress these errors by up to one order of magnitude. Moreover, engineered losses can be used as programmable ``entropy drains'' for autonomous quantum error correction of a logical bosonic qubit beyond break-even, as demonstrated in two recent works~\cite{ni_autonomous_2025, sun_extending_2025}. Beyond quantum computing, driven-dissipative bosonic systems such as Kerr oscillators~\cite{casteels_critical_2017} and single-atom lasers~\cite{sokolova_single_2021,  sokolova_overcoming_2023} can be used as critical quantum sensors for single-photon detection~\cite{fernandez_quantum_2017, dicandia_critical_2023, beaulieu_criticality_2025, petrovnin_microwave_2024}. These are only a few examples of the potential of a device in which long lifetimes and strong, tunable losses coexist.

\chapter{Conclusion}
\label{ch:concl}



Circuit quantum electrodynamics embodies foundational physical models in its operation. From circuit elements wired into artificial atoms and embedded in electromagnetic fields, these systems recreate the dynamics of light-matter interactions on a chip to be used in the laboratory. But circuit QED is more than a mere copy of these interactions. With the highly adaptable properties of superconducting circuits, it is possible to reach beyond the regimes accessible to natural systems, supporting the exploration of the fundamental physics of quantum information~\cite{preskill_physics_2022}.

However, much research remains necessary to realize the full potential of circuit QED. Real quantum systems are challenged by decoherence emerging from interactions with the environment. So control mechanisms, which are essentially engineered environments, also contribute to deteriorate the information they manipulate. This conundrum places circuit QED in a fine balance between control and protection, and the scales often tip toward sacrificing operational capabilities in favor of longer lifetimes. Overcoming these limitations and achieving protected control is essential to progress toward more advanced applications of quantum information processing.

One such point of friction is flux control. Flux lines can inductively couple to superconducting circuits and modify their properties in situ through current bias. The system's eigenstates and eigenvalues are manipulated in real time, introducing an extra set of dynamics that complement those of charge lines. This control significantly broadens the ability of superconducting circuits to employ light-matter interactions to explore new physics. However, flux control comes at a high cost. Flux-sensitive circuits are more vulnerable to noise, which can be carried by the line itself. And the line introduces additional decay channels, limiting the maximum attainable photon lifetimes. Due to these constraints, flux control is often restricted to niche applications or not used at all, curbing the range of dynamics accessible to circuit QED systems.

Flux control imposes a particularly stringent limit on microwave cavities. Cavities are attractive circuit elements due to their high-dimensional Hilbert spaces, and, by tailoring their resonating modes, they can reach especially long photon lifetimes. Together, these properties make them a valuable resource for quantum information processing, either in emulating bosonic dynamics or in encoding qubits for computational tasks. However, coupling them with flux-tunable circuits causes these advantages to collapse. The dispersed fields of three-dimensional cavities couple strongly to the leakage modes of flux lines. With the photon lifetimes compromised, multi-photon states become short-lived, and the large Hilbert space of the cavity becomes practically unusable.

Not surprisingly, bosonic circuit QED has developed for nearly two decades with few applications of flux control. While efforts in combining cavities with flux have re-emerged in the past couple of years~\cite{chapman_high_2023, lu_high_2023}, every step forward faces an uphill battle against cavity leakage, requiring researchers to draw from flux line architectures with specialized capabilities. Overcoming this barrier would equip the field of bosonic circuit QED with a powerful tool to explore the physics of bosonic systems.

In this thesis, I took one more step in this direction. I provided solutions to break the trade-off between flux control and lifetime, demonstrating dynamical light-matter coupling between tunable transmons and protected cavities with direct application to several aspects of quantum information processing.

Each chapter built toward this goal. In Chapter~\ref{ch:bosonic}, I explained the fundamental principles of bosonic circuit QED, focusing on the theoretical and experimental aspects of microwave cavities that were critical to achieving the main results. More importantly, I showed how light-matter interactions are a central element of this field -- it is by strongly coupling to an artificial atom, the transmon, that the Hilbert space of cavities can be fully utilized. The chapter also provided an overall introduction to the field, complementing excellent reviews that already exist in the literature~\cite{blais_circuit_2021, krantz_quantum_2019} with the particular focus of my research.

Chapter~\ref{ch:flux} moved further into the central problem addressed by this thesis. I showed why flux tunability is a valuable resource for quantum information processing: the two central parameters of transmon-cavity interactions, i.e., the detuning $\Delta$ and coupling $g$, can be manipulated in situ by a flux source. The theoretical description shows that implementing this control requires broadband flux lines that are capable of delivering fast-switching step pulses as well as radio-frequency pumps, setting goals for the hardware. Most importantly, in this chapter I analyzed the problem of flux line leakage in $3\textnormal{D}$ superconducting circuits, and how that directly affects the lifetimes of microwave cavities. This discussion presented the challenge to be overcome in the subsequent chapters.

In Chapter~\ref{ch:somerset}, I presented the first solution exploiting fast flux tunability for bosonic quantum information processing. In this work, I used a magnetic flux hose to bias a SQUID-transmon coupled to a microwave cavity. The device was optimized to maximize the cavity lifetime while retaining a minimum coupling with the transmon. This resulted in an experimentally measured cavity quality factor of $Q=9\times10^6$. To the best of my knowledge, this represents an improvement by a factor of $7.5$ over previously published results featuring fast flux tunability~\cite{gargiulo_fast_2021}. The hose provided fast-switching flux control to adjust the transmon-cavity detuning $\Delta$. I demonstrated the usefulness of this capability for quantum information processing by switching between light-matter interaction regimes and mixing techniques that are otherwise mutually exclusive within the same experiment. The resonant regime was used for the fast preparation of Fock states, the strong and weak dispersive regimes for Wigner and characteristic function tomographies, and finally, the transmon was dynamically detuned to show the protection of the cavity state against nonlinear distortions. These results provided a proof-of-principle demonstration for real-time tunable interactions with protected cavities, and revealed the limitations to be addressed in the next iterations.

Further improvements proved challenging . My main goals were to achieve broadband control to tune the light-matter coupling factor $g$ in situ, and to improve the quality of the transmon to access higher-dimensional cavity states with higher precision. Moreover, since the manufacturing of the hose was difficult, I intended to develop a more reproducible solution. The most reasonable approach was to use on-chip flux lines, despite their known challenges. The first attempt at such flux source was designed to use spring-loaded pins to connect the current source to the flux line inside the waveguide (Appendix~\ref{append:pogopin}). The line featured a low-pass filter and a Purcell notch filter that would block the cavity energy from leaking. However, this device faced prohibitive mechanical issues with the connecting pins, which could not provide a stable contact with the thin film of the line. For this reason, this design was abandoned. Nevertheless, it showed that, at least in simulation, the on-chip line could be efficiently filtered to prevent cavity leakage.

Chapter~\ref{ch:lakeside} built upon the lessons learned from the past two projects. I developed a device in which the cavity and the transmon were implemented in a typical $3\textnormal{D}$ stripline geometry, allowing the cavity to reach a high quality factor of $Q = 2.6\times10^7$. In contrast, the on-chip flux line and the readout resonator were implemented in a $2\textnormal{D}$ coplanar waveguide geometry, allowing heavy filtering of these elements to protect the circuit from leakage while offering strong control. Instead of using off-plane connector pins, the flux line was wirebonded to the current source, providing a stable and reproducible connection. With this hybrid architecture, I demonstrated two valuable control primitives for quantum information processing. First, the coupling strength $g$ was modulated by high-frequency flux pumps, and could even be switched off at $g_{eff} = 0$. Flux pumping was also used to drive parametric vacuum Rabi oscillations between the transmon and the cavity with adjustable strength. I then used the frequency control to tune the system into the resonant Jaynes-Cummings regime, enabling access to number-selective transitions in the hybridized system. These transitions were used for arbitrary state preparation in a cavity qudit, as well as for implementing arbitrary Givens rotations between two cavity levels. The results from this chapter provide a possible path for the implementation of exact universal control for bosonic quantum information processing.

These results demonstrate several advances in the tunability of light-matter interactions for bosonic quantum information processing. They provide a high-quality, reproducible solution that can be readily implemented and further improved by other researchers. These results may stimulate developments of flux-tunable systems and their use for novel applications. Throughout this text, I have identified several topics that could benefit from the particular advantages of the presented solutions. Especially, building driven-dissipative systems for quantum critical sensing, for the creation of exceptional points and parity-time symmetric systems, and for the stabilization of quantum error correction codes. These applications can directly benefit from the high $Q$ factor of the cavity, as well from its strong tunable coupling with the environment.

Finally, this thesis has fulfilled its stated goal. By delivering an important capability for superconducting circuits, it provides a stepping stone for the future investigation into novel physics in synthetic light-matter systems.

\appendix
\chapter{Delivering on-chip current bias with pogo pins}
\label{append:pogopin}

As mentioned in the main text, developing a protected on-chip flux line required iterative improvements and experimentation with different architectures. 

The first of such attempts had two particularly new features compared with the magnetic hose. First, it relied on out-of-plane spring-loaded (pogo) pins connecting to the thin film that would carry the current bias on the chip. Second, the flux line was imbued with a thin film low-pass filter to block the propagation of cavity and transmon fields. These two features required thorough consideration of mechanical stability, heating and lifetime in attempts to outperform the results of Chapter~\ref{ch:somerset}.

In this appendix, I want to briefly describe this research route, and explain why it was eventually abandoned. The focus is to highlight what did not work and why, guiding possible developments in this direction. Also, I will point out the positive aspects that were carried out to the subsequent project described in Chapter~\ref{ch:lakeside}.

\section{Design overview}

\begin{figure}
    \centering
    \includegraphics[width=\linewidth]{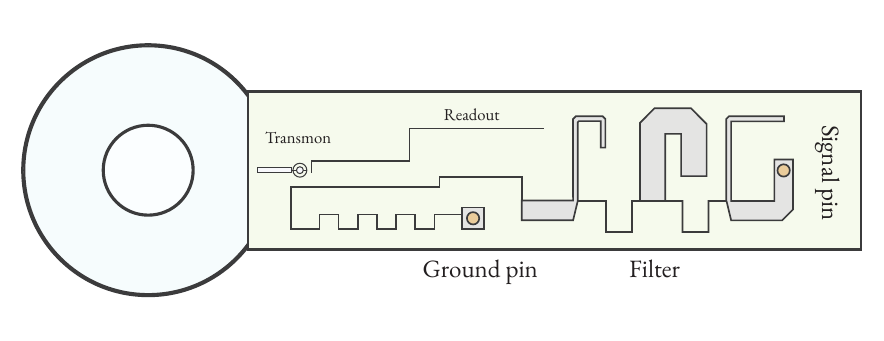}
    \caption[Circuit diagram]{\textbf{Circuit diagram.} The concentric transmon is placed closer to the cavity, and is connected to a short extension that increases their coupling. The flux line can be divided in two main parts: the $3$rd-order lowpass Butterworth filter, placed on the right side of the circuit and connected to the signal pin (shown as an yellow circle), and the flux delivery section, shown in the bottom left of the circuit and connected to the ground pin. The width of the line is decreased to reduce crosstalk with the readout and to increase mutual inductance with the SQUID loop.}
    \label{fig:fullenjoy}
\end{figure}

Figure~\ref{fig:fullenjoy}a shows the top view of the circuit. The cavity is the standard $\lambda/4$ coaxial stub resonator, and the chip is inserted through a perpendicular waveguide as usual. The chip contains a transmon close to the cavity, a readout resonator connecting to the transmon in one of its ends, and the flux line, which is connected to a parallel-stub filter occupying the back of the chip. Let us go through a few of these elements.

\subsection*{Concentric transmon}

In attempts to increase the coherence times of the transmon, we use a concentric transmon design~\cite{braumller_concentric_2016}. This circuit uses a gradiometric SQUID, which is a two-loop element that only responds to the difference of flux acting on each loop. The improved coherences come from the device insensibility to environment noise that acts equally on both loops.

The protection against noise allows the SQUID to have large areas ($56\,000 \,\mu\textnormal{m}^2$ for each loop), and thus couple more strongly with the flux line. We trade the flux strength for a reduced capacitive coupling between the transmon and the flux line, placing them at a relatively large distance of $350\,\mu\textnormal{m}$. This results in a flux tunability of $3.9 \,\Phi_0/\textnormal{mA}$.

The concentric transmon has a small electric dipole by design. However, that becomes a disadvantage here: it reduces the capacitive coupling with the cavity and the readout. To recover reasonable coupling factors ($g/2\pi\approx 12\,\textnormal{MHz}$ with the cavity), the transmon must be very close to the cavity and to the readout. This unfortunately enhances the Purcell decay of the cavity through the readout pin.

\subsection*{Flux line and filters}

The flux line can be divided in two parts: the low-pass filter, located at the back of the waveguide, and the ``coil'', which consists of the thinner section that inductively couples to the concentric transmon.

The low-pass filter is a $3$rd-order Butterworth design implemented as a network of parallel open stubs~\cite{pozar_microwave_2011}. The widths of each line section are designed to amount to a $50\,\Omega$--response that is impedance-matched to the source, connected through a pogo pin. The line impedances strongly depend on the distance between the line and the ground plane defined by the waveguide. To keep these impedances uniform across the structure, the cross-section of the waveguide is changed from a circular to a rectangular shape.

The coil section is made thin to increase the inductive coupling to the transmon at the same time that it reduces the capacitive coupling to other elements. The coil is shorted to the waveguide by a second pogo pin, which defines an implicit Purcell band-stop filter at the cavity frequency (see Figure~\ref{fig:flux_line_strategies}b). The second pogo pin is placed in the middle of the waveguide to prevent crosstalk between the coil section and the back of the filter, which could compromising the filtering strategies.

\begin{figure}
    \centering
    \includegraphics[width=\linewidth]{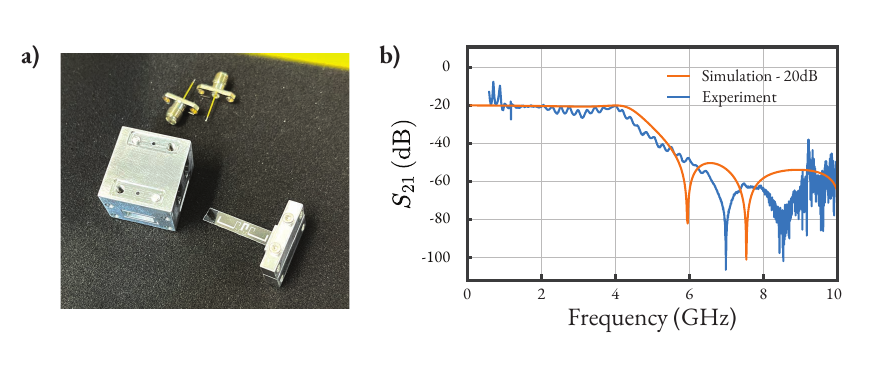}
    \caption[Experimental low-pass filter characterization]{\textbf{Experimental low-pass filter characterization.} a) Picture showing the aluminum package, the filter printed on a sapphire chip held by a clamp, and the two pogo pins connected to SMA ports. b) Comparison between simulation and experimental $S_{21}$ data, showing the expected cutoff at $4\,\textnormal{GHz}$. The experimental $S_{21}$ curve is corrected to compensate for the HEMT on the output line and the $20\,\textnormal{dB}$ total attenuation.}
    \label{fig:filtertest}
\end{figure}

The $S_{21}$ of the low-pass filter was independently tested in a separate device (see Figure~\ref{fig:filtertest}). The response was quite close to the simulations, validating the design methodology.

\subsection*{Pogo pins}

\begin{figure}
    \centering
    \includegraphics[width=\linewidth]{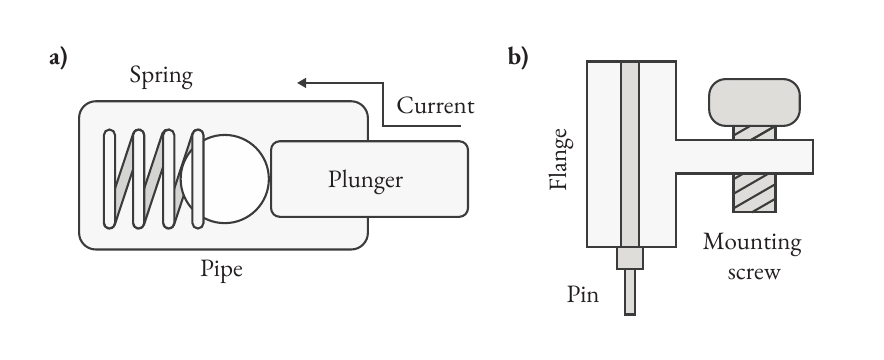}
    \caption[Bespoke spring-loaded superconducting pins]{\textbf{Bespoke spring-loaded superconducting pins.} a) Inner workings of the pogo pin. The current flows from the flux line to the plunger and then to the pipe through $\textnormal{Sn}$-coated surfaces. The spring is pushes the plunger through a insulating ball to prevent from flowing through resistive paths. b) Flange grounding the superconducting pogo pin. The flange is made of superconducting aluminum and has a socket where the pin is secured. The pin has a wider ring that controls the insertion, providing a well-defined electrical length between the flux line and the ground.  }
    \label{fig:pogo}
\end{figure}

Acquiring suitable pogo pins was challenging. Seiken Co., Ltd. has a broad off-the-shelf collection of probes, but all of them are made of conductive materials (e.g. beryllium copper). The pogo pin shorting the coil to the waveguide must be superconducting, since the cavity field has a relatively strong participation in that section. We reached out to Seiken for a bespoke solution and co-developed a pogo pin with a tin ($\textnormal{Sn}$) coating of $0.5-2\,\mu\textnormal{m}$. The critical temperature of the superconducting phase transition of Sn is $3.72\,\textnormal{K}$~\cite{wittig_die_1966}, so the pin will ideally become lossless when mounted at the mixing chamber plate. Moreover, the coating is done on the inner and outer surfaces of both the pipe and the plunger (see Figure~\ref{fig:pogo}a), so the conductive path is perfectly superconducting.

The pin must also be grounded to the waveguide with a well-defined electric path. We create an aluminum flange that can host the pogo pin and be mounted onto the outer surface of the device. The pin has a small ring which can be used to precisely determine its insertion into the flange (see Figure~\ref{fig:pogo}b).

\section{Discussion and experimental tests}

The simulated $Q$ factors are shown in Table~\ref{tab:fullenjoyQs} for each element and for each driving port. There are a few interesting results. The double filtering leads to a very low cavity leakage through the flux line, with $Q$ on the order of $10^{12}$. The second most relevant loss channel for the cavity is the readout pin. While not necessarily a surprise, the $Q= 10^9$ is relatively low considering the readout resonator is not too coupled at $Q= 1.5\times 10^4$. This ratio was obtained with extensive optimization of the readout location and shape, reason for which it is made thin as the flux line and has a meandering path to avoid coupling to other elements. Overall, I attribute this struggle to isolate the elements to a high level of crosstalk due to the proximity between the circuits required by the concentric transmon and the overall stripline architecture, which has more spread-out fields.

The overall simulations are promising, and the filter data in Figure~\ref{fig:filtertest}b indicates the pogo pins can successfully deliver AC signals to the aluminum thin film. However, the first problem appeared when trying to DC bias the flux line. The tests led to unexplainably high heating.

Upon warm-up, the chip was inspected, and we found out the pogo pins were severely scratching the aluminum films and deteriorating the connection. The current was possibly flowing through constricted paths and creating hot spots that would break the superconductivity of the film.

\vspace{0.5cm}
\begin{table}[htp]
\centering
\begin{tabular}{c c c c}
\toprule
\addlinespace
Element & Readout drive $Q$ & Flux line $Q$ & Cavity drive $Q$   \\
\addlinespace
\midrule
\addlinespace	
Transmon & $7\times 10^7$ & $3\times10^9$ & $3\times10^{10}$ \\
\addlinespace
Cavity & $10^9$ & $10^{12}$ & $6\times10^7$  \\
\addlinespace
Readout & $1.5\times10^4$ & $10^6$ & $10^8$ \\
\addlinespace
\addlinespace
\bottomrule
\addlinespace
\end{tabular}
\caption[Simulated $Q$ factors of each circuit element to each control line]{\textbf{Simulated $Q$ factors of each circuit element to each control line.} The values are obtained in finite-element simulations by connecting only one terminal at a time while opening the other lines.}
\label{tab:fullenjoyQs}
\end{table}

To try and solve this issue, the next devices were fabricated using tantalum ($\textnormal{Ta}$) films. These are more resilient and less prone to scratching. For the same reason, they require a stronger contact force to establish a conductive path. This led to another problem: the connection was unstable and highly dependent on mounting. The quality of the connection was monitored by tracking the film resistance using a common multimeter. It was noticed that, even after mounting successfully, the connection could still deteriorate while wiring the device inside the dilution refrigerator. Increasing the pogo pin compression did not solve the problem. In fact, it would cause lateral buckling of the pins, which would slide along the chip and reduce the contact with the film. Moreover, it is reasonable to assume this increased force was causing the chip (which is clamped only on one side, similar to a springboard) to bend down, further deteriorating the connection with the pins.

\section{Conclusion}

The mechanical instabilities and the spurious heating where decisive factors for abandoning this design. There were attempts to solve this issues, for example by placing the pogo pins inside dieletric holders to prevent buckling and by clamping the chip on both sides. However, there was still no concrete evidence this would solve the heating issue. Prof. Y. Nakamura (personal communication, June 19, 2024), who was also working on DC signal delivery through pogo pins at the time, advised that solving this problem could require a more complex solution involving multi-layered thin films. So, instead of insisting on solving the pogo pin problem, I decided to look for designs that could use wirebonding instead.

There were still many interesting takeaways from this device which were carried over to the next iteration. For example, the high cavity $Q$ factor achieved in simulation. In practice, this value is expected be much lower, limited by other loss channels such as surface losses. But the exceedingly high $Q=10^{12}$ relative to the flux line indicates that strong control lines are not incompatible with a strongly protected cavity. For that reason, the idea of explicitly filtering the flux line was brought back in Chapter~\ref{ch:lakeside}.

Overall, this design had many concepts that were physically dissonant: concentric transmon requiring large electric dipole; many stripline elements on the same chip requiring low crosstalk; a narrow stop-band filter relying on the manual mounting of the ground pin. These internal conflicts could never be resolved, only managed into narrow operational sweet spots. It was necessary to come up with a design that would attend to the hardware prerequisites more harmoniously. That was the principle leading to Lakeside and its hybrid $2\textnormal{D}-3\textnormal{D}$ architecture. 

\bookmarksetup{startatroot}
\printbibliography[heading=bibintoc]

\appendix

\end{document}